\documentclass[onecollarge]{svjour2}
\smartqed  
\usepackage{graphicx}
\usepackage{mathptmx}      
\usepackage{subfigure}
\usepackage{amsmath,amssymb}

\def\makeheadbox{{%
}}

\begin{document}

\title{Orbital resonances in discs around braneworld\\Kerr black holes}

\author{Zden\v{e}k Stuchl\'{\i}k \and Andrea Kotrlov\'a}

\authorrunning{Z. Stuchl\'{\i}k \& A. Kotrlov\'a}

\institute{Institute of Physics, Faculty of Philosophy and Science, Silesian University in Opava, Bezru\v{c}ovo n\'am.~13, CZ-746\,01 Opava, Czech Republic \\
              Tel.: +420-553-684-286\\
              Fax: +420-553-716-948\\
              \email{Andrea.Kotrlova@fpf.slu.cz}
              }

\date{}

\maketitle

\begin{abstract}
Rotating black holes in the brany universe of the Randall--Sundrum
type with infinite additional dimension are described by the Kerr
geometry with a tidal charge $b$ representing the interaction of
the brany black hole and the bulk spacetime. For $b<0$ rotating
black holes with dimensionless spin $a>1$ are allowed. We
investigate the role of the tidal charge in the orbital resonance
model of quasiperiodic oscillations (QPOs) in black hole systems.
The orbital Keplerian frequency $\nu_{\mathrm{K}}$ and the radial
and vertical epicyclic frequencies $\nu_{\mathrm{r}}$,
$\nu_{\theta}$ of the equatorial, quasicircular geodetical motion
are given. Their radial profiles related to Keplerian accretion
discs are discussed, assuming the inner edge of the disc located
at the innermost stable circular geodesic. For completeness, naked
singularity spacetimes are considered too. The resonant conditions
are given in three astrophysically relevant situations: for direct
(parametric) resonances of the oscillations with the radial and
vertical epicyclic frequencies, for the relativistic precession
model, and for some trapped oscillations of the warped discs, with
resonant combinational frequencies involving the Keplerian and
radial epicyclic frequencies. It is shown, how the tidal charge
could influence matching of the observational data indicating the
$3\!:\!2$ frequency ratio observed in GRS 1915+105 microquasar
with prediction of the orbital resonance model; limits on allowed
range of the black hole parameters $a$ and $b$ are established.
The ``magic'' dimensionless black hole spin enabling presence of
strong resonant phenomena at the radius, where
$\nu_{\mathrm{K}}\!:\!\nu_{\theta}\!:\!\nu_{\mathrm{r}} =
3\!:\!2\!:\!1$, is determined in dependence on the tidal charge.
Such strong resonances could be relevant even in sources with
highly scattered resonant frequencies, as those expected in
Sgr\,A$^*$. The specific values of the spin and tidal charge are
given also for existence of specific radius where
$\nu_{\mathrm{K}}\!:\!\nu_{\theta}\!:\!\nu_{\mathrm{r}} =
s\!:\!t\!:\!u$ with $5\geq s>t>u$ being small natural numbers. It
is shown that for some ratios such situation is impossible in the
field of black holes. We can conclude that analysing the
microquasars high-frequency QPOs in the framework of orbital
resonance models, we can put relevant limits on the tidal charge
of brany Kerr black holes. 
\end{abstract}

\section{Introduction}

In recent years, one of the most promising approaches to the
higher-dimensional gravity theories seems to be the string theory
and M-theory describing gravity as a truly higher-dimensional
interaction becoming effectively 4D at low enough energies and
these theories inspired braneworld models, where the observable
universe is a 3-brane (domain wall) to which the standard-model
(non-gravitational) matter fields are confined, while gravity
field enters the extra spatial dimensions, the size of which may
be much larger than the Planck length scale $l_\mathrm{P}\sim
10^{-33}~\mathrm{cm}$ \cite{Ark-Dim-Dva:1998:}. The braneworld
models could therefore provide an elegant solution to the
hierarchy problem of the electroweak and quantum gravity scales,
as these scales become to be of the same order
($\sim\mathrm{TeV}$) due to large scale extra dimensions
\cite{Ark-Dim-Dva:1998:}. Therefore, future collider experiments
can test the braneworld models quite well, including the
hypothetical mini black hole production on the TeV-energy scales
\cite{Emp-Mas-Rat:2002:,Dim-Lan:2001:}. On the other hand, the
braneworld models could be observationally tested since they
influence astrophysically important properties of black holes.

Gravity can be localized near the brane at low energies even with
a non-compact, infinite size extra dimension with the warped
spacetime satisfying the 5D Einstein equations with negative
cosmological constant as shown by Randall and Sundrum
\cite{Ran-Sun:1999:}. Then an arbitrary energy-momentum tensor
could be allowed on the brane \cite{Shi-Mae-Sas:2000:}.

The Randall--Sundrum model gives the 4D Einstein gravity in low
energy limit, and the conventional potential of weak, Newtonian
gravity appears on the 3-brane with high accuracy. Significant
deviations from the Einstein gravity occur at very high energies,
e.g., in the very early universe, and in vicinity of compact
objects (see, e.g.,
 \cite{Dad-etal:2000:,Maa:2004:,Ger-Maa:2001:,Ali-Gum:2005:}).
Gravitational collapse of matter trapped on the brane results in
black holes mainly localized on the brane, but their horizon could
be extended into the extra dimension. The high-energy effects
produced by the gravitational collapse are disconnected from the
outside space by the horizon, but they could have a signature on
the brane, influencing properties of black holes \cite{Maa:2004:}.
There are high-energy effects of local character influencing
pressure in collapsing matter, and also non-local corrections of
``back-reaction'' character arising from the influence of the Weyl
curvature of the bulk space on the brane -- the matter on the
brane induces Weyl curvature in the bulk which makes influence on
the structures on the brane due to the bulk graviton stresses
\cite{Maa:2004:}. The combination of high-energy (local) and bulk
stress (non-local) effects alters significantly the matching
problem on the brane, as compared to the 4D Einstein gravity; for
spherical objects, matching no longer leads to a Schwarzschild
exterior in general \cite{Dad-etal:2000:,Ger-Maa:2001:}. Moreover,
the Weyl stresses induced by bulk gravitons imply that the
matching conditions do not have unique solution on the brane; in
fact, knowledge of the 5D Weyl tensor is needed as a minimum
condition for uniqueness \cite{Ger-Maa:2001:}.\footnote{At
present, no exact 5D solution in the braneworld model is known.}

There are two kinds of black hole solutions in the
Randall--Sundrum braneworld model with infinite extension of the
extra dimension. One kind of these solutions looks like black
string from the viewpoint of an observer in the bulk, while being
described by the Schwarzschild metric for matter trapped on the
brane \cite{Cha-Haw-Rea:2000:}. The generalizations to rotating
black string solution \cite{Mod-Pan-Sen:2002:} and solutions with
dilatonic field \cite{Noj-etal:2000:} were also found. However,
the black string solutions have curvature singularities at
infinite extension along the extra dimension \cite{Ali-Gum:2004:}.
There is a proposal that the black hole strings could evolve to
localized black cigar solutions due to the classical instability
near the anti-de~Sitter horizon
\cite{Cha-Haw-Rea:2000:,Gre-Laf:1993:}, but it is not resolved at
present \cite{Hor-Mae:2001:,Ali-Gum:2005:}.

Second kind of these solutions representing a promising way of
generating exact localized solutions in the Randall--Sundrum
brane\-world models was initiated by Maartens and his coworkers
\cite{Maa:2004:,Ger-Maa:2001:,Dad-etal:2000:}. Assuming
spherically symmetric metric induced on the 3-brane, the effective
gravitational field equations on the brane could be solved, giving
Reissner--Nordstr\"{o}m static black hole solutions endowed with a
``tidal'' charge parameter $b$ instead of the standard electric
charge parameter $Q^2$ \cite{Dad-etal:2000:}. The tidal charge
reflects the effects of the Weyl curvature of the bulk space,
i.e., from the 5D graviton stresses with bulk graviton tidal
effect giving the name of the charge
\cite{Dad-etal:2000:,Maa:2004:}. Note that the tidal charge can be
both positive and negative, and there are some indications that
the negative tidal charge should properly represent the
``back-reaction'' effects of the bulk space Weyl tensor on the
brane \cite{Maa:2004:,Dad-etal:2000:,Sas-Shi-Mae:2000:}.

The exact stationary and axisymmetric solutions describing
rotating black holes localized in the Randall--Sundrum braneworld
were derived in \cite{Ali-Gum:2005:}. They are described by the
metric tensor of the Kerr--Newman form with a tidal charge
representing the 5D correction term generated by the 5D Weyl
tensor stresses. The tidal charge has an ``electric'' character
again and arises due to the 5D gravitational coupling between the
brane and the bulk, reflected on the brane through the
``electric'' part of the bulk Weyl tensor \cite{Ali-Gum:2005:}, in
close analogy with the spherically symmetric case
\cite{Dad-etal:2000:}.

When both the tidal and electric charge are present in a brany
black hole, its character is much more complex and usual
Kerr--Newman form of the metric tensor is allowed only in the
approximate case of small values of the rotation parameter $a$.
For linear approximation in $a$, the metric arrives at the usual
Boyer--Lindquist form describing ``tidally'' charged and slowly
rotating brany black holes \cite{Ali-Gum:2005:}. For large
rotational parameters, when the linear approximation is no longer
valid, additional off-diagonal metric components $g_{r\phi}$,
$g_{rt}$ are relevant along with the standard $g_{\phi t}$
component due to the combined effects of the local bulk on the
brane and the dragging effect of rotation, which through the
``squared'' energy momentum tensor on the brane distort the event
horizon that becomes a stack of non-uniformly rotating null
circles having different radii at fixed $\theta$ while going from
the equatorial plane to the poles \cite{Ali-Gum:2005:}. In the
absence of rotation, the metric tensor reduces to the
Reissner--Nordstr\"{o}m form with correction terms of local and
non-local origin \cite{Cha-etal:2001:}.

Here we restrict attention to the Kerr--Newman type of solutions
describing the brany rotating black holes with no electric charge,
when the results obtained in analysing the behaviour of test
particles and photons or test fields around the Kerr--Newman black
holes could be used assuming both positive and negative values of
the brany tidal parameter $b$ (used instead of the charge
parameter $Q^2$) \cite{Mis-Tho-Whe:1973:Gra:}.

It is very important to test the role of the hypothetical tidal
charge, implied by the theory of multidimensional black holes in
the Randall--Sundrum braneworld with non-compactified additional
space dimension, in astrophysical situations, namely in the
accretion processes and related optical phenomena, including the
oscillatory features observed in the black hole systems. There are
two complementary reasons for such studies. First, the
observational data from the black hole systems (both Galactic
binary systems or Sgr\,A$^*$ and active galactic nuclei) could
restrict the allowed values of the tidal charge, giving a relevant
information on the properties of the bulk spacetime and putting
useful additional limits on the elementary particle physics.
Second, the presence of the tidal charge could help much in
detailed understanding of some possible discrepancies in the black
hole parameter estimates coming from observational data that are
obtained using different aspects of modelling accretion phenomena
\cite{Ter-etal:2007:spin-problem,McCli-Nar-Sha:2007:}.

In fact, the black hole parameter estimates come from a variety of
astrophysical observations
\cite{Kli:2000:ARASTRA:,Kli:2004:,McCli-Rem:2004:CompactX-Sources:,Rem:2005:ASTRN:,Rem-McCli:2006:ARASTRA:,McCli-Nar-Sha:2007:}.
 The black hole spin estimates are commonly given by the optical
methods, namely by X-ray line profiles
\cite{Laor:1991:ASTRJ2:,Kar-Vok-Pol:1992:ASTRJ2L:,Dov-Kar-Mar:2004:RAGtime4and5:CrossRef,Fab-Min:2005:XSpectraKerr:Book,Zak:2003:,Zak-Rep:2006:}
 and X-ray continuum spectra
\cite{McCli-etal:2006:astro-ph/0606076:,Mid-etal:2006:,Sha:2006::astro-ph/0508302},
 and by quasiperiodic oscillations, the frequency of which enable,
in principle, the most precise spin estimates, because of high
precision of the frequency measurements.\footnote{But see some
problems connected with the wide variety of the resonance models
\cite{Ter-Abr-Klu:2005:ASTRA:QPOresmodel}.}

Therefore, we discuss here in detail the orbital resonance model
of QPOs, which seems to be the most promising in explaining the
observational data from four microquasars, namely GRO J1655-40,
XTE 1550-564, H 1743-322, GRS 1915+105
\cite{Ter-Abr-Klu:2005:ASTRA:QPOresmodel,Ter:2005:astro-ph0412500:,Tor:2005:ASTRN:,Stu-Sla-Ter:2007:ASTRA:}
 and in Sgr\,A$^*$
\cite{Asch:2004:ASTRA:,Asc-etal:2004:ASTRA:,Ter:2005:astro-ph0412500:},
or some extragalactic sources as NGC 5408 X-1
\cite{Str:2007:astro-ph/0701390:}.

It is well known that in astrophysically relevant situations the
electric charge of a black hole becomes zero or negligible on
short time scales because of its neutralization by accreting
preferentially oppositely charged particles from ionized matter of
the accretion disc
\cite{Zel-Nov:1971:RelAstr:,Mis-Tho-Whe:1973:Gra:,Dam-etal:1978:}.
Clearly, this statement remains true in the braneworld model, and
that is the reason why it is enough to consider properties of
brany Kerr black holes endowed with a tidal charge only. Of
course, the tidal charge reflecting the non-local gravitational
effects of the bulk space is non-negligible in general and it can
have quite strong effect on the physical processes in vicinity of
the black hole. Recently, some authors tested the tidal charge
effects in the weak field limit for optical lensing
\cite{Kee-Pet:2006:brane-lensing:} and relativistic precession or
time delay effect \cite{Boh-etal:2008:SolarSys:}. Here we develop
a framework of testing the tidal charge effects in the strong
field near black holes with accretion discs giving rise to kHz
QPOs.

In Section~\ref{sec:tri}, the Kerr black holes with a tidal
charge, introduced by Aliev and G{\"u}mr{\"u}k{\c c}{\"u}o{\u g}lu
\cite{Ali-Gum:2005:}, are described and their properties are
briefly summarized. In Section~\ref{sec:ctyri}, the Carter
equations of motion are given, the equatorial circular geodesics,
i.e., Keplerian circular orbits reflecting properties of thin
accretion discs, are determined and properties of photon circular
orbits and innermost stable orbits are discussed. In
Section~\ref{sec:pet}, the radial and vertical (latitudinal)
epicyclic frequencies $\nu_{\mathrm{r}}$ and $\nu_{\theta}$,
together with the Keplerian orbital frequency $\nu_{\mathrm{K}}$,
are given. In Section~\ref{sec:sest}, their properties are
discussed, namely their radial profiles through the Keplerian
accretion disc with its inner radius assumed to be located at the
radius of the innermost stable circular geodesic, where the radial
epicyclic frequency vanishes. In Section~\ref{sec:sedm}, we
shortly discuss the resonance conditions for the direct resonance
of the both epicyclic frequencies
($\nu_{\theta}\!:\!\nu_{\mathrm{r}} =  3\!:\!2$) assumed to be in
a parametric resonance \cite{Ter-Abr-Klu:2005:ASTRA:QPOresmodel},
the relativistic precession model \cite{Ste-Vie:1998:ASTRJ2L:}
where we assume a resonance of oscillations with
$\nu_{\mathrm{K}}\!:\!\left(\nu_{\mathrm{K}}-\nu_{\mathrm{r}}\right)=
3\!:\!2$, and the resonance of trapped oscillations assumed in
warped disc as discussed by Kato \cite{Kato:2007:PASJ:}
[$(2\nu_{\mathrm{K}}-\nu_{\mathrm{r}})\!:\!2(\nu_{\mathrm{K}}-\nu_{\mathrm{r}})
= 3\!:\!2$]. In Section~\ref{sec:osm} we determine the ``magic''
(dimensionless) spin of brany Kerr black holes in dependence on
the (dimensionless) tidal charge, enabling presence of strong
resonant phenomena because of the very special frequency ratio
$\nu_{\mathrm{K}}\!:\!\nu_{\theta}\!:\!\nu_{\mathrm{r}} =
3\!:\!2\!:\!1$; possibility of other small integer ratios of the
three frequencies at a shared radius is also discussed. Concluding
remarks on the resonant phenomena in strong gravity of brany black
holes are presented in Section~\ref{sec:devet}.

\section{\label{sec:tri}Braneworld Kerr black holes}

Using the Gauss--Codazzi projective approach, the effective
gravitational field equations on a 3-brane in a 5D bulk spacetime
can be defined \cite{Arn-Des-Mis:1962:,Shi-Mae-Sas:2000:}. The
self-consistent solutions of the effective 4D Einstein equations
on the brane require the knowledge of the non-local gravitational
and energy-momentum terms coming from the bulk spacetime.
Therefore, the brany field equations are not closed in general and
evolution equations into the bulk have to be solved for the
projected bulk curvature and energy-momentum tensors
\cite{Ali-Gum:2004:}. However, in particular cases the
brany-equations system could be made closed assuming a special
ansatz for the induced metric. In this way, both spherically
symmetric \cite{Dad-etal:2000:} and axially symmetric brany black
hole spacetimes \cite{Ali-Gum:2005:} have been found.

Under the assumption of stationary and axisymmetric Kerr--Schild
metric on the brane and supposing empty bulk space and no matter
fields on the brane ($T_{\alpha\beta}=0$) \cite{Ali-Gum:2005:} the
effective Einstein equations on the brane reduce to the form
$R_{\alpha\beta}=-E_{\alpha\beta}$, where
    \begin{equation}
        E_{\alpha\beta}=
        \,^{(5)}C_{ABCD}\,n^A\,n^C\,e_\alpha^B\,e_\beta^D
    \end{equation}
is the projected ``electric'' part of the 5D Weyl tensor
$C_{ABCD}$ used to describe the non-local gravitational effects of
the bulk space onto the brane \cite{Ali-Gum:2005:}.

The line element for the brany rotating black holes can then be
expressed in the standard Boyer--Lindquist coordinates and
geometric units ($c=G=1$) in the form \cite{Ali-Gum:2005:}
    \begin{align}
        \mathrm{d}s^2=&-\left(1-\frac{2Mr-b}{\Sigma}\right)\mathrm{d}t^2-
        \frac{2a(2Mr-b)}{\Sigma}\sin^2{\theta}\,
        \mathrm{d}t\,
        \mathrm{d}\phi \nonumber\\
        &+\frac{\Sigma}{\Delta}\mathrm{d}r^2 + \Sigma\,
        \mathrm{d}\theta^2+\left(r^2+a^2+\frac{2Mr-b}{\Sigma}\,
        a^2\sin^2\theta\right)\sin^2\theta\,\mathrm{d}\phi^2,
    \end{align}
where
    \begin{align}
        \Delta &= r^2+a^2-2Mr+b ,\\
        \Sigma &= r^2+a^2\cos^2\theta .
    \end{align}
We can see that this metric looks exactly like the Kerr--Newman
solution in general relativity \cite{Mis-Tho-Whe:1973:Gra:}, in
which the square of the electric charge $Q^2$ is replaced by a
tidal charge parameter $b$ (or ``brany'' parameter). Since the
metric is asymptotically flat, by passing to the far-field regime
we can interpret the parameter $M$ as the total mass of the black
hole and parameter $a$ as the specific angular momentum (the black
hole spin).

The event horizons of the spacetime are determined by the
condition $\Delta=0$. The radius of the outer event horizon is
given by the relation
\begin{equation}\label{horizont}
r_+=M+\sqrt{M^2-a^2-b} \,.
\end{equation}
The horizon structure depends on the sign of the tidal charge. We
see that, in contrast to its positive values, the negative tidal
charge tends to increase the horizon radius (see, e.g.,
Fig.\,\ref{hor-a-mez}).

The event horizon does exist provided that
\begin{equation}\label{hor-podm}
M^2\geq a^2+b ,
\end{equation}
where the equality corresponds to the family of extreme black
holes. It is clear that the positive tidal charge acts to weaken
the gravitational field and we have the same horizon structure as
in the usual Kerr--Newman solution. However, new interesting
features arise for the negative tidal charge. For $b<0$ and
$a\rightarrow M$ it follows from equation~(\ref{horizont}) that
the horizon radius
\begin{equation}
r_+\rightarrow \left(M+\sqrt{-b}\right) > M ;
\end{equation}
such a situation is not allowed in the framework of general
relativity. From equations~(\ref{horizont}) and~(\ref{hor-podm})
we can see that for $b<0$, the extreme horizon $r_+=M$ corresponds
to a black hole with rotation parameter $a$ greater than its mass
$M$ (e.g., for extreme black hole with $b=-M^2$ we have
$a=\sqrt{2}M$). Thus, the bulk effects on the brane may provide a
mechanism for spinning up the black hole on the brane so that its
rotation parameter exceeds its mass. Such a mechanism is
impossible in general relativity. Further, if the inner horizon
determined by the formula
\begin{equation}
r_-=M-\sqrt{M^2-a^2-b}
\end{equation}
turns out to be negative (it is possible for $b<0$, again), the
physical singularity ($r=0$, $\theta = \pi/2$) is expected to be
of space-like character, contrary to the case of $b>0$, when it is
of time-like character \cite{Dad-etal:2000:}.

\section{\label{sec:ctyri}Geodesic motion}

Motion of a test particle of mass $m$ is given by the standard
geodesic equation
\begin{equation}
    \frac{\mathrm{D} p^\mu}{\mathrm{d} \tau}=0
\end{equation}
accompanied by the normalization condition $p_\mu p^\mu=-m^2$ and
can be treated in full analogy with the Kerr case
\cite{Car:1968:PHYSREV:}. There are three motion constants given
by the spacetime symmetry -- the energy being related to the
Killing vector field $\partial/\partial t$, the axial angular
momentum being related to the Killing vector field
$\partial/\partial \phi$, and the angular momentum constant
related to the hidden symmetry of the Kerr spacetime
\cite{Car:1973:BlaHol:}. The geodesic equations could then be
fully separated and integrated using the Hamilton--Jacobi method
\cite{Car:1973:BlaHol:}.

For the motion restricted to the equatorial plane
($\theta=\pi/2$), the Carter equations take the form
    \begin{align}
        \frac{\mathrm{d} \theta}{\mathrm{d} \lambda} &= 0,\\
        r^2\frac{\mathrm{d} r}{\mathrm{d} \lambda} &= \pm
        \sqrt{R(r)},\\
        r^2\frac{\mathrm{d} \phi}{\mathrm{d} \lambda} &=
        -\left(a E-L\right)+ \frac{a P(r)}{\Delta},\\
        r^2\frac{\mathrm{d} t}{\mathrm{d} \lambda} &=
        -a\left(a E-L\right)+\frac{\left(r^2+a^2\right)
        P(r)}{\Delta},
    \end{align}
where
    \begin{align}
    P(r)&=E\left(r^2+a^2\right)-L a ,\\
    R(r)&=P(r)^2-\Delta\left[m^2 r^2 + (aE-L)^2\right].
    \end{align}

The proper time of the particle $\tau$ is related to the affine
parameter $\lambda$ by $\tau=m \lambda$. The constants of motion
are: energy $E$ and axial angular momentum $L$ of the test
particle in infinity (related to the stationarity and the axial
symmetry of the geometry); for the equatorial motion, the third
constant of motion $Q=0$ \cite{Car:1973:BlaHol:}.

The equatorial circular orbits can most easily be determined by
solving simultaneously the equations
\begin{equation}
    R(r)=0,\qquad \frac{\mathrm{d} R}{\mathrm{d} r}=0.
\end{equation}
The specific energy and the specific angular momentum of the
circular motion at a given radius are then determined by the
relations \cite{Ali-Gum:2005:,Dad-Kal:1977:}
    \begin{align}\label{energie-bezrozm}
   \frac{E}{m} &= \frac{r^2-2 M r+b\pm a\sqrt{M r-b}}{r\left(
   r^2-3 M r+2b\pm 2a\sqrt{M
   r-b}\right)^{1/2}},\\
   \frac{L}{m} &= \pm \frac{\sqrt{M r-b}\left(r^2+a^2\mp 2a\sqrt{M r-b}
   \right)\mp a b}
   {r\left(r^2-3 M r+2b\pm 2a\sqrt{M
   r-b}\right)^{1/2}}.\label{moment-bezrozm}
\end{align}
Here and in the following, the upper sign corresponds to the
corotating orbits ($L>0$), while the lower sign implies
retrograde, counterrotating ($L<0$) motion of the particles.

In the analysis of the epicyclic frequency profiles, it is useful
to relate the profiles to the photon circular geodesic and
innermost stable circular geodesic radii (or innermost bound
circular radii in case of thick disc that is not considered here)
that are relevant in discussions of properties of the accretion
discs and their oscillations. Therefore, we put the limiting radii
in an appropriate form.

\begin{figure*}[!tbp]
\centering
\subfigure{\includegraphics[width=.48\hsize]{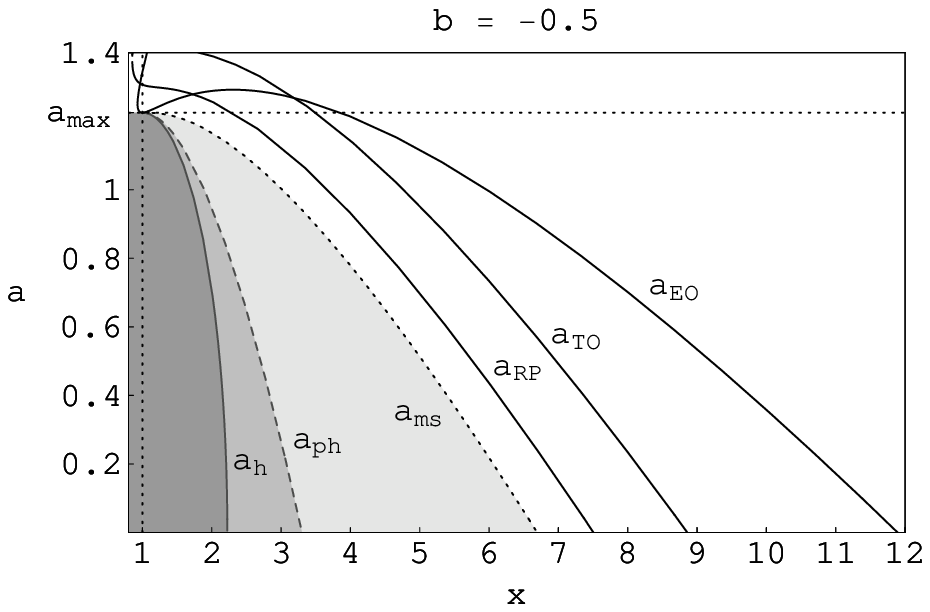}}\quad
\subfigure{\includegraphics[width=.48\hsize]{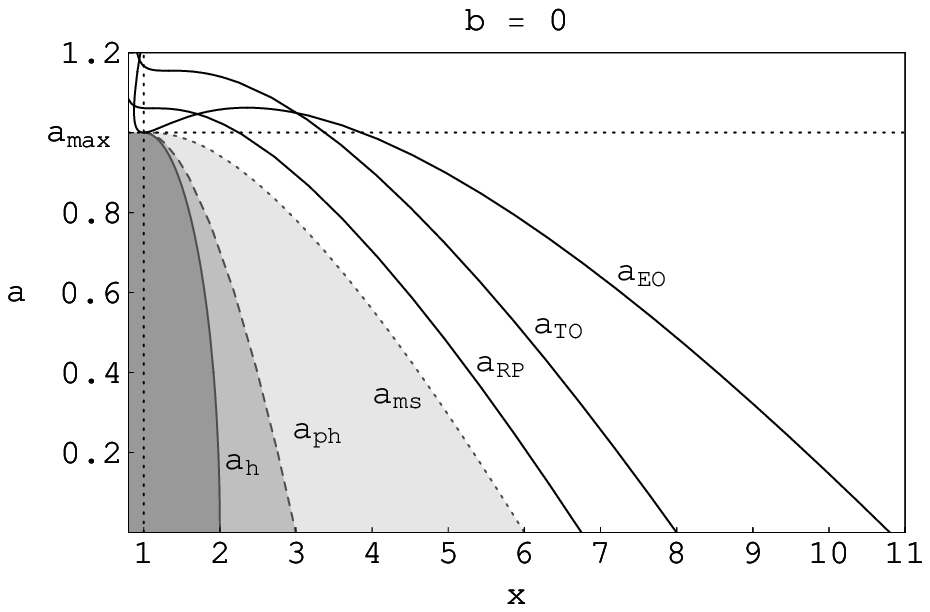}}\\
\subfigure{\includegraphics[width=.48\hsize]{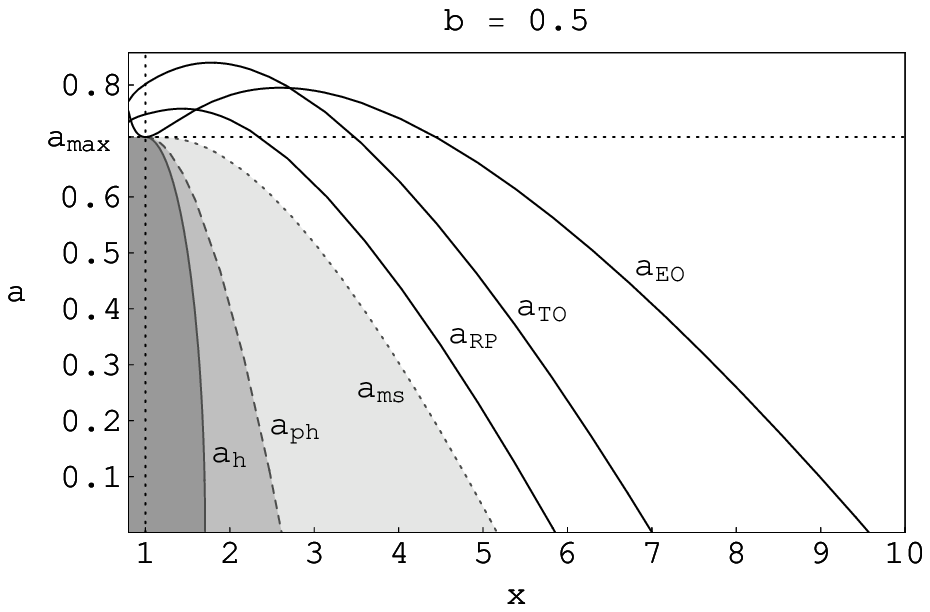}}\quad
\subfigure{\includegraphics[width=.48\hsize]{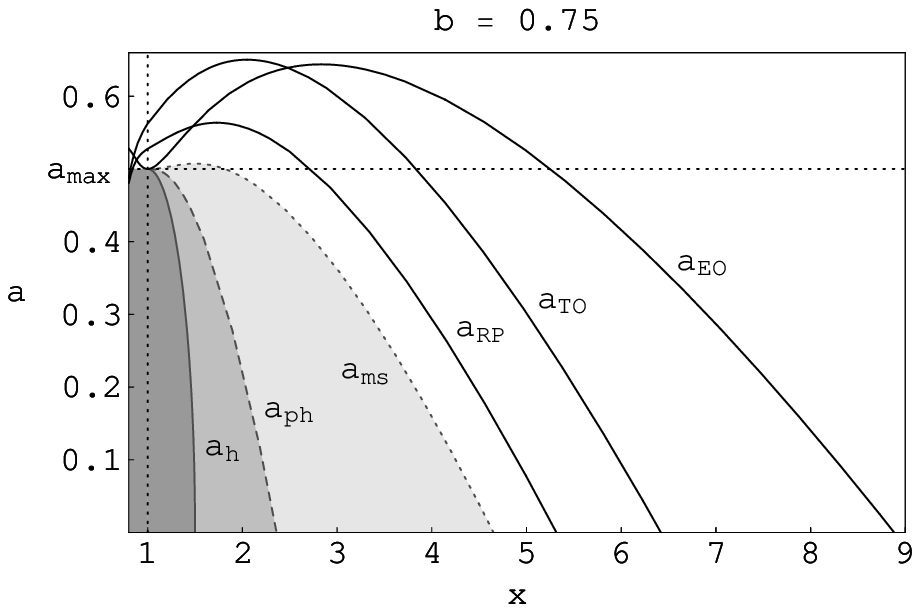}}\\
\subfigure{\includegraphics[width=.48\hsize]{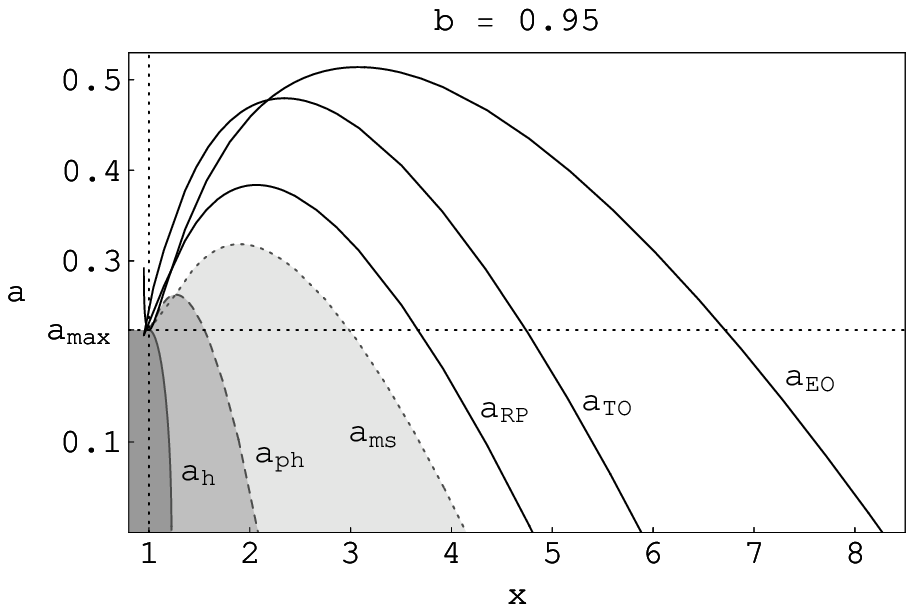}}
\caption{\label{hor-a-mez}The behaviour of functions
$a_{\mathrm{h}}$ (gray solid line), $a_{\mathrm{ph}}$ (gray dashed
line) and $a_{\mathrm{ms}}$ (gray dotted line) that implicitly
determine the radius of the outer event black hole horizon, the
limiting photon orbit and the marginally stable circular orbit in
the equatorial plane ($\theta=\pi/2$) of a rotating black hole
with a fixed value of the tidal charge $b$. The functions
$a_{\mathrm{EO}}=a^{\theta/\mathrm{r}}_{3:2}$,
$a_{\mathrm{RP}}=a^{\mathrm{K}/(\mathrm{K}-\mathrm{r})}_{3:2}$ and
$a_{\mathrm{TO}}=a_{3:2}^{(2\mathrm{K}-\mathrm{r})/(2\mathrm{K}-2\mathrm{r})}$
(black solid lines) represent the radii where the direct Epicyclic
Oscillations resonance
$\nu_{\theta}\!:\!\nu_{\mathrm{r}}=3\!:\!2$, the Relativistic
Precession resonance $\nu_{\mathrm{K}}\!:\!\left(
\nu_{\mathrm{K}}-\nu_{\mathrm{r}}\right)=3\!:\!2$ and the Trapped
Oscillations resonance $(2\nu_{\mathrm{K}}-\nu_{\mathrm{r}}):
2(\nu_{\mathrm{K}}-\nu_{\mathrm{r}})= 3\!:\!2$ occur.}
\end{figure*}

For simplicity, hereafter we use dimensionless radial coordinate
\begin{equation}\label{def-x}
x=r/M
\end{equation}
and putting $M=1$, we also use dimensionless spin $a$ and brany
parameter $b$. The outer event horizon $x_{\mathrm{h}}(a,b)$ is
then implicitly determined by the relation
\begin{equation}\label{hor-impl}
a=a_{\mathrm{h}}(x,b)\equiv\sqrt{2 x-x^2-b }\,.
\end{equation}
Focusing our attention to the corotating orbits, we find the
radius of the photon circular orbit $x_{\mathrm{ph}}(a,b)$ to be
given by the relation
\begin{equation}\label{fot-impl}
a=a_{\mathrm{ph}}(x,b)\equiv\frac{x(3-x)-2b}{2\sqrt{x-b}},
\end{equation}
the radius of the marginally bound orbit $x_{\mathrm{mb}}(a,b)$ to
be given by
\begin{equation}\label{mez-vazana-impl}
a=a_{\mathrm{mb}}(x,b)\equiv\frac{\sqrt{x-b} \left(2x-b\mp
x\sqrt{x}\right)}{x-b},
\end{equation}
and the radius of the marginally stable corotating orbit
$x_{\mathrm{ms}}(a,b)$ by the equation\footnote{The upper sign in
(\ref{mez-vazana-impl}) and (\ref{ms-impl}) is relevant for the
black hole spacetimes, while both signs are relevant for the naked
singularity spacetimes.}
\begin{equation}\label{ms-impl}
a=a_{\mathrm{ms}}(x,b)\equiv\frac{4 (x-b )^{3/2}\mp x \sqrt{3
x^2-2 x(1+2 b )+3 b }}{3 x-4 b };
\end{equation}
for extreme black holes the maximum value of the black hole spin
is
\begin{equation}\label{def-amax}
a_{\mathrm{max}}=\sqrt{1-b} \,,
\end{equation}
thus, e.g., for $b=-1$ we have $a_{\mathrm{max}}=\sqrt{2}$.

\begin{figure*}[!tbp]
\subfigure{\includegraphics[width=.48\hsize]{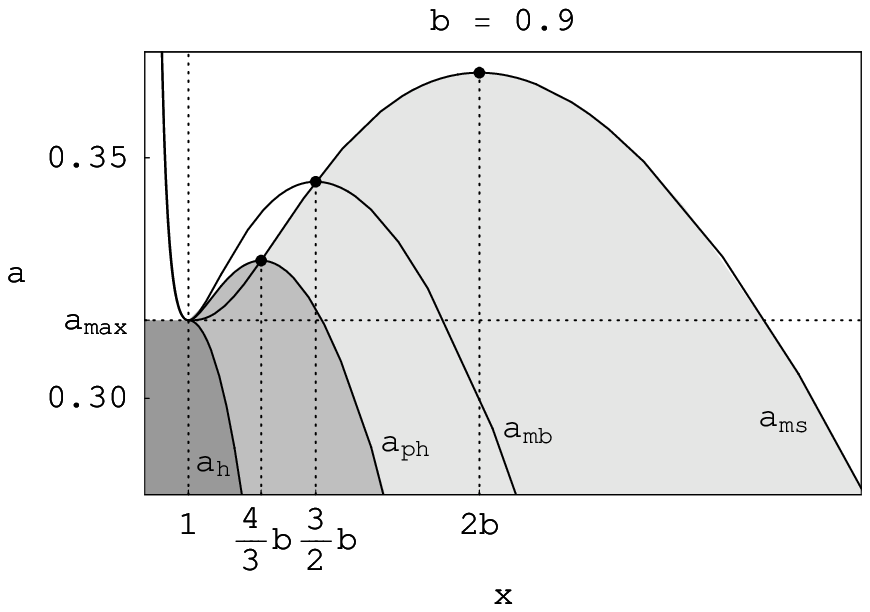}}\quad
\subfigure{\includegraphics[width=.48\hsize]{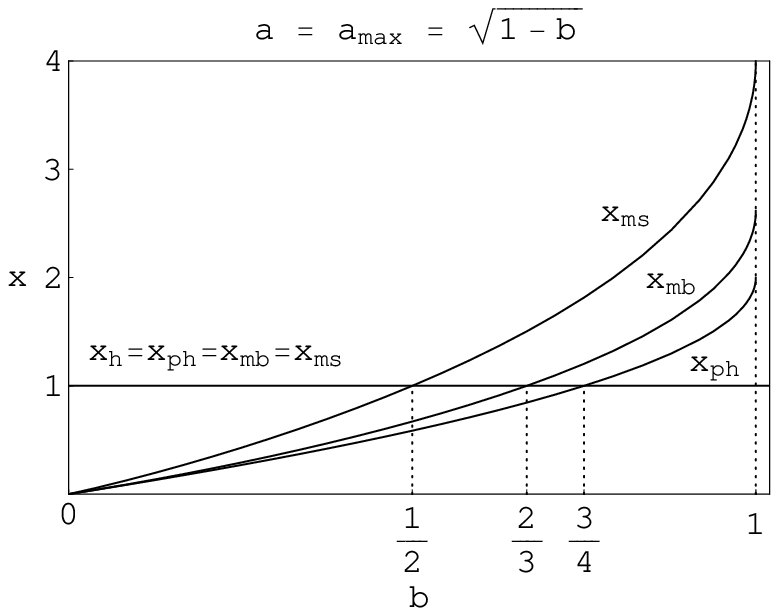}}
\caption{\label{xka-nad-1}\textit{Left panel}: The functions
$a_{\mathrm{h}}$, $a_{\mathrm{ph}}$, $a_{\mathrm{mb}}$ and
$a_{\mathrm{ms}}$ that implicitly determine the radii of the outer
event black hole horizon, the limiting photon orbit and the
marginally bound and stable circular orbits. The \textit{right
panel} illustrates the behaviour of the photon orbit
$x_{\mathrm{ph}}$, marginally bound $x_{\mathrm{mb}}$ and
marginally stable $x_{\mathrm{ms}}$ orbits for the extreme black
holes with $0\leq b\leq 1$. The radii of the orbits that are
situated at $x<1$ are out of our interest being located under the
outer event black hole horizon.}
\end{figure*}

The function $a_{\mathrm{ph}}$ (\ref{fot-impl}) has a local
maximum at $x=1$ for brany parameter $b\leq 3/4$ and a local
maximum at $x=\mathcal{X}_{\mathrm{K}}=4b/3$ for $b>3/4$
corresponding to the naked singularity spacetimes since
$a_{\mathrm{ph}}(x=\mathcal{X}_{\mathrm{K}})> a_{\mathrm{max}}$.
The function $a_{\mathrm{mb}}$ (\ref{mez-vazana-impl}) has a local
maximum at $x=1$ for $b\leq 2/3$ and at $x=3b/2$ for $b>2/3$.  The
function $a_{\mathrm{ms}}$ (\ref{ms-impl}) has a local maximum at
$x=1$ for $b\leq 1/2$ and at $x=2b$ for $b>1/2$ (this local
maximum appears again only for naked singularities, since
$a>a_{\mathrm{max}}$). So, for $x>1$ the functions
$a_{\mathrm{ph}}(x,b)$, $a_{\mathrm{mb}}(x,b)$ and
$a_{\mathrm{ms}}(x,b)$ are not monotonically decreasing functions
of radius for the whole range of the brany tidal charge parameter
$b$, as usual in Kerr spacetimes. This special behaviour of
$a_{\mathrm{ph}}$, $a_{\mathrm{mb}}$ and $a_{\mathrm{ms}}$ implies
that for some values of the brany parameter $b$ of extreme
braneworld Kerr black holes the radial Boyer--Lindquist
coordinates of the photon orbit $x_{\mathrm{ph}}$, the marginally
bound $x_{\mathrm{mb}}$ and marginally stable $x_{\mathrm{ms}}$
orbits do not merge with the black hole horizon radial coordinate
at $x_{\mathrm{h}}=1$, as usual in Kerr spacetimes and are shifted
to higher radii. There is
\begin{align}
 x_{\mathrm{ph}}(a_{\mathrm{ph}},b)>1&\quad\textrm{for}\quad
 3/4<b<1\,,\label{hrb-mez-foton}\\
 x_{\mathrm{mb}}(a_{\mathrm{mb}},b)>1&\quad\textrm{for}\quad
 2/3<b<1\,,\label{hrb-mez-vazana}\\
 x_{\mathrm{ms}}(a_{\mathrm{ms}},b)>1&\quad\textrm{for}\quad
 1/2<b<1\, .\label{hrb-mez}
\end{align}
For typical values of the tidal charge $b$ the functions
$a_{\mathrm{ph}}(x,b)$ and $a_{\mathrm{ms}}(x,b)$ are illustrated
in Fig.\,\ref{hor-a-mez}. The behaviour of the photon and
marginally bound and stable orbits for the extreme black holes
with $0\leq b\leq 1$ is illustrated in Fig.\,\ref{xka-nad-1}.

\begin{figure*}[!tbp]
\centering
\includegraphics[width=.68\hsize]{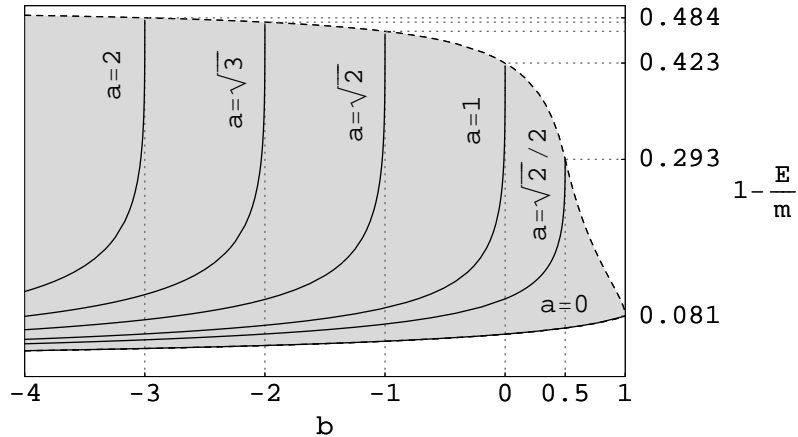}
\caption{\label{vazb-energ}The specific binding energy per unit
mass ($E_{\mathrm{b}}=1-E(x=x_{\mathrm{ms}})/m$) of a particle at
the marginally stable direct orbit $x=x_{\mathrm{ms}}$ as a
function of the brany parameter $b$. The binding energy profile in
thin Keplerian disc is given for appropriately chosen values of
black hole spin $a$ (full lines). The dashed line corresponds to
the binding energy for maximally rotating (extreme) braneworld
black holes. We can see that for extreme black hole with $b=-3$
and $a=2$ the specific binding energy $E_{\mathrm{b}}\simeq
48.4\,\%$, while for extreme Kerr black hole with $b=0$, $a=1$ we
have $E_{\mathrm{b}}\simeq 42.3\,\%$ and for extreme case of the
Reissner--Nordstr\"{o}m black hole where $b=1$ and $a=0$ there is
$E_{\mathrm{b}}\simeq 8.1\,\%$. So for the class of corotating
orbits, the negative tidal charge tends to increase the efficiency
of an accretion disc around a maximally rotating braneworld black
hole.}
\end{figure*}

It is evident (see Fig.\,\ref{hor-a-mez}) that the positive tidal
charge will play the same role in its effect on the circular
orbits as the electric charge in the Kerr--Newman spacetime -- the
radius of the circular photon orbit, as well as the radii of the
innermost bound and the innermost stable circular orbits move
towards the event horizon as the positive tidal charge increases
for both direct and retrograde orbits. For the negative tidal
charge the distance of the radii of the limiting photon orbit, the
innermost bound and the innermost stable circular orbits from the
event horizon enlarge as the absolute value of $b$ increases for
both direct and retrograde motions of the particles
\cite{Ali-Gum:2005:}. Further, for the class of direct orbits, the
negative tidal charge tends to increase the efficiency of
accretion processes in accretion disc around a maximally rotating
braneworld black hole (the specific binding energy of a particle
at the marginally stable direct orbit is given for appropriately
chosen values of $a$ in Fig.\,\ref{vazb-energ}
\cite{Ali-Gum:2005:}). The specific binding energy, i.e.,
$E_\mathrm{b}=1 - E (x=x_{\mathrm{ms}},a,b)$, decreases with
descending brany parameter $b$ with spin $a$ being fixed; its
maximum is reached when $b$ gives the extreme black hole
(Fig.\,\ref{vazb-energ}). On the other hand, the binding energy of
extreme black holes grows with descending brany parameter $b$. For
example, for $b=1$ ($a=0$) $E_\mathrm{b}=0.081$, for $b=0.5$
($a=\sqrt{2}/2$) $E_\mathrm{b}=0.293$, for $b=0$ ($a=1$)
$E_\mathrm{b}=0.423$, for $b=-1$ ($a=\sqrt{2}$)
$E_\mathrm{b}=0.465$, for $b=-3$ ($a=2$) $E_\mathrm{b}=0.484$.

\section{\label{sec:pet}Epicyclic oscillations of Keplerian motion}

It is well known that for oscillations of both thin Keplerian
\cite{Kat-Fuk-Min:1998:BHAccDis:,Klu-etal:Mexico:2007:} and
toroidal discs \cite{Rez-etal:2003:MONNR:} around black holes
(neutron stars) the orbital Keplerian frequency $\nu_{\mathrm{K}}$
and the related radial and vertical epicyclic frequencies
$\nu_{\mathrm{r}}$ and $\nu_{\theta}$ of geodetical quasicircular
motion are relevant and observable directly or through some
combinational frequencies
\cite{Ter-Abr-Klu:2005:ASTRA:QPOresmodel,Tor-Stu:2005:RAGtime6and7:CrossRef,Ter-Stu:2005:ASTRA:,Stu-Kot-Tor:2007:RAGtime8and9CrossRef:MrmQPO}.
 Of course, for extended tori, the eigenfrequencies of their
oscillations are shifted from the epicyclic frequencies in
dependence on the thickness of the torus
\cite{Sra:2005:ASTRN:,Bla-etal:2006:ASTRJ2:}. Similarly, due to
non-linear resonant phenomena, the oscillatory eigenfrequencies
could be shifted from the values corresponding to the geodetical
epicyclic frequencies in dependence on the oscillatory amplitude
\cite{Lan-Lif:1976:Mech:}. It is expected that shift of this kind
is observed in neutron star systems
\cite{Abr-etal:2005:RAGtime6and7:CrossRef,Abr-etal:2005:ASTRN:},
while in microquasars, i.e., binary black hole systems, the
observed frequency scatter is negligible and the geodetical
epicyclic frequencies should be relevant. Here, we restrict our
attention to the geodetical epicyclic oscillations of Keplerian
discs in microquasars. Such an approach is quite correct for
efficiently radiating discs \cite{Nov-Tho:1973:BlaHol:} in high
accretion rates. However, in low accretion rates the accretion
discs radiate inefficiently, and due to the pressure effects they
become thick in the innermost part being advection dominated (so
called ADAFs \cite{Nar-Yi:1994:ASTRJ2:,Abr-etal:ADAFs:1995:}). The
eigenfrequencies of oscillations of such toroidal structures
deviate from the geodetical epicyclic frequencies up to 20\,\% for
relative high thickness of the tori, in realistic disc
configurations it is expected to be $< 10\,\%$
\cite{Sra:2005:ASTRN:,Bla-etal:2006:ASTRJ2:}. Note that the
epicyclic oscillatory frequencies could be also efficiently
influenced by the strong magnetic field of neutron stars, as shown
in~\cite{Bak-etal:2008:magn:}. Of course, no strong magnetic field
can be related to black holes \cite{Mis-Tho-Whe:1973:Gra:}.

In the case of the Kerr black holes with the brany tidal charge
$b$, the formulae of the test particle geodetical circular motion
and its epicyclic oscillations, obtained by Aliev and Galtsov
\cite{Ali-Gal:1981:}, could be directly applied. We can write down
the following relations for the orbital and epicyclic frequencies:
\begin{align}\label{def-epi-rad}
\nu^2_{\mathrm{r}}&=\alpha_{\mathrm{r}}\,\nu_{\mathrm{K}}^2,\\
\nu^2_{\theta}&=\alpha_{\theta}\,\nu_{\mathrm{K}}^2,\label{def-epi-ver}
\end{align}
where the Keplerian frequency reads
\begin{equation}\label{def-Kep}
\nu_{\mathrm{K}}=\frac{1}{2\pi}\left(\frac{\mathrm{G}M
}{r_{\mathrm{G}}^3}\right)^{1/2}\frac{\sqrt{x-b }}{x^2+a \sqrt{x-b
}}=\frac{1}{2\pi}\frac{\mathrm{c}^3}{\mathrm{G}M}\frac{\sqrt{x-b
}}{x^2+a \sqrt{x-b }},
\end{equation}
and the dimensionless quantities determining the epicyclic
frequencies are given by
\begin{align}
\alpha_{\mathrm{r}}(x,a,b)&=\frac{4
\left(b-x-a^2\right)}{x^2}+\frac{8 a \sqrt{x-b
}}{x^2}+\frac{x(x-2)+a^2+b }{x (x-b )},\\
\alpha_{\theta}(x,a,b)&= 1+\frac{2 a^2}{x^2}-\frac{2 a \sqrt{x-b
}}{x^2}-\frac{2 a\sqrt{x-b }}{x (x-b)}+\frac{a^2}{x (x-b )},
\end{align}
which reduce to the standard relations for quasicircular geodesics
in Kerr metric \cite{Ter-Stu:2005:ASTRA:} for $b=0$.

In the limit of the Reissner--Nordstr\"{o}m like static braneworld
black hole ($a = 0$), we arrive at
\begin{align}
\alpha_{\mathrm{r}}(x,b)&=\frac{4
\left(b-x\right)}{x^2}+\frac{x(x-2)+b }{x (x-b )},\\
\alpha_{\theta}(x,b)&= 1,
\end{align}
so that $\nu_{\mathrm{K}}(x,b)=\nu_{\theta}(x,b)$ due to the
spherical symmetry of the spacetime.

\begin{figure*}[!tbp]
\subfigure{\includegraphics[width=.48\hsize]{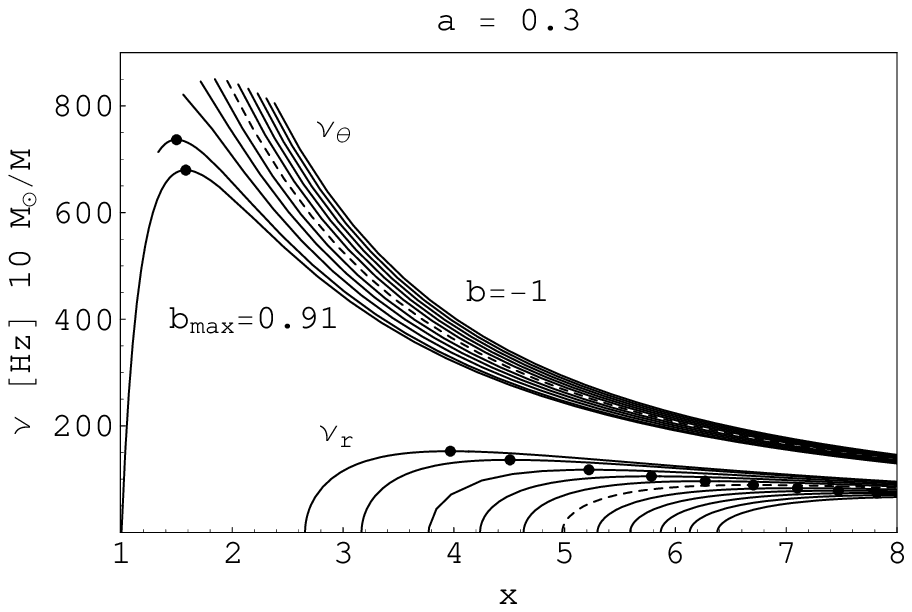}}\quad
\subfigure{\includegraphics[width=.48\hsize]{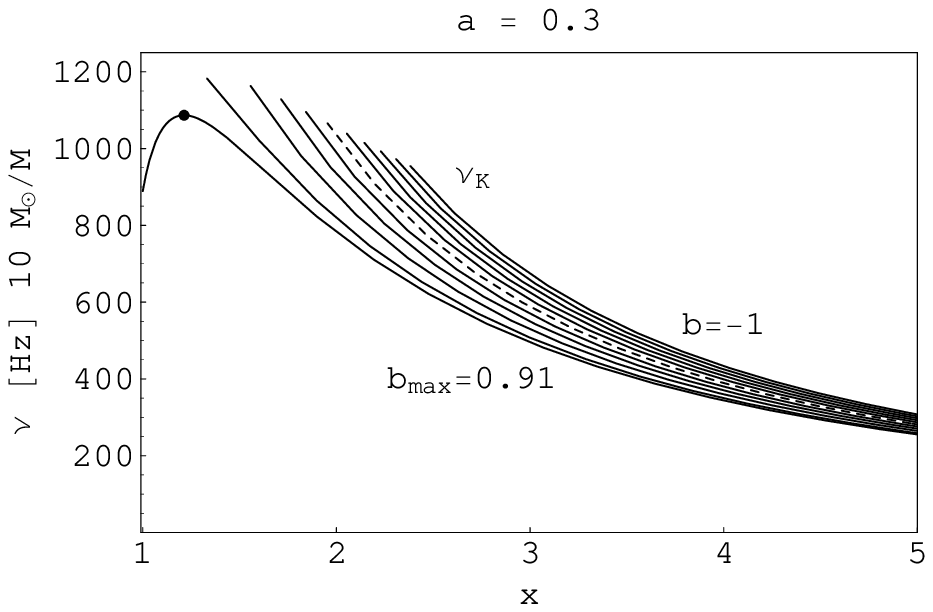}}
\caption{\label{frekv-obecne}The behaviour of the two epicyclic
frequencies $\nu_{\mathrm{r}}$, $\nu_{\theta}$ (\textit{left
panel}), and Keplerian frequency $\nu_{\mathrm{K}}$ (\textit{right
panel}) in the field of braneworld Kerr black holes with fixed
value of the black hole spin $a=0.3$ and various values of the
tidal charge parameter $b$. The curves are spaced by $0.2$ in $b$
and they are plotted from the outer event black hole horizon
$x_{\mathrm{h}}$. The dashed lines represent Kerr spacetime with
zero tidal charge.}
\end{figure*}

In the field of brany Kerr black holes ($a\neq 0$), there is (see
Fig.\,\ref{frekv-obecne})
\begin{equation}
\nu_{\mathrm{K}}(x,a,b)>\nu_{\theta}(x,a,b)>\nu_{\mathrm{r}}(x,a,b),
\end{equation}
however, this statement is not generally correct in the case of
brany Kerr naked singularities. In the next section we show that
the case $\nu_{\theta}(x,a,b)\leq\nu_{\mathrm{r}}(x,a,b)$ is also
possible.

The properties of $\nu_{\mathrm{K}}$, $\nu_{\theta}$,
$\nu_{\mathrm{r}}$ for Kerr black hole spacetimes are reviewed,
e.g., in \cite{Kat-Fuk-Min:1998:BHAccDis:} and for both Kerr black
hole and Kerr naked singularity spacetimes in
\cite{Ter-Stu:2005:ASTRA:}. We can summarize that in Kerr
spacetime with zero tidal charge ($b=0$)
\begin{itemize}
\item the Keplerian frequency is a monotonically decreasing
    function of radius for the whole range of black hole rotational
    parameter $a\in (-1,1)$ in astrophysically relevant radii above
    the photon circular orbit;
\item for slowly rotating black holes the vertical
    epicyclic frequency is a monotonically decreasing function of
    radius in the same radial range as well; however, for rapidly
    rotating black holes this function has a maximum;
\item the radial
    epicyclic frequency has a local maximum for all $a\in (-1,1)$, and
    vanishes at the innermost stable circular geodesic;
\item for Kerr naked singularities the behaviour of the epicyclic
    frequencies is different; a detailed analysis
    \cite{Ter-Stu:2005:ASTRA:} shows that the vertical frequency
    can have two local extrema, and the radial one even three local
    extrema.
\end{itemize}

In the next section we discuss the behaviour of the fundamental
orbital frequencies for Keplerian motion in the field of both
brany Kerr black holes and brany Kerr naked singularities.

We express the frequency as
$\nu\,[\mathrm{Hz}]\,10\,\mathrm{M}_{\odot}/M$ in every
quantitative plot of frequency dependence on radial coordinate
$x$, i.e., displayed value is the frequency relevant for a central
object with a mass of $10\,\mathrm{M}_{\odot}$, which could be
simply rescaled for another mass by just dividing the displayed
value by the respective mass in units of ten solar mass.

\section{\label{sec:sest}Properties of the Keplerian and epicyclic frequencies}

First, it is important to find the range of relevance for the
functions $\nu_{\mathrm{K}}(x,a,b)$, $\nu_{\theta}(x,a,b)$,
$\nu_{\mathrm{r}}(x,a,b)$ above the event horizon $x_{\mathrm{h}}$
for black holes, and above the ring singularity located at $x = 0$
($\theta=\pi/2)$ for naked singularities.

Stable circular geodesics, relevant for the Keplerian, thin
accretion discs exist for $x>x_{\mathrm{ms}}(a,b)$, where
$x_{\mathrm{ms}}(a,b)$ denotes the radius of the marginally stable
orbit, determined (in an implicit form) by the relation
(\ref{ms-impl}), which coincides with the condition
\begin{equation}\label{podm-mez}
    \alpha_{\mathrm{r}}(x,a,b)=0 .
\end{equation}
For toroidal, thick accretion discs the unstable circular
geodesics can be relevant in the range $x_{\mathrm{mb}} \leq
x_{\mathrm{in}} < x < x_{\mathrm{ms}}$, being stabilized by
pressure gradients in the tori. The radius of the marginally bound
circular geodesic $x_{\mathrm{mb}}$, implicitly determined by the
equation (\ref{mez-vazana-impl}), is the lower limit for the inner
edge of thick discs
\cite{Koz-Jar-Abr:1978:ASTRA:,Kro-Haw:2002:ASTRJ2:}.

Clearly, the Keplerian orbital frequency is well defined up to $x
= x_{\mathrm{ph}}$. However, $\nu_{\mathrm{r}}$ is well defined,
if $\alpha_{\mathrm{r}}\geq 0$, i.e., at $x\geq x_{\mathrm{ms}}$,
and $\nu_{\mathrm{r}}=0$ at $x_{\mathrm{ms}}$. We can also show
that for $x\geq x_{\mathrm{ph}}$, there is $\alpha_{\theta}\geq
0$; i.e., the vertical frequency $\nu_{\theta}$ is well defined at
$x> x_{\mathrm{ph}}$.

From Fig.\,\ref{frekv-obecne}, we can conclude that not only the
both epicyclic frequencies but even the Keplerian frequency can
have a maximum located above the outer event black hole horizon;
this kind of behaviour is not allowed in the Kerr spacetimes. In
the next subsection we will discuss, if the maximum of
$\nu_{\mathrm{K}}(x,a,b)$ could be located above the marginally
stable or the limiting photon circular orbit.

\subsection{\label{sec:sest:1}Local extrema of the Keplerian frequency}

Denoting by $\mathcal{X}_{\mathrm{K}}$ the local extrema of the
Keplerian frequency $\nu_{\mathrm{K}}$, we can give the extrema by
the condition
\begin{equation}\label{podm-max-Kep}
    \frac{\partial \nu_{\mathrm{K}}}{\partial x}=0.
\end{equation}
From (\ref{def-Kep}), we find that the corresponding
derivative\footnote{After introducing $'$ as
$\mathrm{d}/\mathrm{d} x$.} is
\begin{equation}\label{der-Kepl}
\nu_{\mathrm{K}}'=\frac{1}{2\pi}\sqrt{\frac{\mathrm{G}
M}{r_{\mathrm{G}}^3}}\frac{x(4b-3x)}{2\sqrt{x-b}(x^2+a\sqrt{x-b})^2}=
\frac{x(4b-3x)\nu_{\mathrm{K}}}{2(x-b)(x^2+a\sqrt{x-b})},
\end{equation}
and relation (\ref{podm-max-Kep}) implies that the Keplerian
frequency has a local extremum located at
\begin{equation}\label{}
\mathcal{X}_{\mathrm{K}}= \frac{4}{3}b .
\end{equation}
The second derivative at $x=\mathcal{X}_{\mathrm{K}}$
\begin{equation}
\nu_{\mathrm{K}}''=-\frac{1}{2\pi}\sqrt{\frac{\mathrm{G}
M}{r_{\mathrm{G}}^3}}\frac{162\sqrt{3}}{\sqrt{b}\left(3\sqrt{3}a+
16b^{3/2}\right)^2}
\end{equation}
is always negative, thus the Keplerian frequency has a local
maximum at $x=\mathcal{X}_{\mathrm{K}}$, independently of the spin
parameter $a$.

Generally, the maximum is located at or above the outer event
black hole horizon if the condition
\begin{equation}
\mathcal{X}_{\mathrm{K}}\geq 1
\end{equation}
is satisfied that implies the relevant range of the tidal charge
parameter
\begin{equation}\label{Kep-int-b}
0.75\leq b\leq 1
\end{equation}
and from relation (\ref{def-amax}) we conclude that the possible
values of the black hole spin are allowed at the interval
\begin{equation}\label{Kep-int-a}
0\leq a \leq 0.5\,.
\end{equation}
The case of $a=0.5$ and $b=0.75$ corresponds to the maximally
rotating (extreme) braneworld Kerr black hole.

From relations (\ref{hor-impl}) and (\ref{fot-impl}) we obtain
\begin{align}\label{max-na-hor}
    a_{\mathrm{h}}(x=\mathcal{X}_{\mathrm{K}})&=\frac{1}{3}\sqrt{b(15-16b)}\,,\\
    a_{\mathrm{ph}}(x=\mathcal{X}_{\mathrm{K}})&=\sqrt{3b}\left(1-
    \frac{8}{9}b\right),\label{max-na-fot}
\end{align}
which implicitly determine that the maximum of the Keplerian
frequency radial profile is situated at the radius coinciding with
the radius of the black hole horizon
$\mathcal{X}_{\mathrm{K}}=x_{\mathrm{h}}$ (\ref{max-na-hor}) or
the circular photon orbit
$\mathcal{X}_{\mathrm{K}}=x_{\mathrm{ph}}$ (\ref{max-na-fot}).

\begin{figure*}[!tbp]
\subfigure{\includegraphics[width=.48\hsize]{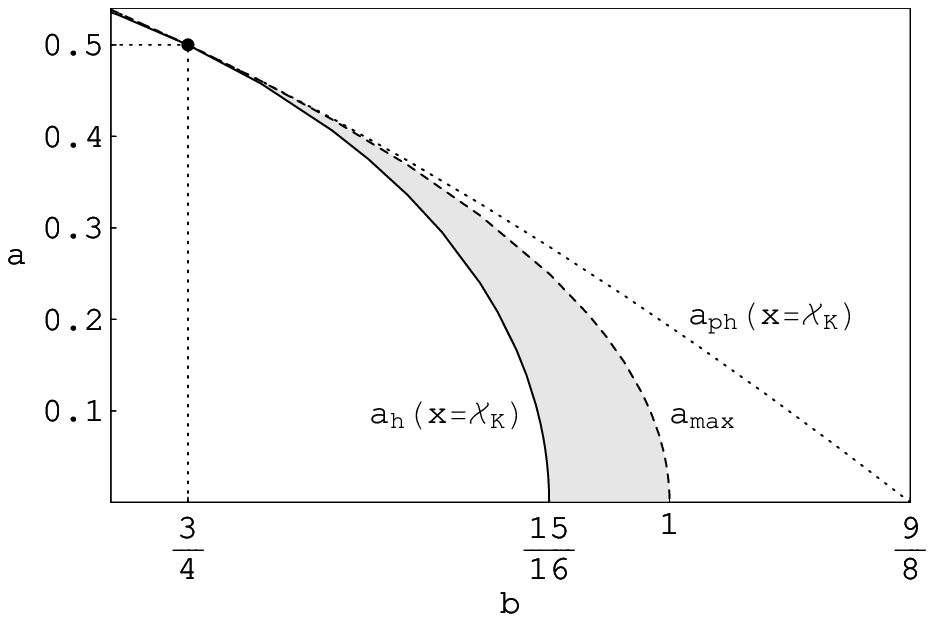}}\quad
\subfigure{\includegraphics[width=.48\hsize]{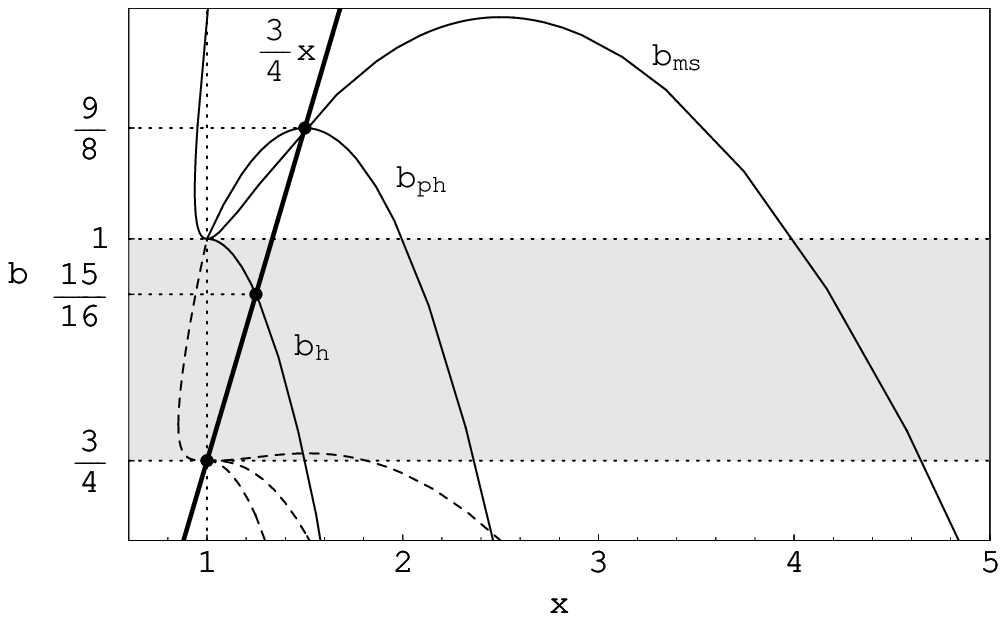}}
\caption{\label{ex-kep}\textit{Left panel}: the functions
$a_{\mathrm{h}}(x=\mathcal{X}_{\mathrm{K}})$ (solid line),
$a_{\mathrm{ph}}(x=\mathcal{X}_{\mathrm{K}})$ (dotted line)
determining that the maximum of the Keplerian frequency is
situated exactly at the black hole horizon radius
$\mathcal{X}_{\mathrm{K}}=x_{\mathrm{h}}$, or at the photon orbit
$\mathcal{X}_{\mathrm{K}}=x_{\mathrm{ph}}$. Dashed line represents
maximum possible value of the black hole spin corresponding to the
concrete value of the brany parameter $b$, so the area above
$a_{\mathrm{max}}$ belongs to naked singularities. The gray area
illustrates all possible combinations of the black hole spin $a$
and the tidal charge $b$ for which the Keplerian frequency has its
maximum located at $x_{\mathrm{h}}\leq\mathcal{X}_{\mathrm{K}}\leq
x_{\mathrm{ph}}$. The \textit{right panel} displays the functions
$b_{\mathrm{h}}$, $b_{\mathrm{ph}}$ and $b_{\mathrm{ms}}$
implicitly determining the location of the black hole horizon, the
limiting photon orbit and the marginally stable orbit for $a=0$
(solid lines) and $a=0.5$ (dashed lines). Thick line represents
the maximum of the Keplerian frequency $b=3x/4$.}
\end{figure*}

\begin{figure*}[!tbp]
\subfigure[][]{\includegraphics[width=.48\hsize]{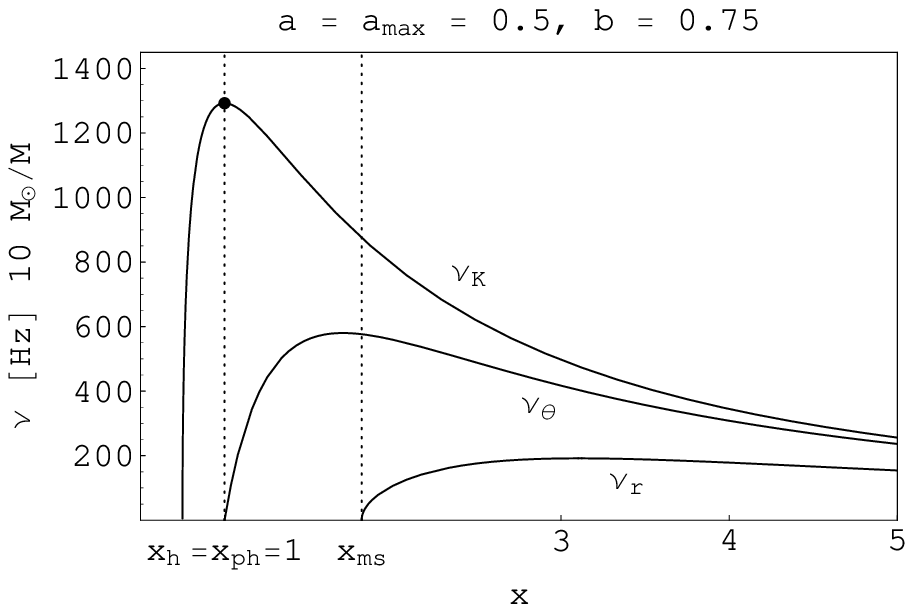}}\quad
\subfigure[][]{\includegraphics[width=.48\hsize]{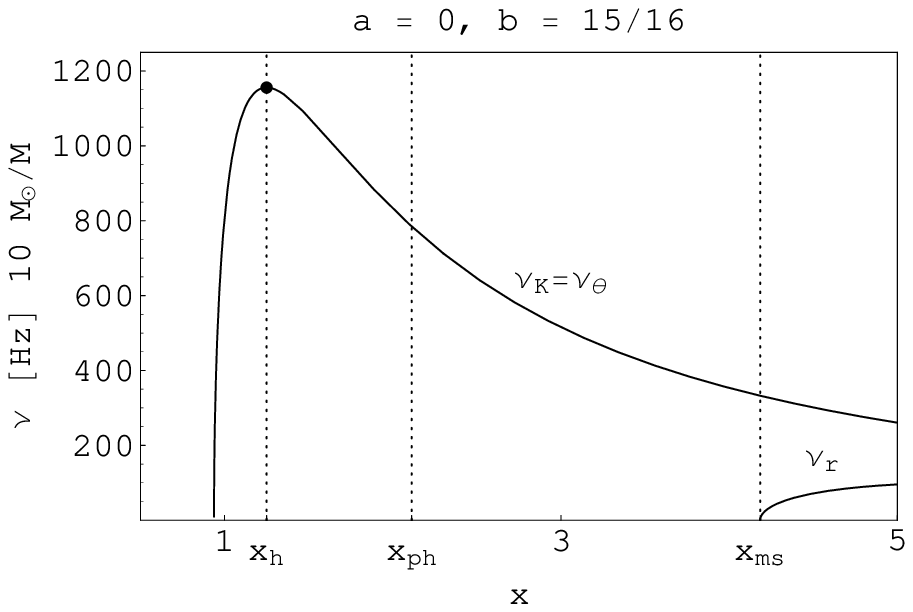}}\\
\subfigure[][]{\includegraphics[width=.48\hsize]{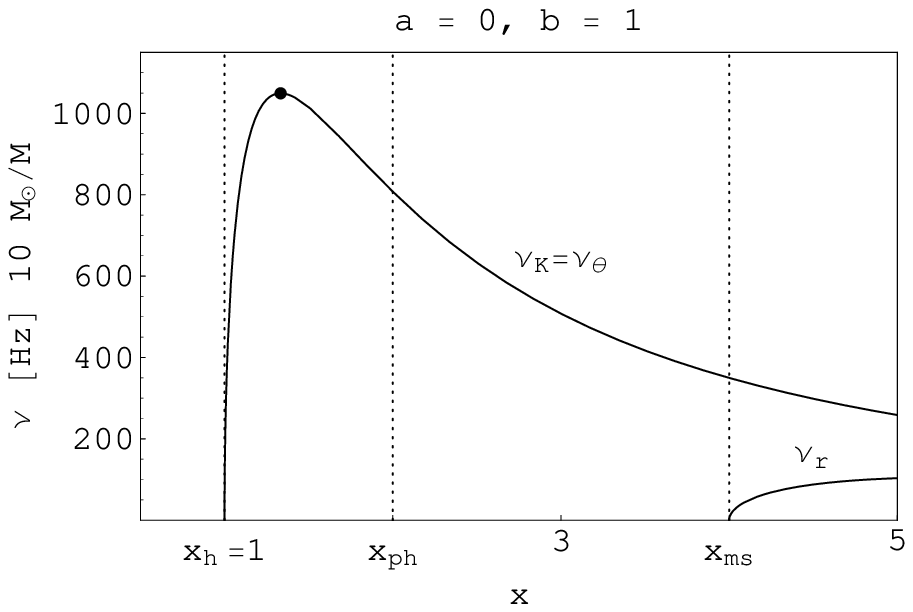}}\quad
\subfigure[][]{\includegraphics[width=.48\hsize]{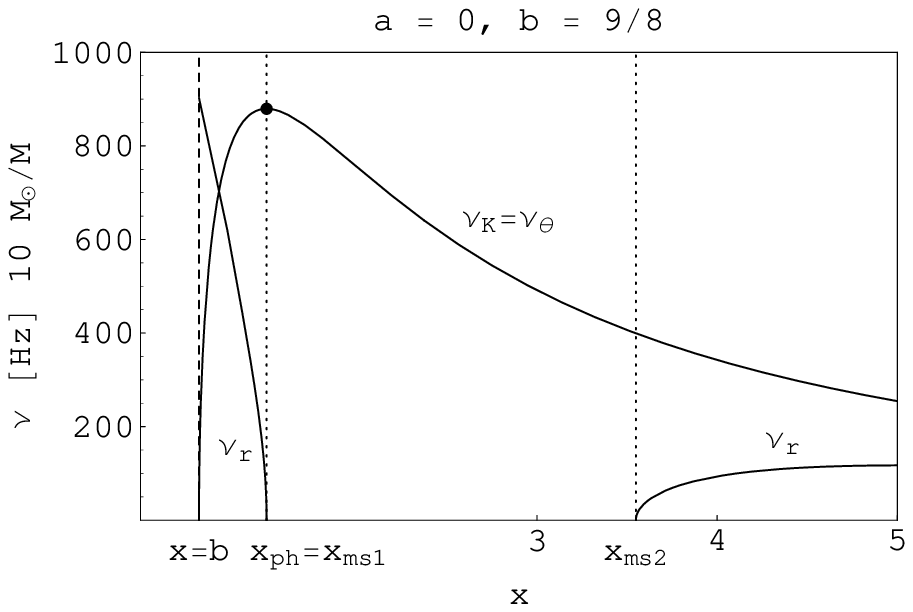}}
\caption{\label{ex-kep-rezy}The Keplerian and epicyclic
frequencies for various values of the black hole spin $a$ and the
tidal charge $b$: (a) the only case when the Keplerian frequency
has its maximum located exactly at
$\mathcal{X}_{\mathrm{K}}=x_{\mathrm{h}}=x_{\mathrm{ph}}$;
(b)\,--\,(d) represent braneworld Reissner--Nordstr\"{o}m
spacetime where $\nu_{\mathrm{K}}=\nu_{\theta}$; (a),\,(c)
correspond to the extreme black holes, (d) to a naked singularity
spacetime (notice that the condition $x>b$ has to be satisfied).}
\end{figure*}

The functions $a_{\mathrm{h}}(x=\mathcal{X}_{\mathrm{K}})$,
$a_{\mathrm{ph}}(x=\mathcal{X}_{\mathrm{K}})$ are shown in the
left panel of Fig.\,\ref{ex-kep}. We can see that for brany Kerr
black holes all possible values of the tidal charge parameter and
black hole spin imply the condition
\begin{equation}
    a_{\mathrm{ph}}(x=\mathcal{X}_{\mathrm{K}})\geq
a_{\mathrm{max}},
\end{equation}
thus the maximum of the Keplerian frequency could never be located
above the photon orbit $x_{\mathrm{ph}}$ (and the marginally
stable orbit $x_{\mathrm{ms}}$). Only for maximally rotating black
hole with $b=0.75$ and $a=a_{\mathrm{max}}=0.5$, the maximum is
situated exactly at the Boyer--Lindquist coordinate radius of the
limiting photon orbit that merges with the radius of the black
hole horizon, so
$\mathcal{X}_{\mathrm{K}}=x_{\mathrm{ph}}=x_{\mathrm{h}}=1$ (see
Fig.\,\ref{ex-kep-rezy}).\footnote{However, note the behaviour of
extreme Kerr black hole at $x=1$, where the same coordinate
corresponds to an infinitely long throat of the proper radial
distance, with different positions of the horizon and the circular
photon, marginally bound and marginally stable orbits
\cite{Bar:1973:BlaHol:}.}

We can conclude that for brany parameter from interval
(\ref{Kep-int-b}) and black hole spin from interval
(\ref{Kep-int-a}), the Keplerian frequency has its maximum located
between the black hole horizon and the photon circular orbit
\begin{equation}
x_\mathrm{h}\leq\mathcal{X}_{\mathrm{K}}\leq x_{\mathrm{ph}}.
\end{equation}

Clearly, the maximum of the Keplerian frequency is physically
irrelevant for all the brany Kerr black holes. In astrophysically
relevant radii above the photon orbit, $x>x_{\mathrm{ph}}$, the
Keplerian frequency is a monotonically decreasing function of
radius for the whole range of the brany tidal charge parameter
$b$, as in the standard Kerr spacetimes. In the brany Kerr naked
singularity spacetimes, the situation is more complicated, because
of the complexity of the behaviour of the  functions
$a_{\mathrm{ph}}(x,b)$ and $a_{\mathrm{ms}}(x,b)$. The situation
is clearly illustrated in Fig.\,\ref{ex-kep}.

\subsection{\label{sec:sest:2}Local extrema of epicyclic frequencies}

It is important to determine for a given spacetime (with
parameters $a$ and $b$), if profiles of both the epicyclic
frequencies have a local extrema in the region of relevance
($x>x_{\mathrm{ms}}(a,b)$), since in such a case the resonant
radii might not be given uniquely (i.e., the same frequency ratio
can appear at different resonant radii, with different
frequencies), and the analysis of resonant phenomena must then be
done very carefully \cite{Ter-Stu:2005:ASTRA:}.

The local extrema of the radial and vertical epicyclic frequencies
$\mathcal{X}_{\mathrm{r}}$, $\mathcal{X}_\theta$ are given by the
condition
\begin{equation}\label{podm-max-epi}
    \frac{\partial \nu_{i}}{\partial x}=0\quad\textrm{for}\quad \mathcal
    {X}_{i},\quad
    \textrm{where}\quad
    i\in \{\mathrm{r},\theta\}.
\end{equation}
Using~(\ref{def-epi-rad}) and~(\ref{def-epi-ver}), the
corresponding derivatives can be given in the form
\begin{align}\label{der-epi}
\nu_{i}'&=\sqrt{\alpha_{i}}\left(\nu_{\mathrm{K}}'+\frac{\alpha_{i}'}{2
\alpha_{i}}\nu_{\mathrm{K}}\right),\\
\alpha_{i}'&=\frac{\beta_{i}}{x^3 (x-b )^{5/2}},
\end{align}
where $\nu_{\mathrm{K}}'$ is given by~(\ref{der-Kepl}), and
\begin{align}
\beta_{\mathrm{r}}(x,a,b)&=-4 a (3 x-4 b ) (x-b)^2 \nonumber\\
&\quad +\sqrt{x-b } \left[a^2 \left(6 x^2-15 x b +8 b
^2\right)-\left(
x^3b-6x^3+18 x^2 b -21 x b ^2+8 b ^3\right)\right],\\
\beta_{\theta}(x,a,b)&= a \left(x-a \sqrt{x-b }-b \right) \left(6
x^2-9 x b +4 b ^2\right).
\end{align}
Relations~(\ref{podm-max-epi}) and~(\ref{der-epi}) imply the
condition determining extrema $\mathcal{X}_{i}(a,b)$ of the
epicyclic frequency profiles
\begin{equation}\label{podm-max-epi-impl}
\beta_{i}(x,a,b)=-\frac{2\nu_{\mathrm{K}}'}{\nu_{\mathrm{K}}}\,x^3
(x-b)^{5/2}\, \alpha_{i}(x,a,b), \qquad i\in
\{\mathrm{r},\theta\}.
\end{equation}
We have checked that in the case of counterrotating orbits ($a <
0$) the extrema $\mathcal{X}_{\theta}$ are located under the
photon circular orbit and the extrema $\mathcal{X}_{\mathrm{r}}$
are just extensions of the $\mathcal{X}_{\mathrm{r}}$ for
corotating case. Therefore, we focus on the case of corotating
orbits ($a > 0$) in the next discussion.

\begin{figure*}[!tbp]
\subfigure{\includegraphics[width=.48\hsize]{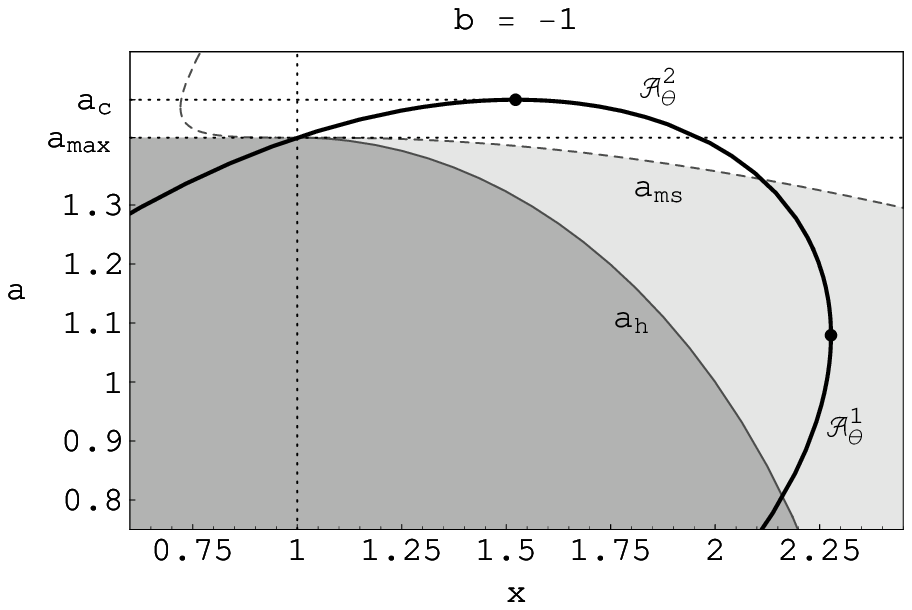}}\quad
\subfigure{\includegraphics[width=.48\hsize]{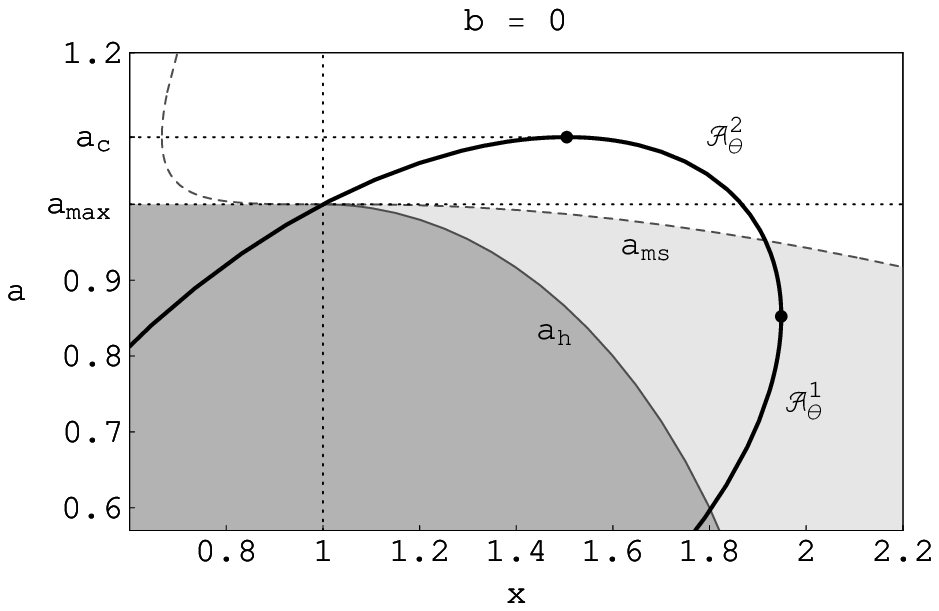}}\\
\subfigure{\includegraphics[width=.48\hsize]{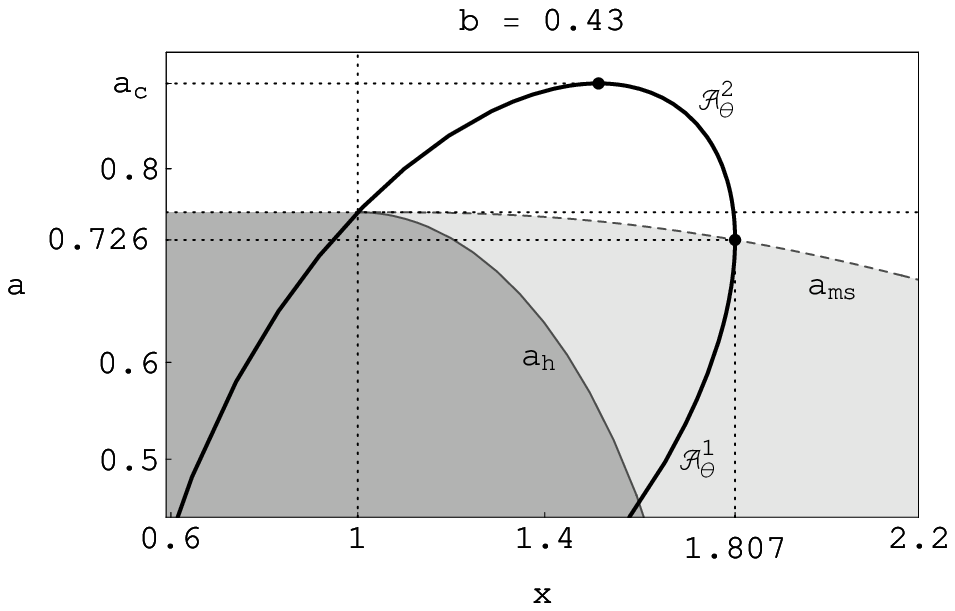}}\quad
\subfigure{\includegraphics[width=.48\hsize]{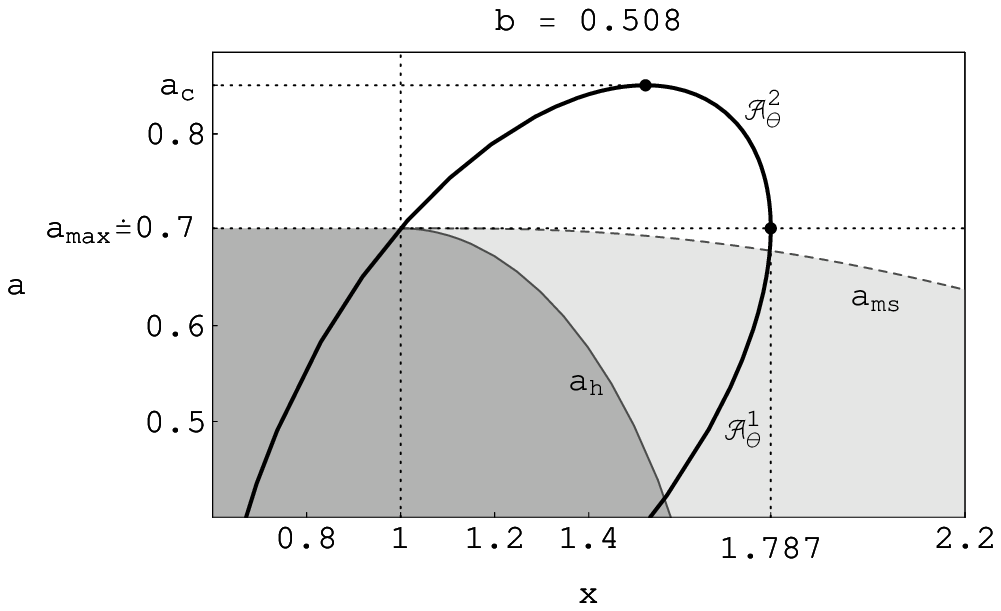}}\\
\subfigure{\includegraphics[width=.48\hsize]{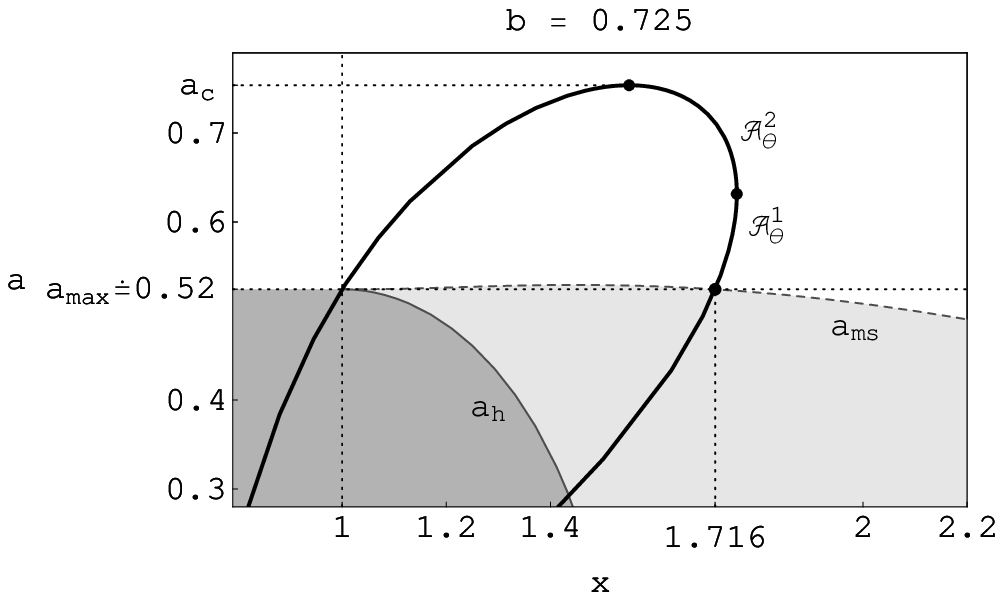}}\quad
\subfigure{\includegraphics[width=.48\hsize]{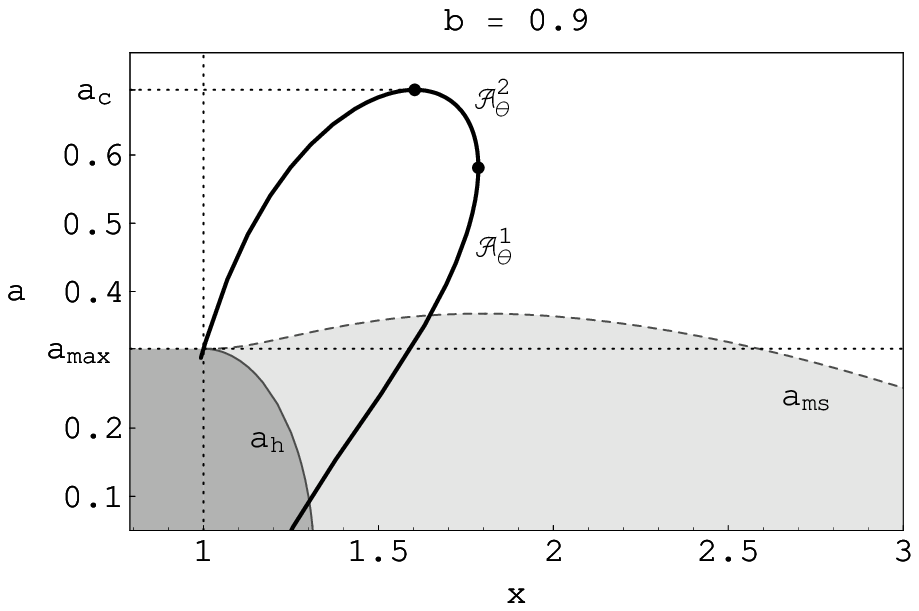}}
\caption{\label{ex-epi-ver} The functions $\mathcal {A}_\theta^1
(x=\mathcal{X}_{\theta},b)$, $\mathcal {A}_\theta^2
(x=\mathcal{X}_{\theta},b)$, implicitly determining the locations
$\mathcal{X}_{\theta}$ of the vertical epicyclic frequency local
extrema for typical values of the brany parameter $b$.}
\end{figure*}

In Fig.~\ref{ex-epi-rad} (\ref{ex-epi-ver}) we show for various
values of brany parameter $b$ curves $\mathcal {A}_\mathrm{r}^k
(x=\mathcal{X}_{\mathrm{r}},b)$ ($\mathcal {A}_\theta^k
(x=\mathcal{X}_{\theta},b)$), $k\in\{1,2\}$ implicitly determined
by the relations~(\ref{podm-max-epi-impl}); index $k$ denotes
different branches of the solution of~(\ref{podm-max-epi-impl}).

The marginally stable orbit radius $x_{\mathrm{ms}}$ (where
$\alpha_{\mathrm{r}}=0$) falls with tidal charge $b$ growing and
spin $a$ being fixed.

\subsubsection{Radial epicyclic frequency}

\begin{figure*}[!tbp]
\subfigure{\includegraphics[width=.48\hsize]{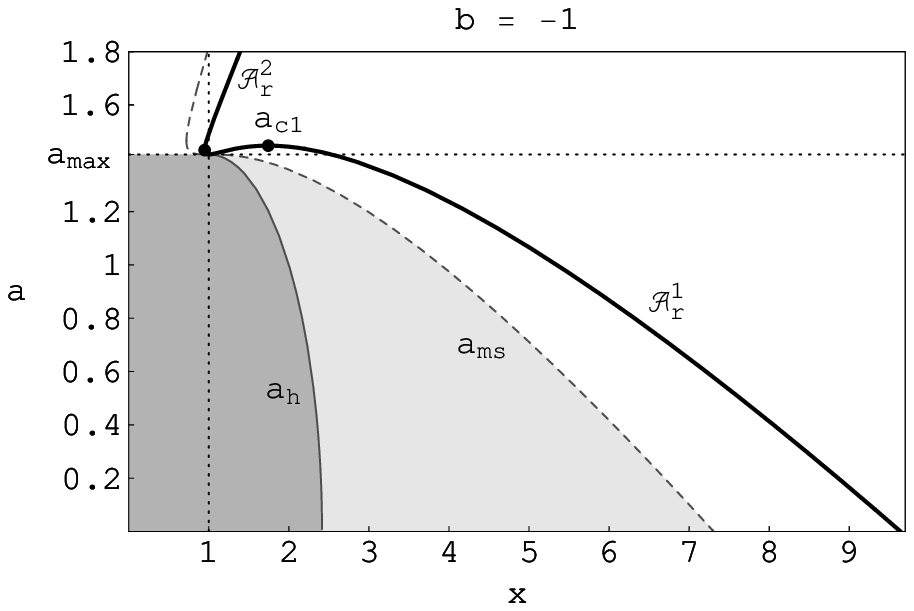}}\quad
\subfigure{\includegraphics[width=.48\hsize]{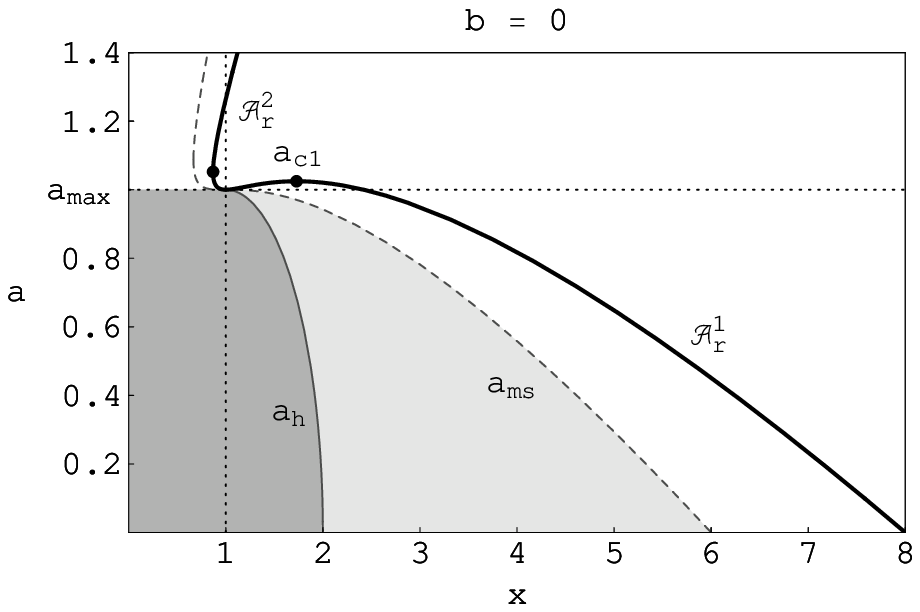}}\\
\subfigure{\includegraphics[width=.48\hsize]{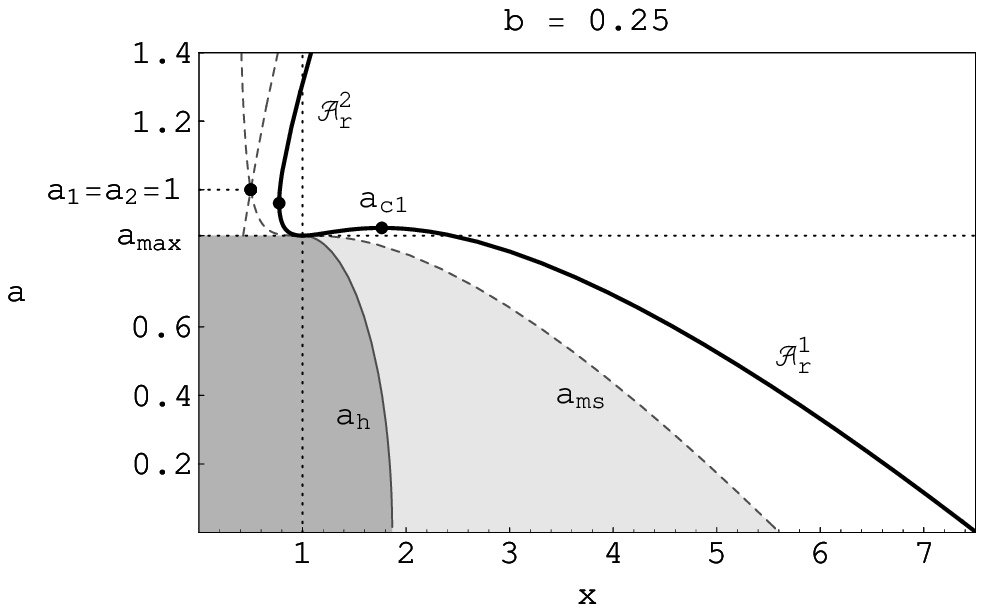}}\quad
\subfigure{\includegraphics[width=.48\hsize]{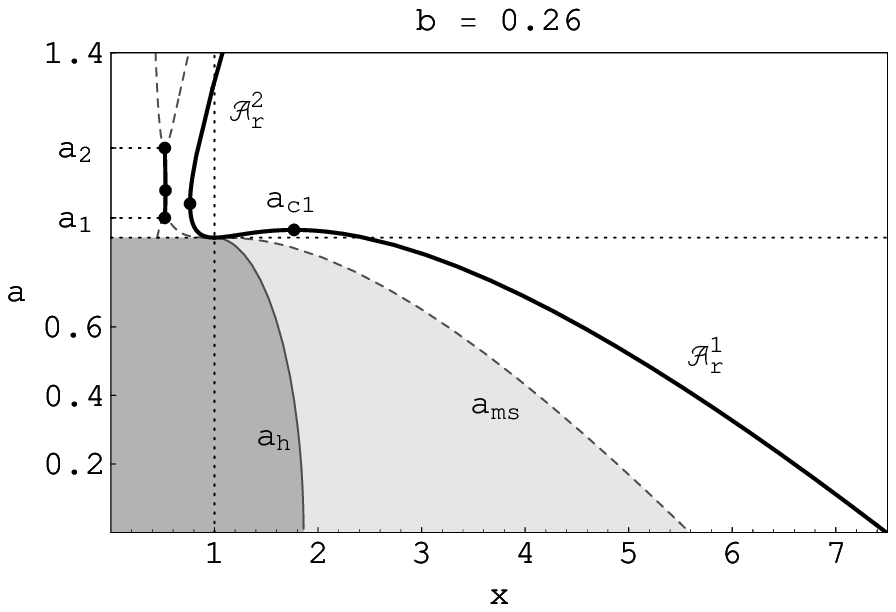}}\\
\subfigure{\includegraphics[width=.48\hsize]{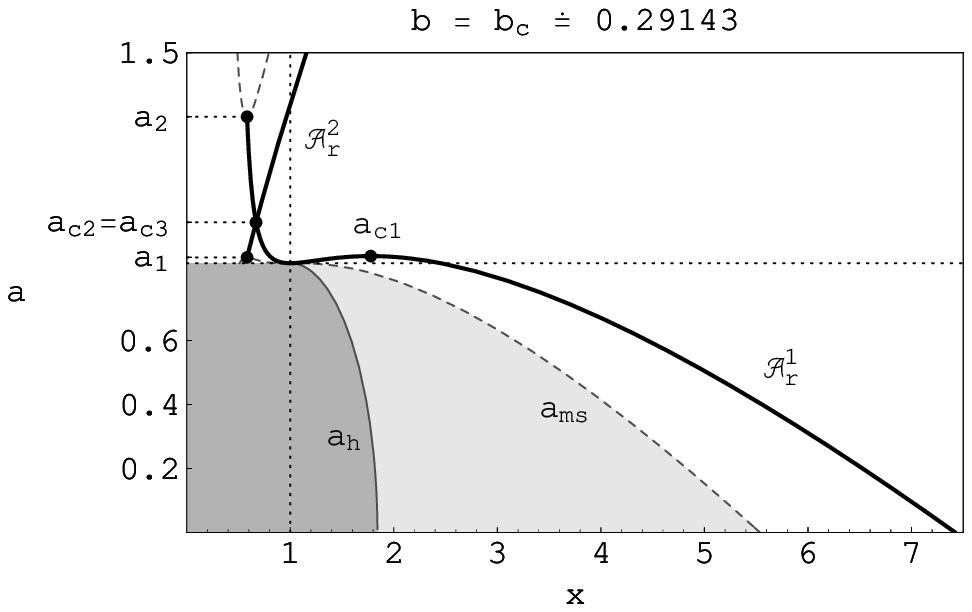}}\quad
\subfigure{\includegraphics[width=.48\hsize]{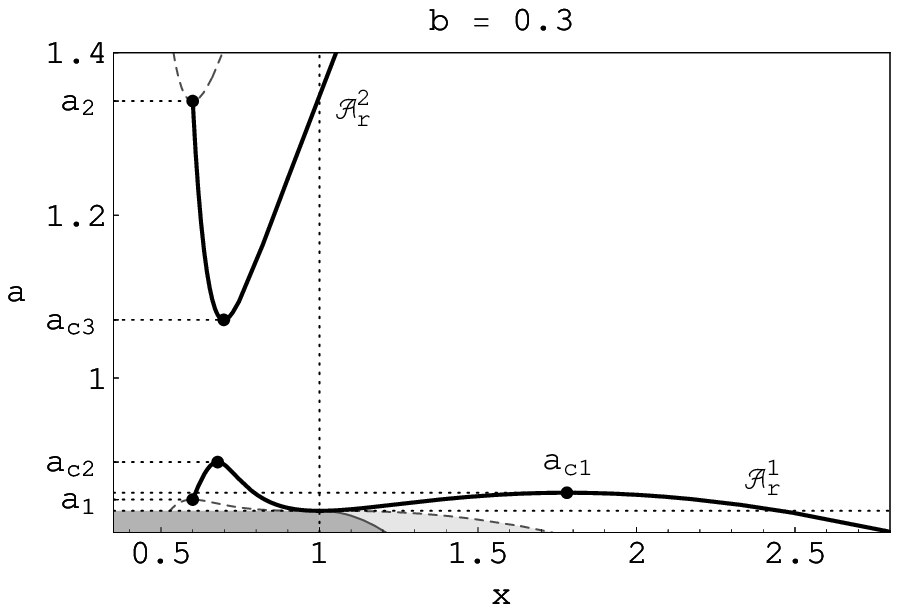}}\\
\subfigure{\includegraphics[width=.48\hsize]{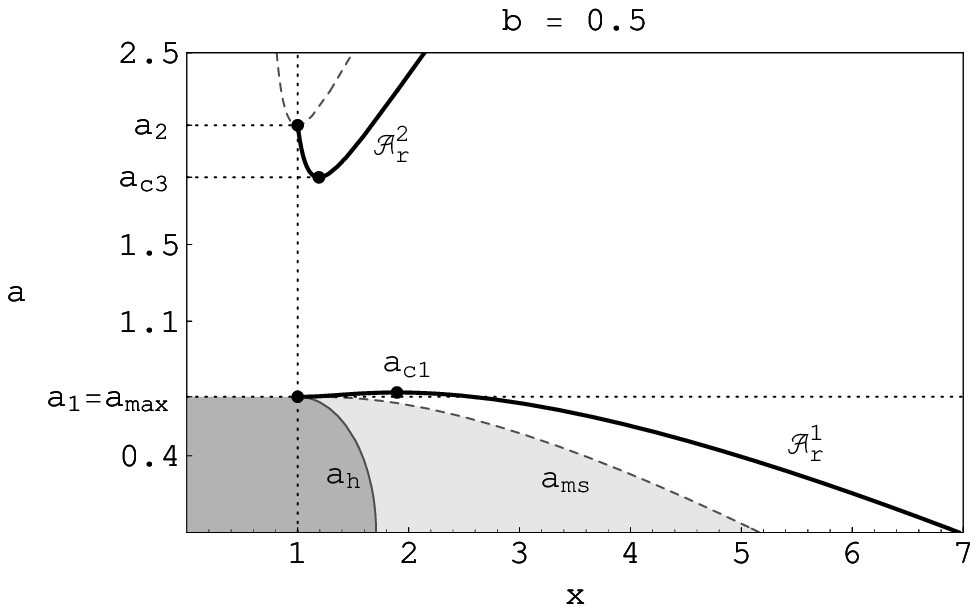}}\quad
\subfigure{\includegraphics[width=.48\hsize]{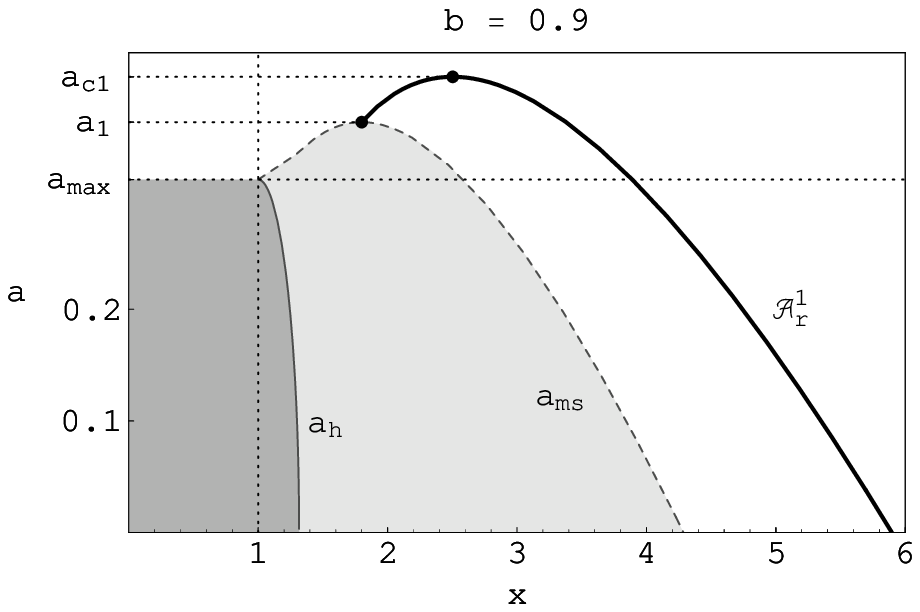}}
\caption{\label{ex-epi-rad}The functions $\mathcal
{A}_\mathrm{r}^1 (x=\mathcal{X}_{\mathrm{r}},b)$, $\mathcal
{A}_\mathrm{r}^2 (x=\mathcal{X}_{\mathrm{r}},b)$, implicitly
determining the locations $\mathcal{X}_{\mathrm{r}}$ of the radial
epicyclic frequency local extrema for various values of brany
parameter $b$.}
\end{figure*}

For all possible values of the brany parameter $b$, the radial
epicyclic frequency $\nu_{\mathrm{r}}$ has one local maximum for
braneworld Kerr black holes with rotational parameter restricted
by
\begin{equation}
0 \leq a\leq a_{\mathrm{max}}(b).
\end{equation}
The local maximum is always located above the marginally stable
orbit $x_{\mathrm{ms}}$ (see Fig.\,\ref{ex-epi-rad}).

In the case of naked singularities, the situation is complicated
and the discussion can be separated into three parts according to
the parameter $b$.

\begin{figure*}[!tbp]
\subfigure{\includegraphics[width=.31\hsize]{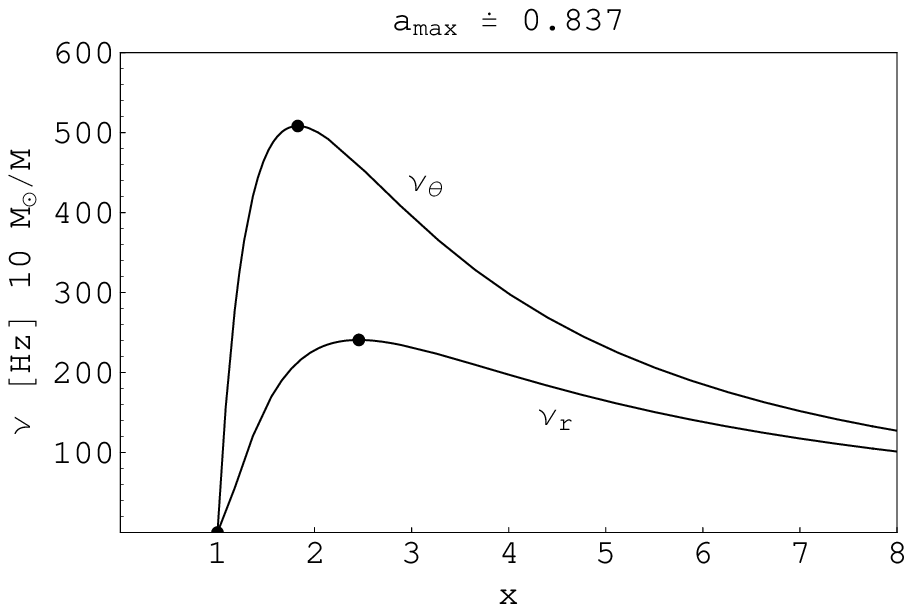}}\quad
\subfigure{\includegraphics[width=.31\hsize]{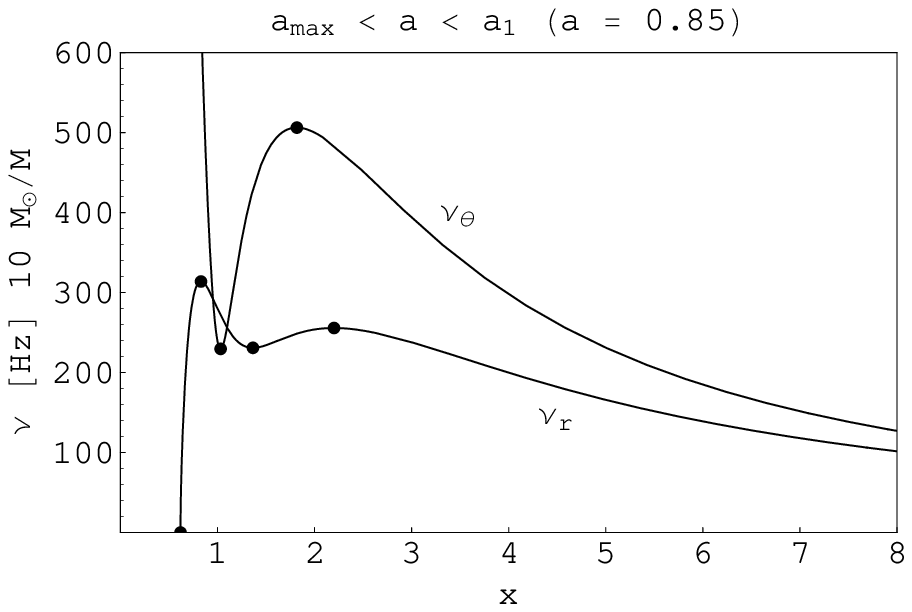}}\quad
\subfigure{\includegraphics[width=.31\hsize]{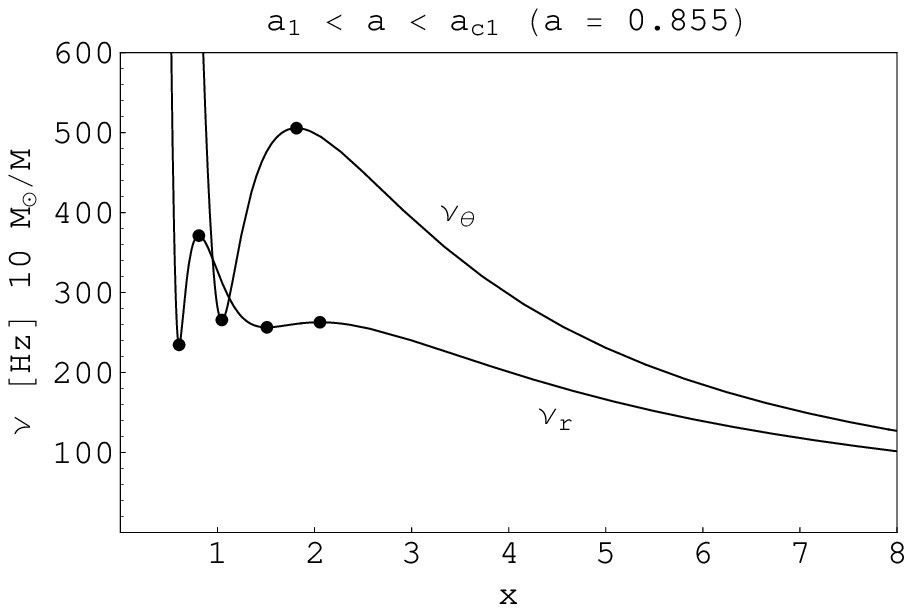}}\\
\subfigure{\includegraphics[width=.31\hsize]{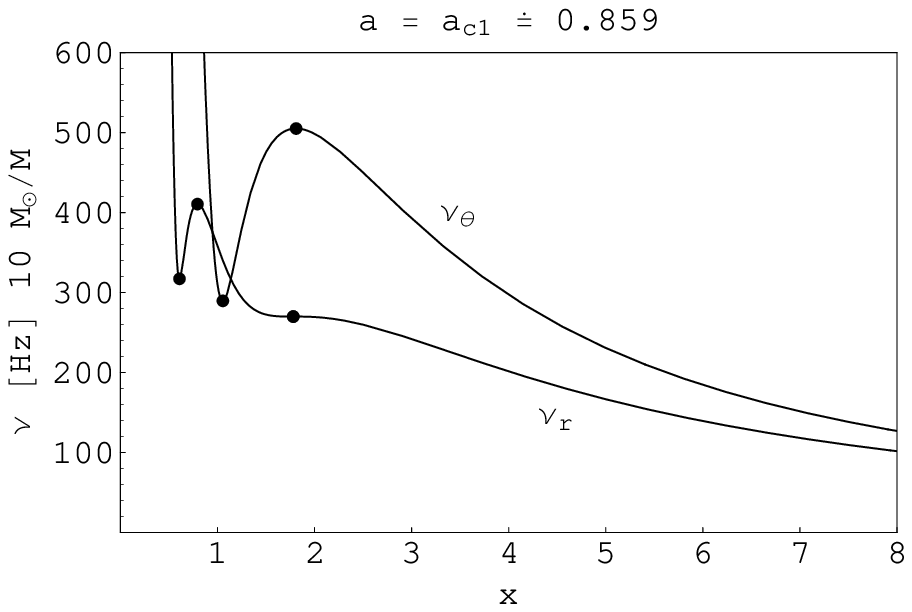}}\quad
\subfigure{\includegraphics[width=.31\hsize]{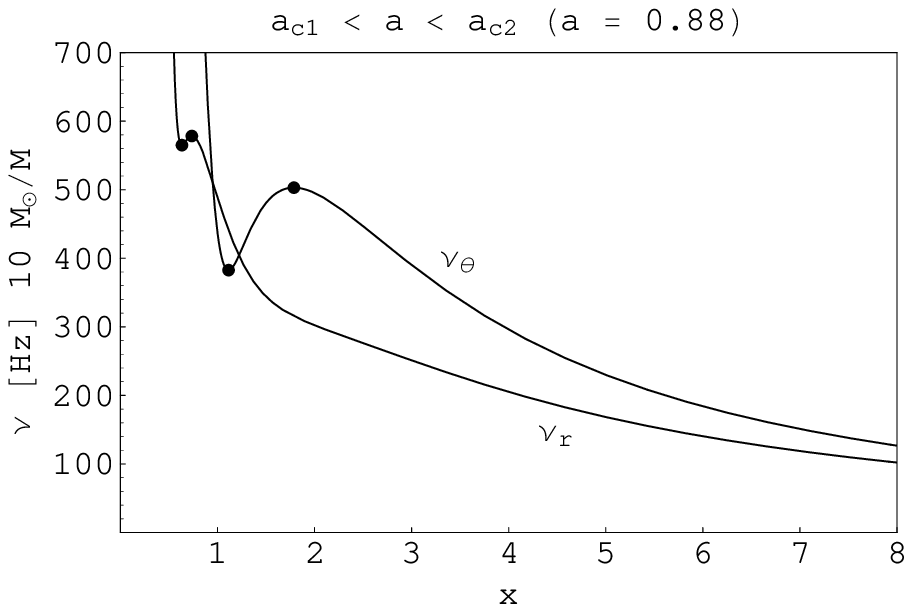}}\quad
\subfigure{\includegraphics[width=.31\hsize]{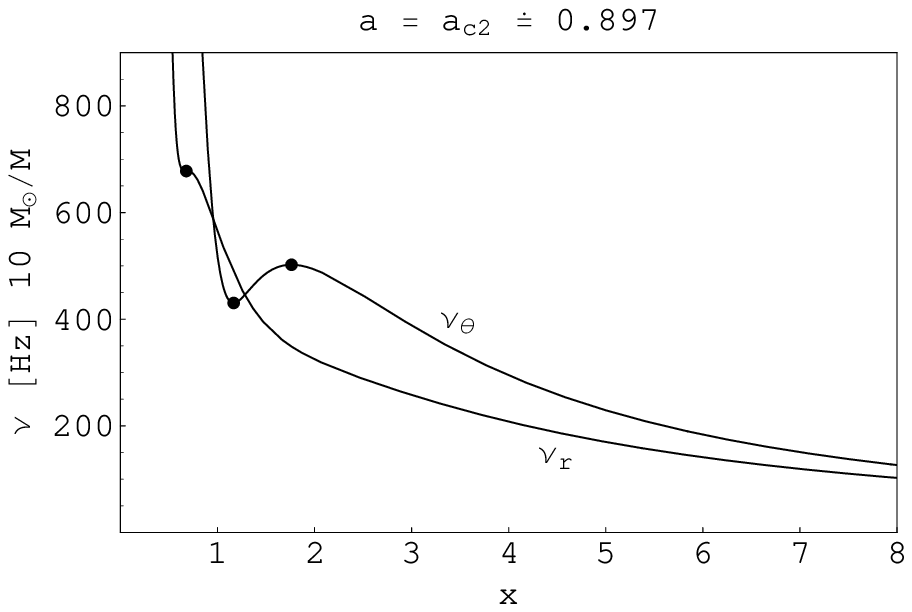}}\\
\subfigure{\includegraphics[width=.31\hsize]{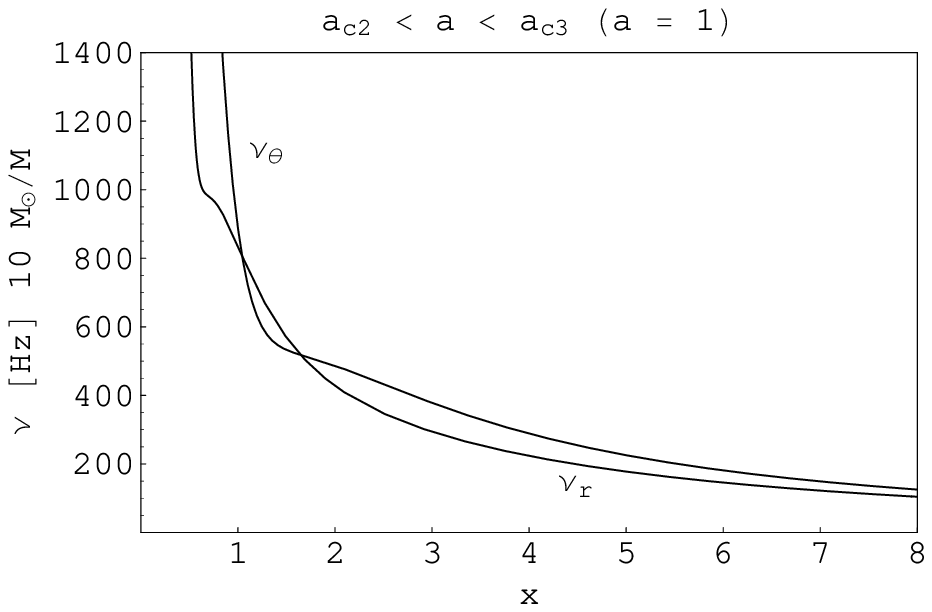}}\quad
\subfigure{\includegraphics[width=.31\hsize]{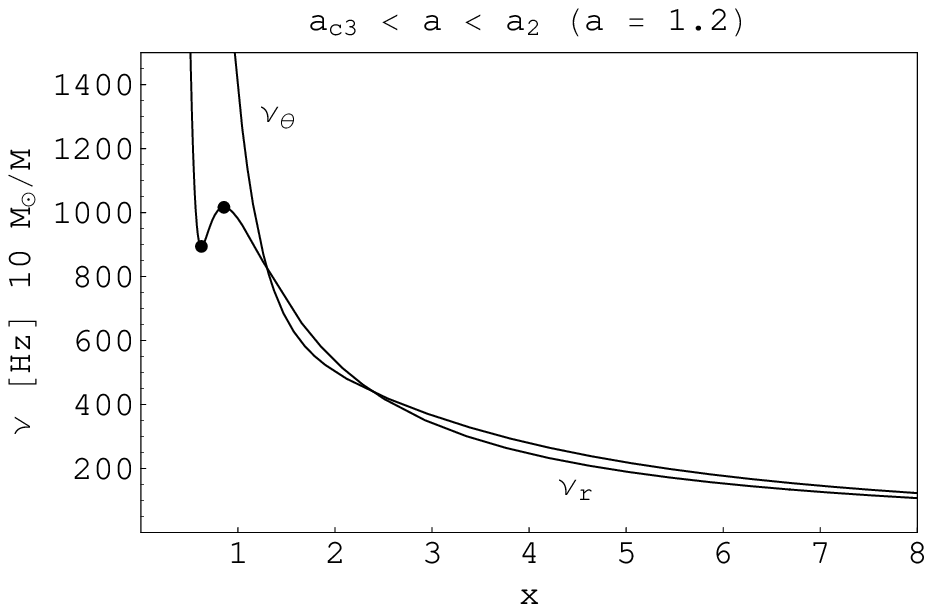}}\quad
\subfigure{\includegraphics[width=.31\hsize]{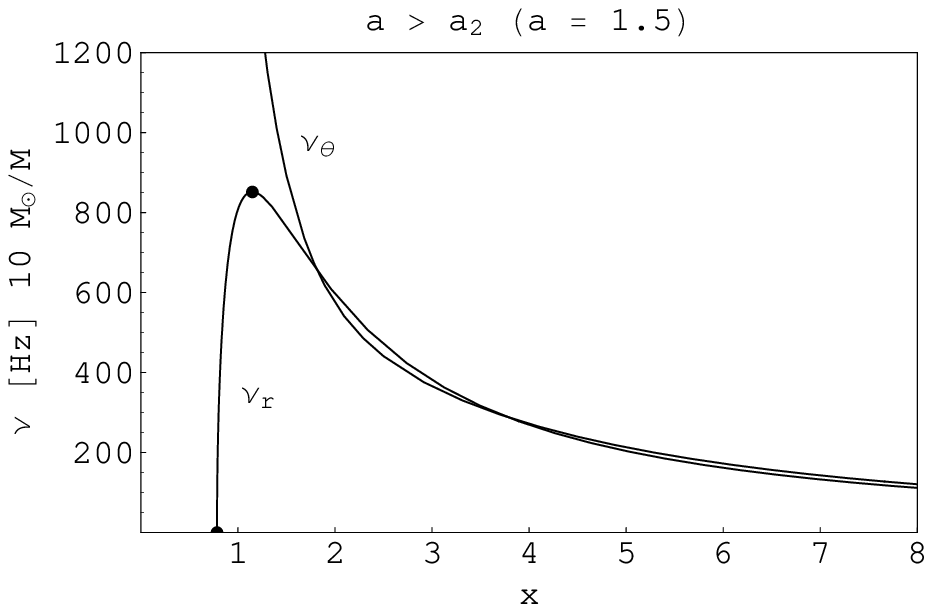}}
\caption{\label{frkv-rezy}The behaviour of the epicyclic
frequencies for $b=0.3$ and for some representative values of
rotational parameter $a$ in braneworld Kerr naked singularity
spacetimes. For comparison, an extreme braneworld Kerr black hole
case is included.}
\end{figure*}

\begin{enumerate}
    \item[(a)] $b < b_{\mathrm{c}}\doteq 0.29143$

        In the field of such brany Kerr naked singularities the radial
        epicyclic frequency has two local maxima and one local
        minimum for
        \begin{equation}
        a_{\mathrm{max}}<a<a_{\mathrm{c1}(\mathrm{r})},
        \end{equation}
        where $a_{\mathrm{c1}(\mathrm{r})}$ corresponds to the local
        maximum of $\mathcal {A}_\mathrm{r}^1
        (x=\mathcal{X}_{\mathrm{r}},b)$, and one local maximum for
        \begin{equation}
        a\geq a_{\mathrm{c1}(\mathrm{r})},
        \end{equation}
        as usual in Kerr naked singularity spacetimes
        \cite{Ter-Stu:2005:ASTRA:}. However, for values of spin $a$
        satisfying the condition
        \begin{equation}\label{interval-a-min}
        a_1< a < a_2,
        \end{equation}
        where $a_1$ is given by the equation
        \begin{equation}\label{rce-a1}
        \mathcal {A}_\mathrm{r}^1
        (x=\mathcal{X}_{\mathrm{r}},b)=a_{\mathrm{ms}}
        \end{equation}
        and $a_2$ is given by
        \begin{equation}\label{rce-a2}
        \mathcal {A}_\mathrm{r}^2
        (x=\mathcal{X}_{\mathrm{r}},b)=a_{\mathrm{ms}},
        \end{equation}
        the radial epicyclic frequency has one extra local minimum,
        because in brany Kerr naked singularity spacetimes with these special values
        of $a$ there exists no marginally stable orbit (as given by the
        condition (\ref{podm-mez}), i.e., $\nu_{\mathrm{r}}=0$).
        Therefore, at $x>0$ the radial epicyclic frequency is for $a$ from
        the interval~(\ref{interval-a-min}) always greater than zero.

        For braneworld Kerr naked singularities with $b>b_{\mathrm{c}}$,
        the situation is even more complicated, as we can see in
        Fig.\,\ref{ex-epi-rad}. We denote $a_{\mathrm{c1}(\mathrm{r})}$
        and $a_{\mathrm{c2}(\mathrm{r})}$ the local maxima of $\mathcal
        {A}_\mathrm{r}^1 (x=\mathcal{X}_{\mathrm{r}},b)$, and
        $a_{\mathrm{c3}(\mathrm{r})}$ the local minimum of $\mathcal
        {A}_\mathrm{r}^2 (x=\mathcal{X}_{\mathrm{r}},b)$.

    \item[(b)] $b_{\mathrm{c}} < b < 0.5$

        In such a case, we consider two possible relations
        of the spin parameter, namely
        $a_{\mathrm{c2}(\mathrm{r})}>a_{\mathrm{c1}(\mathrm{r})}$
        (or $a_{\mathrm{c2}(\mathrm{r})}<a_{\mathrm{c1}(\mathrm{r})}$).
        Then we find that the radial epicyclic frequency has for
        $a_{\mathrm{max}}<a\leq a_1$ two local maxima and one local
        minimum, for $a_1<a< a_{\mathrm{c1}(\mathrm{r})}$ ($a_1<a<
        a_{\mathrm{c2}(\mathrm{r})}$) two local maxima and also two local
        minima, for $a_{\mathrm{c1}(\mathrm{r})} \leq a<
        a_{\mathrm{c2}(\mathrm{r})}$ ($a_{\mathrm{c2}(\mathrm{r})} \leq a<
        a_{\mathrm{c1}(\mathrm{r})}$) one local maximum and one local
        minimum, for $a_{\mathrm{c2}(\mathrm{r})} \leq a \leq
        a_{\mathrm{c3}(\mathrm{r})}$ ($a_{\mathrm{c1}(\mathrm{r})} \leq a
        \leq a_{\mathrm{c3}(\mathrm{r})}$) the radial epicyclic frequency
        is a monotonically decreasing function of the radial coordinate
        without any extrema, for $a_{\mathrm{c3}(\mathrm{r})} < a <a_2$ it
        has again one local maximum and one local minimum, and finally for
        $a \geq a_2$ it has only one local maximum as in the black hole
        spacetimes ($a_1$ and $a_2$ are given by~(\ref{rce-a1})
        and~(\ref{rce-a2}), the condition
        $a_{\mathrm{max}}<a_1<a_{\mathrm{c1}(\mathrm{r})}$ is always
        satisfied). Notice that for $a_1 < a < a_2$, $\nu_{\mathrm{r}}$
        could not be equal to zero (see Fig.\,\ref{frkv-rezy}).

        For $b=0.5$, there is
        $a_1=a_{\mathrm{c2}(\mathrm{r})}=a_{\mathrm{max}}=\sqrt{2}/2$,
        and for $a_{\mathrm{max}}<a < a_{\mathrm{c1}(\mathrm{r})}$ the
        radial epicyclic frequency has one local maximum and one local
        minimum, for $a_{\mathrm{c1}(\mathrm{r})} \leq a \leq
        a_{\mathrm{c3}(\mathrm{r})}$ it is a monotonically decreasing
        function of the radial coordinate without any extrema, for
        $a_{\mathrm{c3}(\mathrm{r})} < a <a_2$ it has again one local
        maximum and one local minimum, and for $a \geq a_2$ it has only
        one local maximum.

    \item[(c)] $b > 0.5$

        In the case of braneworld Kerr naked singularities with
        such brany parameters the behaviour of the radial epicyclic frequency
        is different due
        to the effect related to the loci of the marginally stable orbits
        as described in Section~\ref{sec:ctyri} (see relation
        (\ref{hrb-mez})). For $a_{\mathrm{max}}<a \leq a_1$,
        $\nu_{\mathrm{r}}$ has one local maximum, for $a_1<a <
        a_{\mathrm{c1}(\mathrm{r})}$ it has one local maximum and one
        local minimum, for $a_{\mathrm{c1}(\mathrm{r})} \leq a \leq
        a_{\mathrm{c3}(\mathrm{r})}$ it is a monotonically decreasing
        function of the radial coordinate without any extrema, for
        $a_{\mathrm{c3}(\mathrm{r})} < a <a_2$ it has again one local
        maximum and one local minimum, and finally for $a \geq a_2$ it has
        only one local maximum.
\end{enumerate}

\subsubsection{Vertical epicyclic frequency}

Qualitatively different types of its behaviour in dependence on
the brany parameter $b$ are illustrated in Fig.\,\ref{ex-epi-ver}.

The vertical epicyclic frequency has a local maximum (at $x
> x_{\mathrm{ms}}$) only for rapidly rotating black holes
with (see Fig.\,\ref{ex-epi-ver})
\begin{equation}
a_{\mathrm{ms}(\theta)}< a < a_{\mathrm{max}}
\qquad\textrm{and}\qquad b < 0.725,
\end{equation}
where $a_{\mathrm{ms}(\theta)}$ is the solution of the equation
\begin{equation}
a_{\mathrm{ms}}= \mathcal {A}_\theta^k,
\end{equation}
for $k=1$ or 2. There is $a_{\mathrm{ms}(\theta)}=
a_{\mathrm{max}}(b=0.725)\doteq 0.524$. (For maximally rotating
black hole with $b=0.725$ and $a_{\mathrm{max}}\doteq 0.524$, the
local maximum is located exactly at the radius of the marginally
stable orbit, $\mathcal{X}_{\theta} = x_{\mathrm{ms}}$). Recall
that in the black hole case the local maximum of $\nu_\theta
(x,a,b)$ is relevant in resonant effects at $x > x_{\mathrm{ms}}$.
For $b>0.725$ the vertical epicyclic frequency is a monotonically
decreasing function of radius for the whole range of black hole
rotational parameter $a$.

In the braneworld Kerr naked singularity spacetimes, the function
$\nu_\theta$ has a local minimum and a local maximum for
\begin{equation}
a_{\mathrm{max}}< a < a_{\mathrm{c}(\theta)},
\end{equation}
and has no astrophysically relevant local extrema for
\begin{equation}
a \geq a_{\mathrm{c}(\theta)}.
\end{equation}

We can summarize that for braneworld black holes with any value of
$b$ the radial epicyclic frequency profile has a local maximum and
zero point at $x=x_{\mathrm{ms}}$, as in the case of standard Kerr
black holes. The vertical epicyclic frequency has a local maximum
for spin $a$ close to extreme black hole states for values of
$b<0.725$ as in the standard Kerr case, but it is purely
monotonically decreasing function of radius for black holes with
$b \geq 0.725$.

The behaviour of the epicyclic frequencies substantially differs
for braneworld Kerr naked singularities in comparison with
braneworld Kerr black holes. Examples of the behaviour of the
epicyclic frequencies in Kerr naked singularity spacetimes with
$b=0.3$ are given in Fig.\,\ref{frkv-rezy}.

\begin{figure*}[!tbp]
\subfigure{\includegraphics[width=.31\hsize]{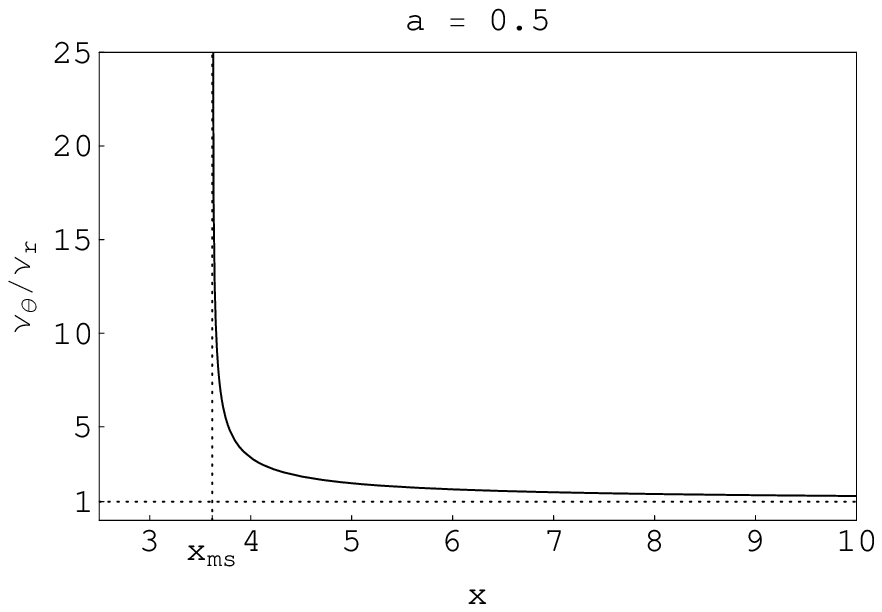}}\quad
\subfigure{\includegraphics[width=.31\hsize]{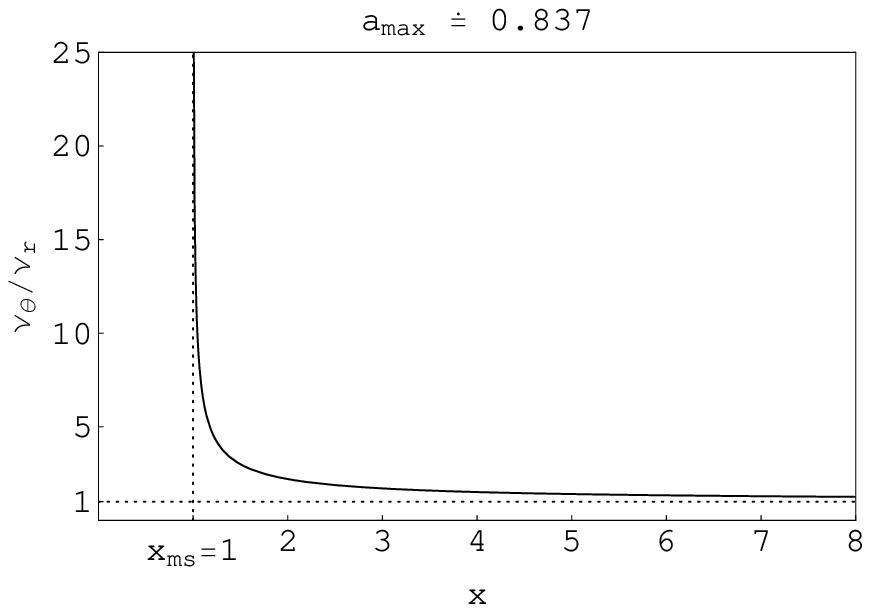}}\quad
\subfigure{\includegraphics[width=.31\hsize]{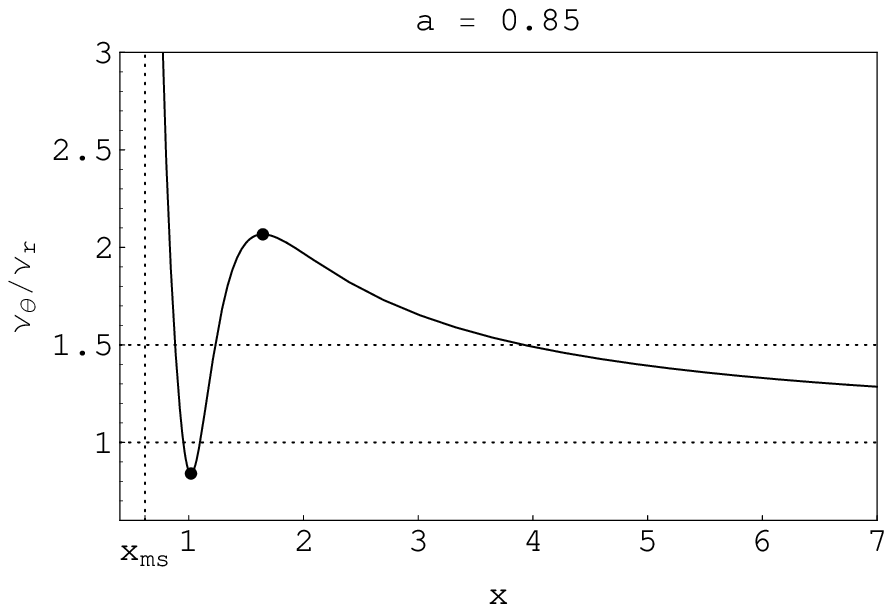}}\\
\subfigure{\includegraphics[width=.31\hsize]{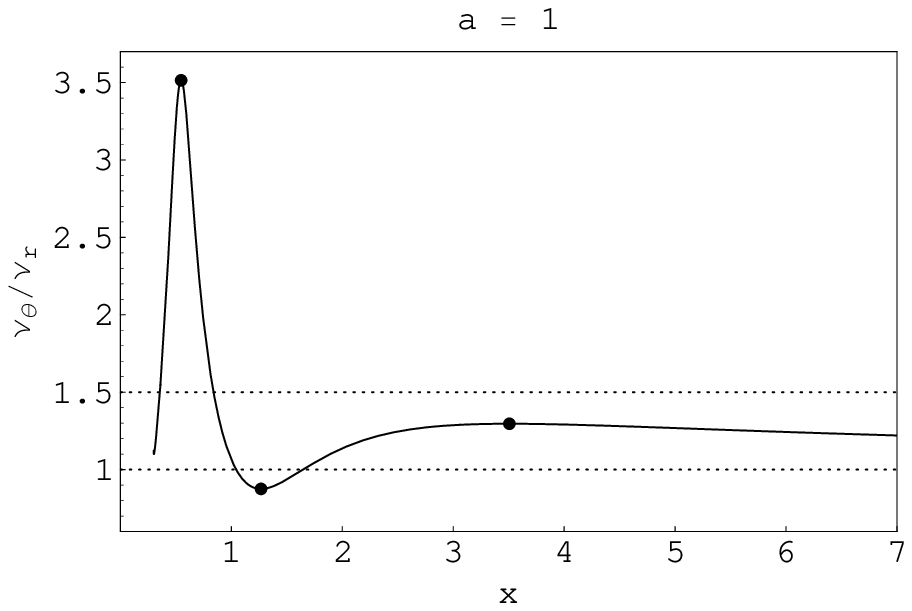}}\quad
\subfigure{\includegraphics[width=.31\hsize]{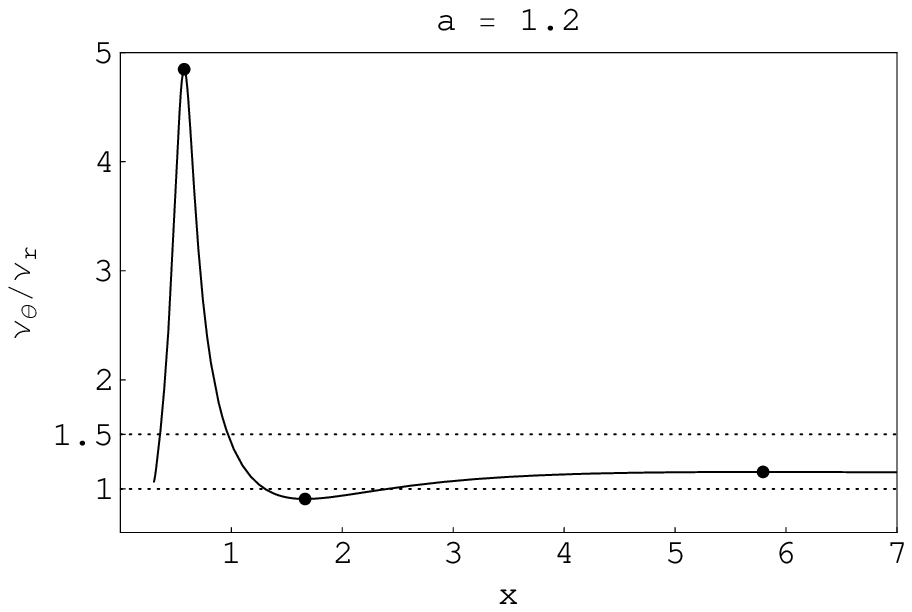}}\quad
\subfigure{\includegraphics[width=.31\hsize]{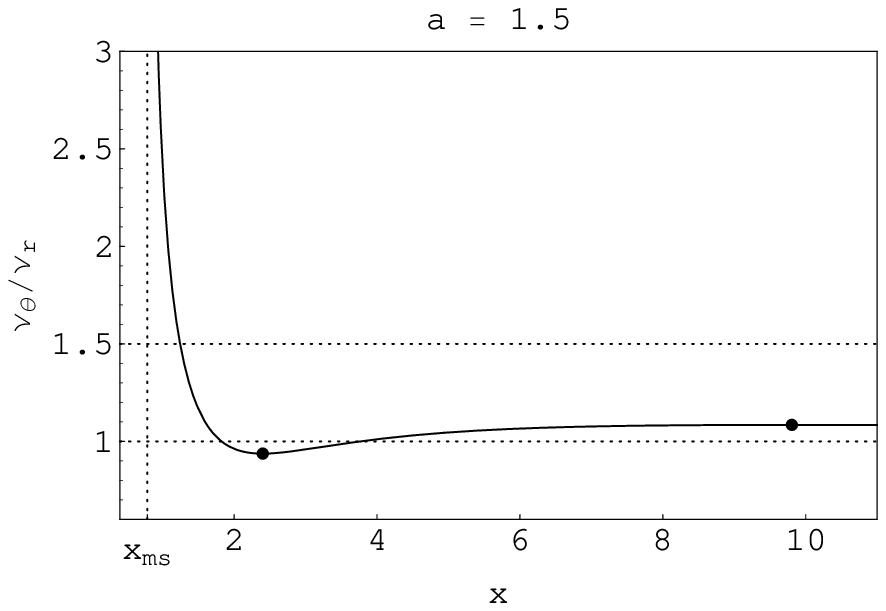}}
\caption{\label{pomery-frkv}The behaviour of ratio
$\nu_\theta/\nu_\mathrm{r}(x)$ of the epicyclic frequencies for
braneworld Kerr black hole and naked singularity spacetimes with
$b=0.3$. The marginally stable orbit $x_\mathrm{ms}$ is denoted by
a dotted vertical line (if this orbit exists at the given
spacetime).}
\end{figure*}

\subsection{\label{sec:sest:3}Ratio of epicyclic frequencies}

The ratio of epicyclic frequencies $\nu_\theta$ and
$\nu_{\mathrm{r}}$ is crucial for the orbital resonance models of
QPOs
\cite{Abr-etal:2004:RAGtime4and5:CrossRef,Kat:2004:PUBASJ:QPOsmodel}.
It is well known
\cite{Kat-Fuk-Min:1998:BHAccDis:,Ter-Stu:2005:ASTRA:} that for the
Kerr black holes ($-1 \leq a \leq 1$) the inequality
\begin{equation}\label{frkv-pomer}
\nu_{\mathrm{r}}(x,a) < \nu_{\theta}(x,a)
\end{equation}
holds, i.e., the equation $\nu_{\mathrm{r}}(x,a) =
\nu_{\theta}(x,a)$ does not have any real solution in the whole
range of black hole rotational parameter $a\in(-1, 1)$ and
\begin{equation}
\frac{\nu_{\theta}}{\nu_{\mathrm{r}}} >1
\end{equation}
for any Kerr black hole. Furthermore, this ratio is a monotonic
function of radius for any fixed $a\in(-1, 1)$
\cite{Ter-Stu:2005:ASTRA:}. These statements are valid also for
any brany Kerr black hole (i.e., for $b$ fixed and $a\leq
a_{\mathrm{max}}$).

However, the situation is different for Kerr naked singularities.
For $b=0$ and $a> 1$, the epicyclic frequencies $\nu_\theta$,
$\nu_{\mathrm{r}}$ can satisfy the equality condition
\begin{equation}
\nu_{\theta}(a,x)=\nu_{\mathrm{r}}(a,x)
\end{equation}
giving a possibility of strong resonant phenomenon, which could
occur at the critical radius
\begin{equation}\label{x-sr}
x_{\mathrm{sr}} = a^2\qquad (a\geq 1).
\end{equation}
This means that for any Kerr naked singularity the epicyclic
frequency ratio $\nu_\theta/\nu_{\mathrm{r}}$ is a non-monotonic
function that reaches value 1 at the point given by~(\ref{x-sr})
\cite{Ter-Stu:2005:ASTRA:}.

Furthermore, for brany Kerr naked singularities with $b>0$ the
epicyclic frequencies can satisfy even the condition (see
Figs~\ref{frkv-rezy} and \ref{pomery-frkv})
\begin{equation}
\nu_{\theta}(x,a,b)\leq\nu_{\mathrm{r}}(x,a,b),
\end{equation}
that is not allowed in Kerr naked singularities. For naked
singularities with $b<0$ the relation
\begin{equation}
\nu_{\theta}(x,a,b)>\nu_{\mathrm{r}}(x,a,b)
\end{equation}
is valid, as in the Kerr black hole spacetimes.

\subsection{\label{sec:sest:4}Ratio of Keplerian and
epicyclic frequencies}

We have to consider the possibility of
$\nu_{\mathrm{K}}<\nu_{\mathrm{r}}\left(\nu_{\theta}\right)$,
since such a situation could imply change in the maximum frequency
supposed to be given by $\nu_{\mathrm{K}}$ at $x_{\mathrm{ms}}$,
observable in the field of a black hole (or a naked singularity).
We have to look for the possibility to satisfy the condition
$\nu_{\mathrm{K}}=\nu_{\mathrm{r}}\left(\nu_{\theta}\right)$
(i.e., $\alpha_{\mathrm{r}}(\alpha_{\theta})=1$).

For the radial epicyclic frequency we arrive to the condition
\begin{equation}\label{podminka-rovnost2-res}
a=a_{\mathrm{sr}}^{\mathrm{r}}(x,b)\equiv\frac{4 (x-b)^{3/2}\pm x
\sqrt{(3-4 b) b+(3 b-2) x}}{3 x-4 b}.
\end{equation}
For the vertical epicyclic frequency there are two solutions of
the equation $\alpha_{\theta}=1$, the first one reads
\begin{equation}\label{podminka-rovnost1-res}
a=a_{\mathrm{sr}}^{\theta}(x,b)\equiv\frac{2(2x-b)
\sqrt{x-b}}{3x-2b}
\end{equation}
and the second one is $a=0$, representing the
Reissner--Nordstr\"{o}m braneworld solution, where
$\nu_{\theta}=\nu_{\mathrm{K}}$ due to the spherical symmetry of
the spacetime.

\begin{figure*}[!tbp]
\centering
\includegraphics[width=.68\hsize]{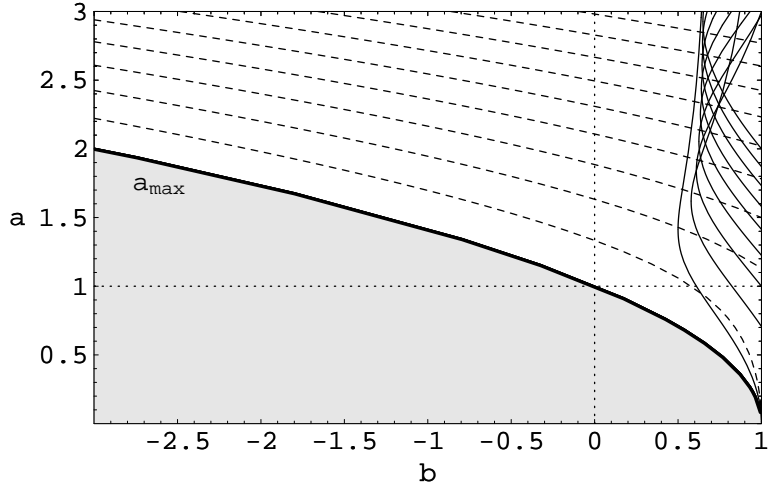}
\caption{\label{podminka-alfa-je-1} The functions
$a_{\mathrm{sr}}^{\mathrm{r}}(x,b)$ (solid lines) and
$a_{\mathrm{sr}}^{\theta}(x,b)$ (dashed lines) that represent the
values of the black hole spin $a$ and the brany parameter $b$ for
which the conditions
$\nu_{\mathrm{r}}(x,a,b)=\nu_{\mathrm{K}}(x,a,b)$ and
$\nu_{\theta}(x,a,b)=\nu_{\mathrm{K}}(x,a,b)$ can be satisfied.
The curves are plotted for
$x=1,\,1.5,\,2,\,2.5,\,3,\,3.5,\,4,\,4.5,\,5$. The black thick
line represents the function $a_{\mathrm{max}}$, so only the gray
area belongs to the black hole spacetimes. We can see that the
relations $\nu_{\theta}(x,a,b)\geq\nu_{\mathrm{K}}(x,a,b)$ and
$\nu_{\mathrm{r}}(x,a,b)\geq\nu_{\mathrm{K}}(x,a,b)$ are relevant
only in naked singularity spacetimes, where even the case with
$\nu_{\mathrm{K}}(x,a,b)=\nu_{\theta}(x,a,b)=\nu_{\mathrm{r}}(x,a,b)$
is possible (where the lines $a_{\mathrm{sr}}^{\mathrm{r}}(x,b)$
and $a_{\mathrm{sr}}^{\theta}(x,b)$ cross each other).}
\end{figure*}

The relations~(\ref{podminka-rovnost2-res})
and~(\ref{podminka-rovnost1-res}) are illustrated in
Fig.\,\ref{podminka-alfa-je-1}. We can see that
\begin{equation}
a_{\mathrm{sr}}^{\mathrm{r}}(x,b)>a_{\mathrm{max}}\qquad
\mathrm{and} \qquad
a_{\mathrm{sr}}^{\theta}(x,b)>a_{\mathrm{max}},
\end{equation}
so we can conclude that for braneworld black holes the condition
\begin{equation}
\nu_{\mathrm{K}}>\nu_{\theta}>\nu_{\mathrm{r}}
\end{equation}
is satisfied, similarly to the case of standard Kerr black holes.
On the other hand, in brany Kerr naked singularity spacetimes, the
epicyclic frequencies could overcome the Keplerian frequency at
small radii (see Fig.\,\ref{ex-kep-rezy}\,(d)). We not go into
detailed discussion since we concentrate on the black hole case.

\section{\label{sec:sedm}Resonance conditions}

The orbital resonance models for QPOs proposed in
\cite{Abr-Klu:2001:ASTRA:,Abr-etal:2004:RAGtime4and5:CrossRef} are
particularly based on resonance between oscillations with the
epicyclic frequencies which are excited at a well defined
resonance radius $x_{n:m}$ given by the condition
\begin{equation}
\frac{\nu_{\theta}}{\nu_{\mathrm{r}}}(a,b,x_{n:m})=\frac{n}{m},
\end{equation}
where $n\!:\!m$ is (most often) $3\!:\!2$ in the case of
\textit{parametric resonance} (the effect itself is described by
the Mathieu equation discussed by Landau and Lifshitz
\cite{Lan-Lif:1976:Mech:}) and arbitrary rational ratio of two
small integral numbers (1, 2, 3,\,\dots) in the case of
\textit{forced resonances}
\cite{Abr-etal:2004:RAGtime4and5:CrossRef}. Another, so called
\emph{``Keplerian'' resonance} model, takes into account possible
parametric or forced resonances between oscillations with radial
epicyclic frequency $\nu_\mathrm{r}$ and Keplerian orbital
frequency $\nu_\mathrm{K}$.

For a particular resonance $n\!:\!m$, the equation
\begin{equation}
\label{ratios} n\nu_{\mathrm{r}} = m\nu_{\mathrm{v}};
\quad\nu_\mathrm{v}\in\{\nu_\theta,\, \nu_\mathrm{K}\}
\end{equation}
determines the dimensionless resonance radius $x_{n:m}$ as a
function of the dimensionless spin $a$ in the case of direct
resonances that can be easily extended to the resonances with
combinational frequencies
\cite{Stu-Kot-Tor:2007:RAGtime8and9CrossRef:MrmQPO}. From the
known mass of the central black hole (e.g., low-mass in the case
of binary systems or hi-mass in the case of supermassive black
holes), the observed twin peak frequencies ($\nu_{\mathrm{upp}}$,
$\nu_{\mathrm{down}}$), and the equations
(\ref{def-epi-rad})\,--\,(\ref{def-Kep}), (\ref{ratios}) imply the
black hole spin, consistent with different types of resonances
with the beat frequencies taken into account. This procedure was
first applied to the microquasar GRO~1655-40 by Abramowicz and
Klu{\'z}niak \cite{Abr-Klu:2001:ASTRA:}, more recently to the
other three microquasars \cite{Ter-Abr-Klu:2005:ASTRA:QPOresmodel}
and also to the Galaxy center black hole Sgr\,A$^*$
\cite{Ter:2005:astro-ph0412500:}.

The twin peak QPOs were observed in four microquasars, namely GRO
1655-40, XTE 1550-564, H 1743-322, GRS 1915+105
\cite{Ter-Abr-Klu:2005:ASTRA:QPOresmodel}. In all of the four
cases, the frequency ratio of the twin peaks is very close to
$3\!:\!2$. The very probable interpretation of observed twin peak
kHz QPOs is the $3\!:\!2$ parametric resonance, however, generally
it is not unlikely that more than one resonance could be excited
in the disc at the same time (or in different times) under
different internal conditions. Indeed, observations of the kHz
QPOs in the microquasar GRS 1915+105, and of the QPOs in
extragalactic sources NGC 4051, MCG-6-30-15
\cite{Lac-Cze-Abr:2006:astro-ph0607594:} and NGC 5408 X-1
\cite{Str:2007:astro-ph/0701390:}, and the Galaxy center
Sgr\,A$^*$ \cite{Asc-etal:2004:ASTRA:} show a variety of QPOs with
frequency ratios differing from the $3\!:\!2$ ratio.

The resonances could be parametric or forced and of different
versions according to the epicyclic (Keplerian) frequencies
entering the resonance directly, or in some combinational form. In
principle, for any case of the resonance model version, one can
determine both the spin and mass of the black hole just from the
eventually observed set of frequencies. However, the obvious
difficulty would be to identify the right combination of
resonances and its relation to the observed frequency set. Within
the range of black hole mass allowed by observations, each set of
twin peak frequencies puts limit on the black hole spin. Of
course, the resonance model versions are consistent with
observations, if the allowed spin ranges are overlapping each
other. Clearly, two or more twin peaks then generally make the
spin measurement more precise.

Here we consider the versions of the resonance model explaining
the twin peaks with the $3\!:\!2$ frequency ratio. We take into
account both the direct and simple combinational resonances.

The resonant conditions determining implicitly the resonant radius
$x_{n:m}$ must be related to the radius of the innermost stable
circular geodesic $x_{\mathrm{ms}}$ giving the inner edge of
Keplerian discs. Therefore, for all the relevant resonance radii,
there must be $x_{n:m} \geq x_{\mathrm{ms}}$, where
$x_{\mathrm{ms}}$ is implicitly given by~(\ref{ms-impl}).

First, we investigate radial coordinate where the ratio
\begin{equation}
\frac{\nu_{\mathrm{upp}}}{\nu_{\mathrm{down}}}=\frac{3}{2}
\end{equation}
occurs for the simple case of the parametric resonance between the
radial and vertical epicyclic oscillations.

The result is given in the way relating the dimensionless spin $a$
and the dimensionless resonance radius $x$ for frequency ratio
$n\!:\!m = 3\!:\!2$
        \begin{multline}\label{a-param}
        a=a^{\theta/\mathrm{r}}_{3:2}(x,b)\equiv a_{\mathrm{EO}}(x,b)\equiv
        \frac{1}{39 x-44 b}\,\Big\{ 4 (11 x-10 b ) \sqrt{x-b
        }\\
        -\sqrt{(5 x+4 b ) \left[39 x^3-2 x^2 (17+22 b )+43 x
        b -4 b ^2\right]}    \Big\}.
        \end{multline}

The behaviour of the function $a^{\theta/\mathrm{r}}_{3:2}$
representing the direct resonance
$\nu_{\theta}\!:\!\nu_{\mathrm{r}}=3\!:\!2$ is for various values
of the brany parameter $b$ illustrated in Fig.\,\ref{hor-a-mez}.
We can see that for all considered values of $b$ the condition
$x_{3:2} > x_{\mathrm{ms}}$ is always satisfied.

Another possibility, applied in the context of both black hole and
neutron star systems producing kHz QPOs is the relativistic
precession model \cite{Ste-Vie:1998:ASTRJ2L:}. Now, we can assume
a resonance of the oscillations with the Keplerian frequency
$\nu_{\mathrm{K}}\equiv \nu_{\mathrm{upp}}$ and the relativistic
precession frequency
$\nu_{\mathrm{P}}=\nu_{\mathrm{K}}-\nu_{\mathrm{r}}\equiv\nu_{\mathrm{down}}$.
Considering the condition
$\nu_\mathrm{K}\!:\!(\nu_\mathrm{K}-\nu_\mathrm{r})=3\!:\!2$, we
arrive to the relation\footnote{Clearly, we obtain the same
relation also for the direct ``Keplerian'' resonance
$\nu_\mathrm{K}\!:\!\nu_\mathrm{r}=3\!:\!1$.}
        \begin{equation}
        a=a^{\mathrm{K}/(\mathrm{K}-\mathrm{r})}_{3:2}(x,b)
        \equiv a_{\mathrm{RP}}(x,b)\equiv
        \frac{12 (x-b )^{3/2}-x
        \sqrt{6 x (4 x-3) -29 x b+27 b -4 b ^2}}{3 (3 x-4 b
        )}.
        \end{equation}
This is again illustrated in Fig.\,\ref{hor-a-mez}.

\begin{figure*}[!tbp]
\subfigure{\includegraphics[width=.31\hsize]{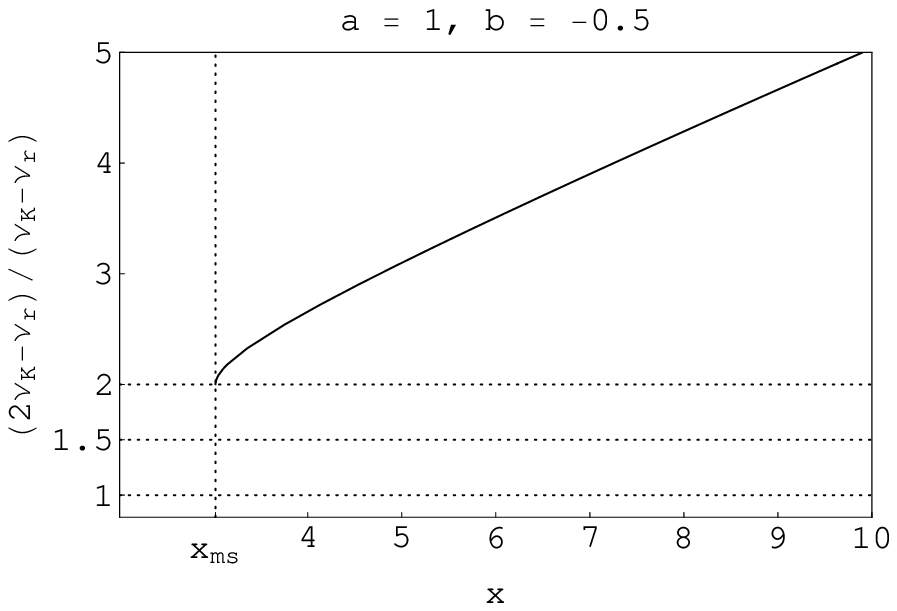}}\quad
\subfigure{\includegraphics[width=.31\hsize]{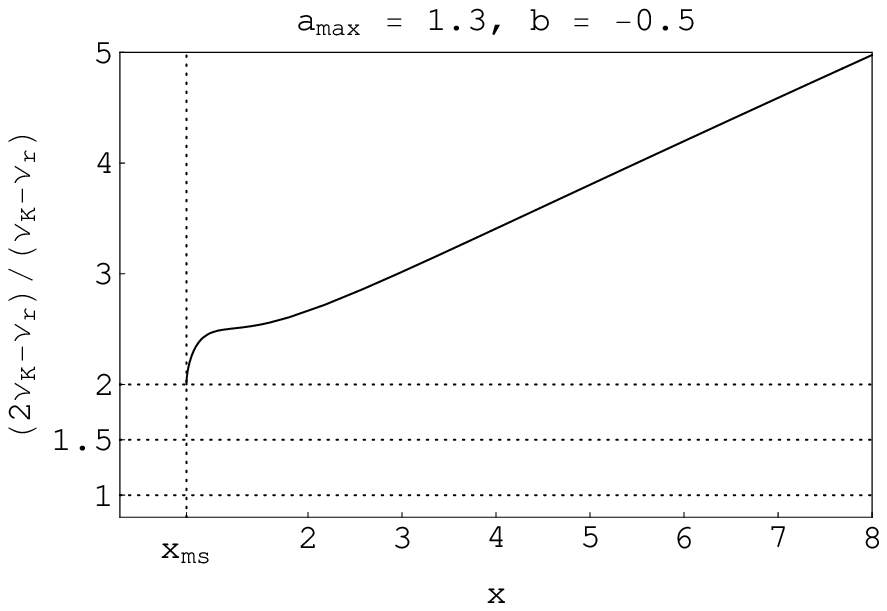}}\quad
\subfigure{\includegraphics[width=.31\hsize]{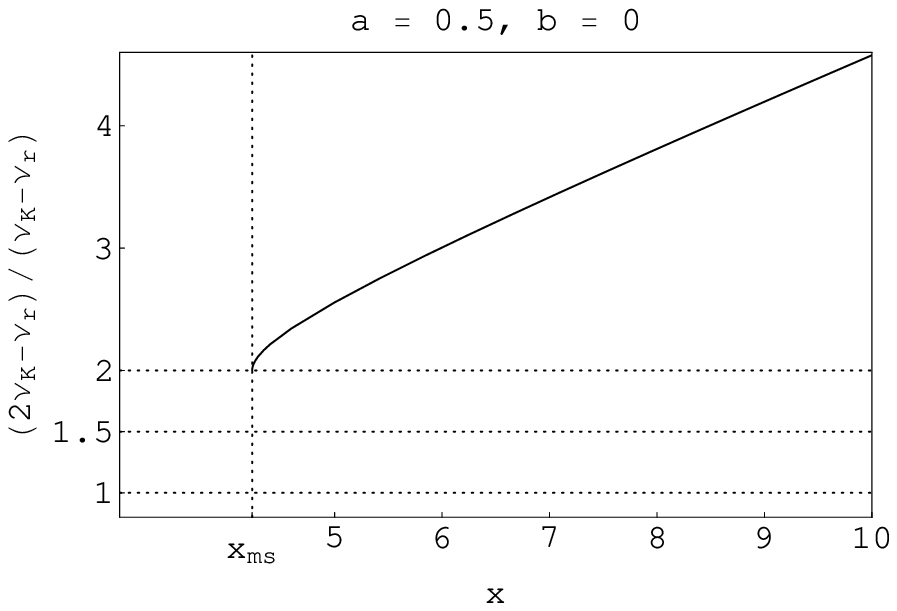}}\\
\subfigure{\includegraphics[width=.31\hsize]{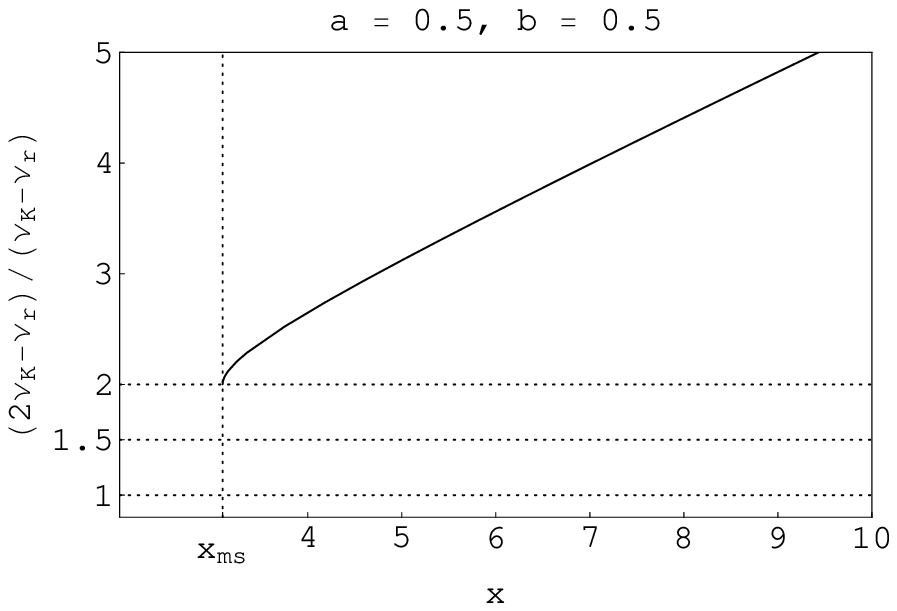}}\quad
\subfigure{\includegraphics[width=.31\hsize]{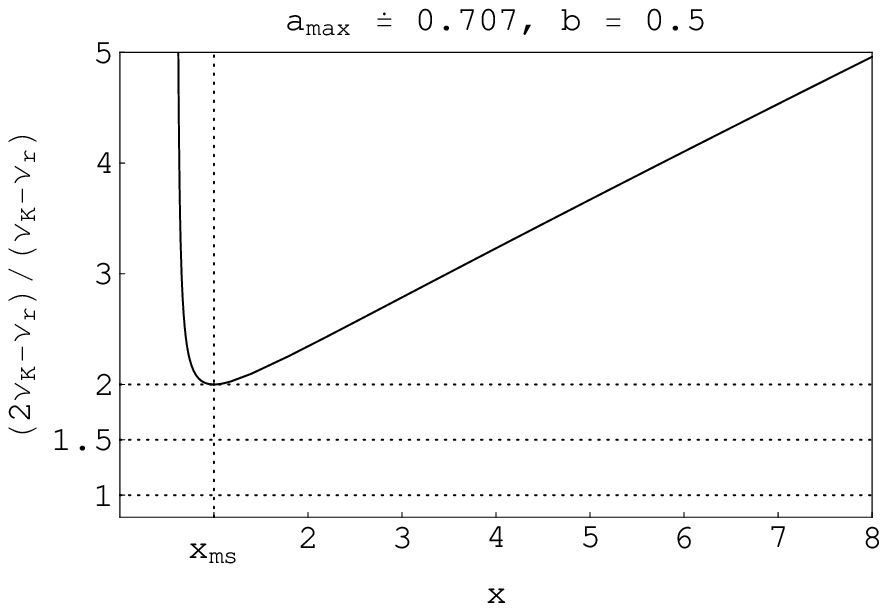}}\quad
\subfigure{\includegraphics[width=.31\hsize]{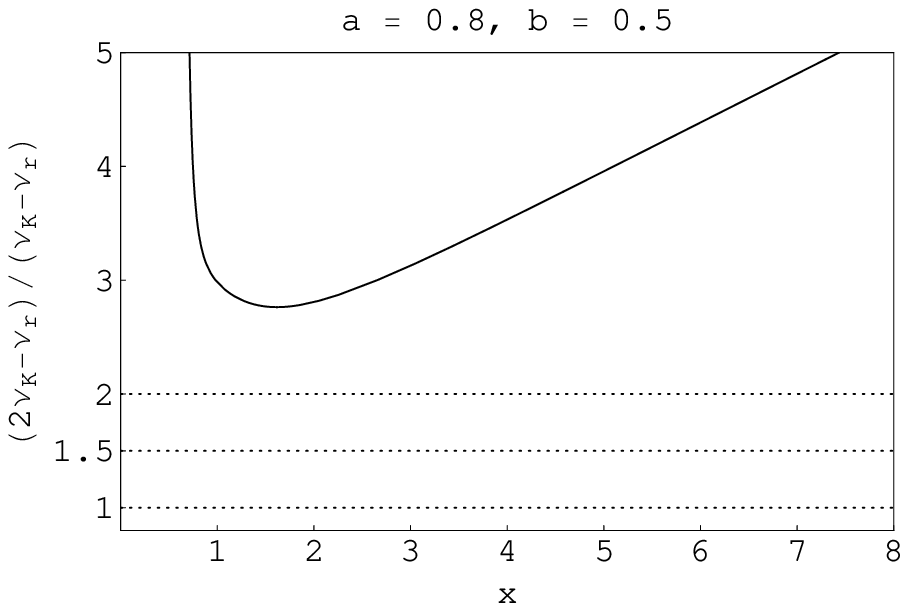}}
\caption{\label{warp} The behaviour of ratio
$(2\nu_{\mathrm{K}}-\nu_{\mathrm{r}})/(\nu_{\mathrm{K}}-\nu_{\mathrm{r}})$
representing the resonance of trapped oscillations assumed in
warped disc \cite{Kato:2007:PASJ:}. The marginally stable orbit
$x_\mathrm{ms}$ is denoted by a dotted vertical line (if this
orbit exists at the given spacetime).}
\end{figure*}

On the other hand, in the framework of the warp disc oscillations,
the frequencies of which are given by combinations of the
Keplerian and epicyclic frequencies, resonant phenomena could be
relevant too. Usually, the inertial-acoustic and g-mode
oscillations and their resonances are relevant
\cite{Kato:2007:PASJ:}. Here, we give as an example the study of
the simple frequency relation
\begin{equation}\label{warp-pomer}
    \frac{2\nu_{\mathrm{K}}-\nu_{\mathrm{r}}}
    {\nu_{\mathrm{K}}-\nu_{\mathrm{r}}}=\frac{3}{2}\,.
\end{equation}
However, there exists no solution for this ratio in both Kerr
black hole and Kerr naked singularity spacetimes for any value of
brany parameter $b$. The lowest relevant frequency ratio is
$(2\nu_{\mathrm{K}}-\nu_{\mathrm{r}})\!:\!(\nu_{\mathrm{K}}-\nu_{\mathrm{r}})
= 2\!:\!1$ (see Fig.\,\ref{warp}). Therefore, only the
modification to the relation of observed frequencies
\cite{Kato:2007:PASJ:}
\begin{equation}\label{warp-upr}
    \frac{2\nu_{\mathrm{K}}-\nu_{\mathrm{r}}}
    {2\left(\nu_{\mathrm{K}}-\nu_{\mathrm{r}}\right)}=\frac{3}{2}
\end{equation}
could be relevant. Then we have
      \begin{equation}
        a=a_{3:2}^{(2\mathrm{K}-\mathrm{r})/(2\mathrm{K}-2\mathrm{r})}(x,b)
        \equiv a_{\mathrm{TO}}(x,b)\equiv\frac{8
(x-b)^{3/2}\pm x\sqrt{9 x^2-(9 b+8) x-4 (b-3) b}}{2(3 x-4 b)}.
        \end{equation}

Again, we illustrate this function in Fig.\,\ref{hor-a-mez}. For
all of the tested values of the brany parameter $b$, the resonance
radius of the trapped oscillations lies between the resonance
radius of the relativistic precession oscillations and the radial
and vertical epicyclic oscillations.

\subsection{\label{sec:sedm:2}Application to microquasar GRS 1915+105}

The frequency ratio of the upper twin peak QPOs observed in
microquasar GRS 1915+105 is very close to $3\!:\!2$
\cite{Ter-Abr-Klu:2005:ASTRA:QPOresmodel}:
\begin{align}\label{GRO-up}
      \nu_{\mathrm{upp}} &= (168\pm 3)\,\mathrm{Hz},\\
      \nu_{\mathrm{down}} &= (113\pm 5)\,\mathrm{Hz}. \label{GRO-down}
\end{align}

\begin{figure*}[!tbp]
\subfigure{\includegraphics[width=.48\hsize]{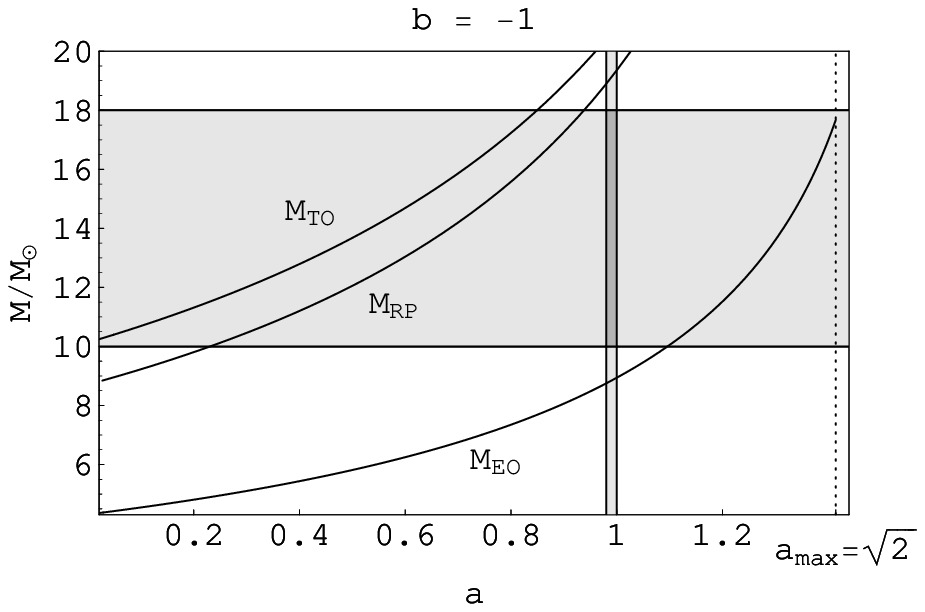}}\quad
\subfigure{\includegraphics[width=.48\hsize]{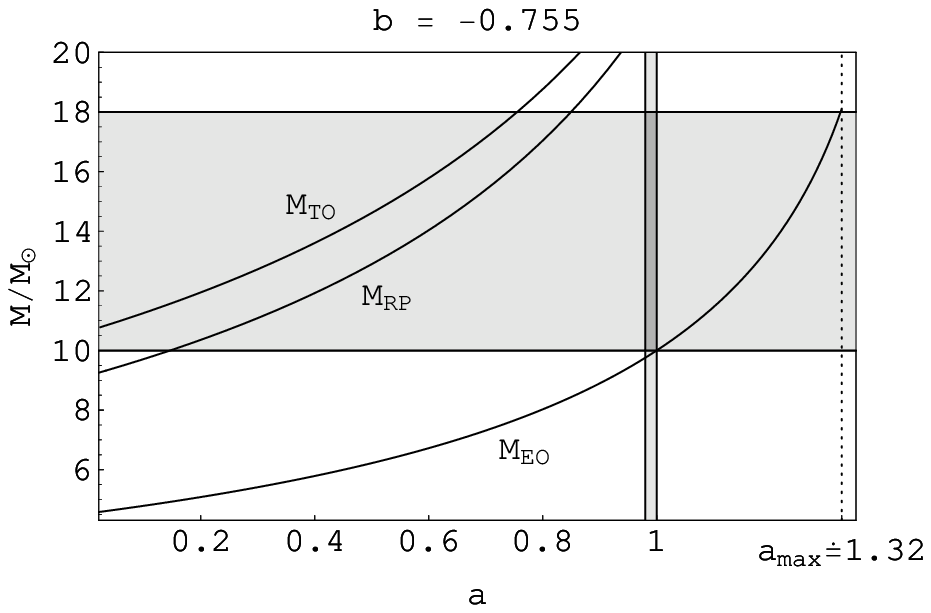}}\\
\subfigure{\includegraphics[width=.48\hsize]{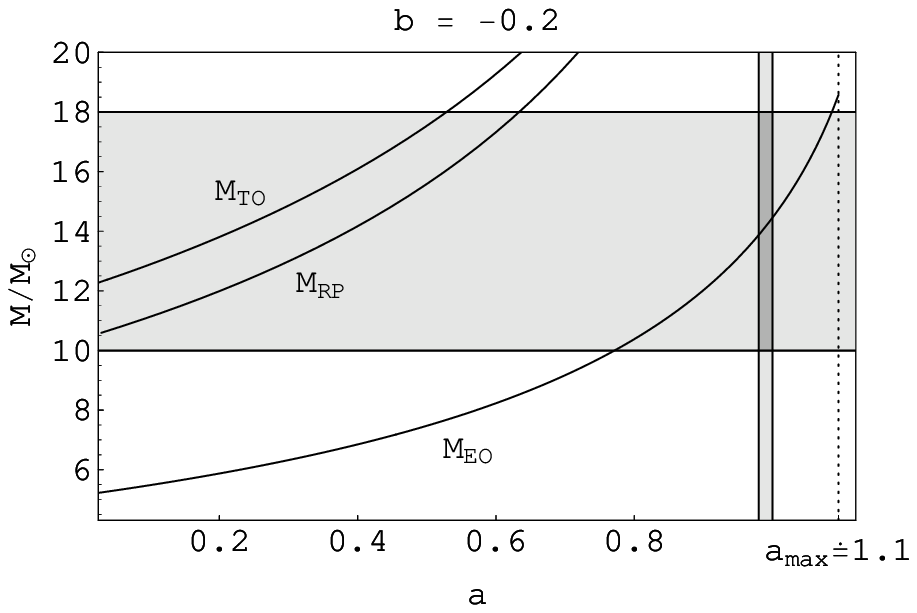}}\quad
\subfigure{\includegraphics[width=.48\hsize]{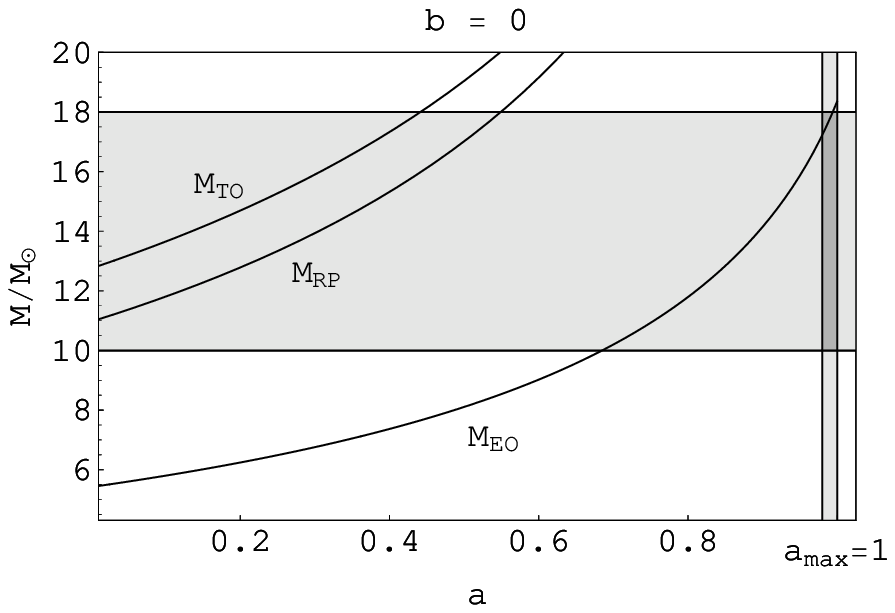}}\\
\subfigure{\includegraphics[width=.48\hsize]{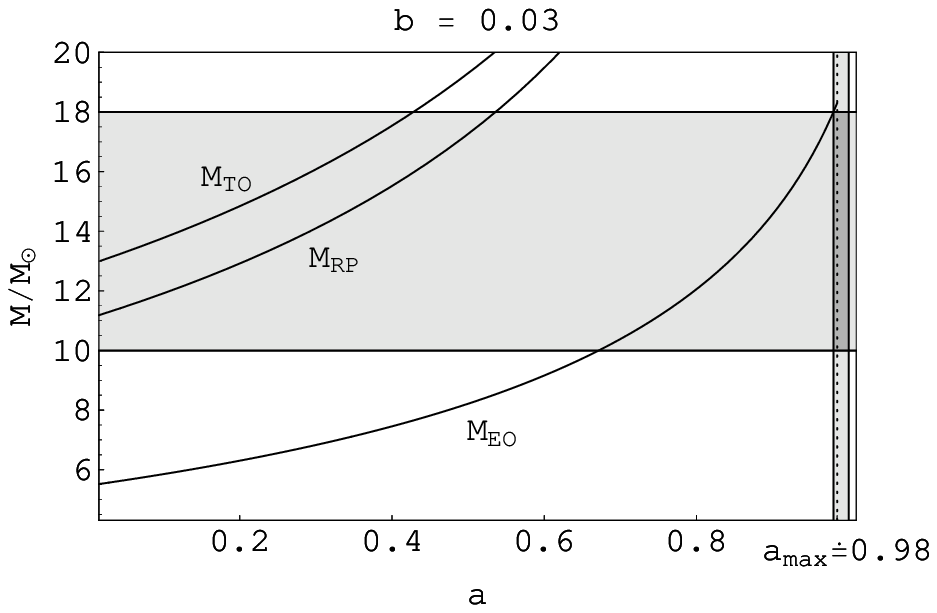}}\quad
\subfigure{\includegraphics[width=.48\hsize]{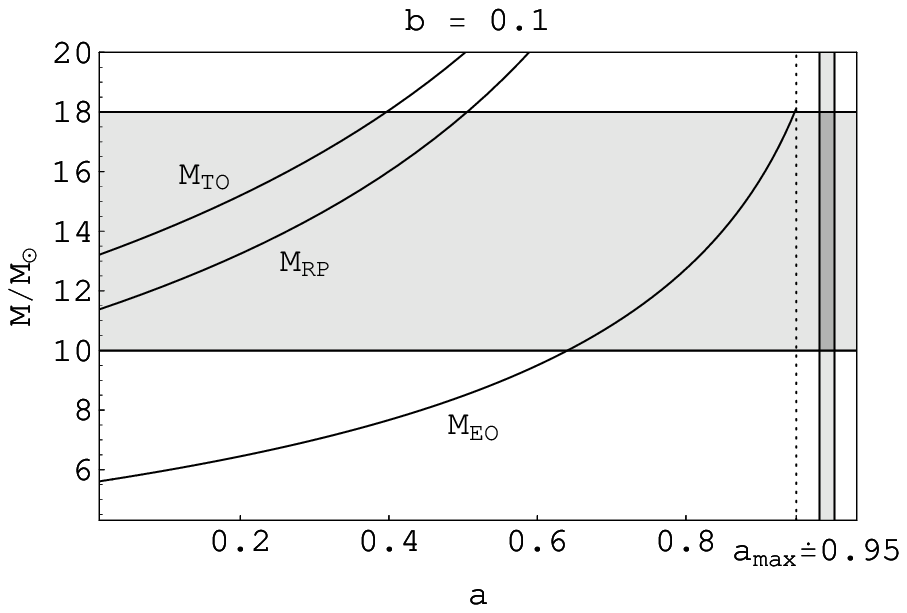}}
\caption{\label{GRS}Possible combinations of mass and black hole
spin predicted by the standard parametric epicyclic oscillations
resonance model $M_{\mathrm{EO}}(a,b)$ with
$\nu_{\theta}\!:\!\nu_{\mathrm{r}}=3\!:\!2$, the relativistic
precession model $M_{\mathrm{RP}}(a,b)$ with $\nu_{\mathrm{K}}
\!:\! \left( \nu_{\mathrm{K}}-\nu_{\mathrm{r}}\right)=3\!:\!2$ and
modified trapped oscillations model $M_{\mathrm{TO}}(a,b)$ with
$(2\nu_{\mathrm{K}}-\nu_{\mathrm{r}})\!:\!
2(\nu_{\mathrm{K}}-\nu_{\mathrm{r}})=3\!:\!2$ for the
high-frequency QPOs observed in the spectra of the microquasar GRS
1915+105. Shaded regions indicate the likely ranges for the mass
(inferred from optical measurements of radial curves
\cite{McCli-etal:2006:astro-ph/0606076:}) and the dimensionless
spin (inferred from the X-ray spectral data fitting
\cite{McCli-Rem:2004:CompactX-Sources:}) of GRS 1915+105.}
\end{figure*}

From the known limits on the mass of the black hole in GRS
1915+105 \cite{McCli-Rem:2004:CompactX-Sources:}
\begin{equation}
    10.0\,\mathrm{M}_{\odot} < M < 18.0\,\mathrm{M}_{\odot},
\end{equation}
the observed twin peak frequencies (\ref{GRO-up}),
(\ref{GRO-down}), and the equations
(\ref{def-epi-rad})\,--\,(\ref{def-Kep}), (\ref{ratios}) imply the
black hole spin consistent with different types of resonances.
Assuming the very probable interpretation of observed twin peak
kHz QPOs in microquasar as the $3\!:\!2$ standard parametric
resonance
\begin{equation}
\frac{\nu_{\theta}}{\nu_{\mathrm{r}}}(a,b,x_{3:2})=\frac{3}{2}
\end{equation}
and identifying
\begin{equation}
\nu_{\mathrm{upp}}\equiv \nu_{\theta},
\end{equation}
we can express the black hole mass in the form
\begin{equation}
M_{\mathrm{EO}}(x,a,b)=
\frac{\mathrm{c}^3}{2\pi\mathrm{G}}\frac{\sqrt{x-b }}{x^2+a
\sqrt{x-b }}\frac{\sqrt{\alpha_{\theta}} }{\nu_{\mathrm{upp}}},
\end{equation}
where $a=a^{\theta/\mathrm{r}}_{3:2}(x,b)$ is given by the
relation~(\ref{a-param}). Putting
$\nu_{\mathrm{upp}}=\nu_{\theta}$, we obtain the mass dependence
of the spin assuming $b=\mathrm{const}$ (see Fig.\,\ref{GRS}). In
a similar way, we obtain the relations $M(x,a,b)$ for the
relativistic precession model
\begin{equation}
M_{\mathrm{RP}}(x,a,b)=
\frac{\mathrm{c}^3}{2\pi\mathrm{G}}\frac{\sqrt{x-b }}{x^2+a
\sqrt{x-b }}\frac{1}{\nu_{\mathrm{upp}}}
\end{equation}
or Kato's warped disc trapped oscillation model
\begin{equation}
M_{\mathrm{TO}}(x,a,b)=
\frac{\mathrm{c}^3}{2\pi\mathrm{G}}\frac{\sqrt{x-b }}{x^2+a
\sqrt{x-b}}\frac{2-\sqrt{\alpha_{\mathrm{r}}}}{\nu_{\mathrm{upp}}},
\end{equation}
putting $\nu_{\mathrm{upp}}\equiv \nu_{\mathrm{K}}$ or
$\nu_{\mathrm{upp}}\equiv 2\nu_{\mathrm{K}}-\nu_{\mathrm{r}}$,
respectively.

Fig.\,\ref{GRS} shows the predictions of the $3\!:\!2$ parametric
resonance model as well as the predictions of the relativistic
precession and trapped oscillations models in the mass-spin plane.
It demonstrates possible combinations of mass and black hole spin
of GRS 1915+105 as they are predicted for various values of brany
parameter $b$, considering the allowed range of the black hole
mass.

It follows immediately from Fig.\,\ref{GRS} that for all
considered values of the brany parameter $b$ the relativistic
precession and the trapped oscillations models do not meet the
observational data of GRS 1915+105 for the same spin as the
parametric resonance model of epicyclic oscillations. Clearly,
they can be applied for object with substantially lower spin.

According to the spectral analysis of the X-ray continuum by
McClintock et al. \cite{McCli-etal:2006:astro-ph/0606076:}, the
compact primary of the binary X-ray source GRS 1915+105 is a
rapidly-rotating Kerr black hole with a lower limit on its
dimensionless spin of
\begin{equation}
a > 0.98 .
\end{equation}
We use this spin limit in Fig.\,\ref{GRS}.

Notice, however, that although the spectral fitting analysis has
been done by McClintock et al.
\cite{McCli-etal:2006:astro-ph/0606076:} very carefully, the spin
estimate is valid only in the Kerr spacetime. We can suppose that
for braneworld Kerr black hole with non-zero brany parameter $b$,
the spin estimates may be shifted to higher (lower) values of $a$
due to the influence of the brany tidal charge $b<0$ ($b>0$) on
the optical phenomena near a rotating black hole
\cite{Sch-Stu:2007:RAGtime8and9CrossRef:SpBranKBH,Sch-Stu:2007:RAGtime8and9CrossRef:OEBraK}.
 Therefore, using the optical phenomena modelling with brany
parameter $b$ included, we could expect stronger limits on allowed
values of $b$.

\section{\label{sec:osm}Strong resonant phenomena -- ``magic'' spin}

Generally, the resonances could be excited at different radii of
the accretion disc under different internal conditions; such a
situation is discussed in detail by Stuchl\'{\i}k et al.
 \cite{Stu-Kot-Tor:2007:RAGtime8and9CrossRef:MrmQPO}. However, we
have shown \cite{SKT:2007:kratky:} that for special resonant
values of dimensionless black hole spin $a$ strong resonant
phenomena could occur when different resonances can be excited at
the same radius, as cooperative phenomena between the resonances
may work in such situations.

There exists a possibility of direct resonances of oscillations
with all of the three orbital frequencies, characterized by a
triple frequency ratio set
\begin{equation}\label{pomer}
             \nu_{\mathrm{K}}:\nu_{\theta}:\nu_{\mathrm{r}} = s:t:u
\end{equation}
with $s>t>u$ being small integers. The frequency set ratio
(\ref{pomer}) can be realized only for special resonant values of
the black hole spin $a$. The black hole mass is then related to
the magnitude of the frequencies.

Assuming two resonances
$\nu_{\mathrm{K}}\!:\!\nu_{\theta}=s\!:\!t$ and
$\nu_{\mathrm{K}}\!:\!\nu_{\mathrm{r}}=s\!:\!u$ occurring at the
same $x$, we arrive to the conditions
\begin{align}
    \alpha_{\theta}(a,b,x)&=\left(\frac{t}{s}\right)^2,\\
    \alpha_{\mathrm{r}}(a,b,x)&=\left(\frac{u}{s}\right)^2
\end{align}
that have to be solved simultaneously for $x$, $a$ and $b$. The
solution is given by the condition
\begin{equation}
    a^{\theta}(x,b,t/s)=a^{\mathrm{r}}(x,b,u/s),
\end{equation}
where
\begin{equation}\label{spin}
    a^{\theta}(x,b,t/s)=
\frac{\sqrt{x-b}}{3 x-2 b}\left\{2 x-b\pm\sqrt{b^2-2 b x
\left[2+x\left(\left(t/s\right)^2-1\right)\right]+x^2 \left[4+3
x\left(\left(t/s\right)^2-1\right)\right]}\right\},
\end{equation}
\begin{multline}
    a^{\mathrm{r}}(x,b,u/s)=\frac{1}{3 x-4 b} \,\Bigg\{4 (x-b)^{3/2}\\
\pm
    x \sqrt{b\left[3-4 b\left(u/s\right)^2\right]+
     x\left[7b\left(u/s\right)^2-2\left(2b+1\right)\right]-3  x^2
\left[\left(u/s\right)^2-1\right]}\Bigg\}.
\end{multline}

For Kerr spacetime with $b=0$ the explicit solution determining
the relevant radius for any triple frequency set ratio
$s\!:\!t\!:\!u$ takes the form \cite{SKT:2007:kratky:}
\begin{equation}
x(s,t,u)=\frac{6 s^2}{6 s^2\pm2 \sqrt{2} \sqrt{(t-u) (t+u) \left(3
s^2-t^2-2 u^2\right)}- \left(t^2+5 u^2\right)}.
\end{equation}

\begin{figure*}[!tbp]
\centering
\includegraphics[width=.68\hsize]{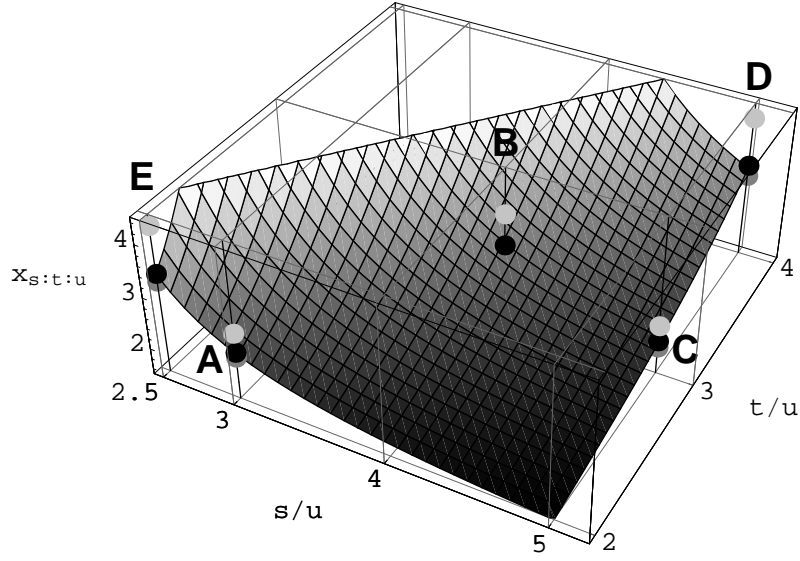}
\caption{\label{x-stu}The function $x(s/u,t/u)$ determining the
triple frequency ratio set
$\nu_{\mathrm{K}}\!:\!\nu_{\theta}\!:\!\nu_{\mathrm{r}} =
s\!:\!t\!:\!u$ at the same radius in Kerr black hole spacetime
($b=0$). The black points represent the ratios: $s\!:\!t\!:\!u=
3\!:\!2\!:\!1$ (A), $4\!:\!3\!:\!1$ (B), $5\!:\!3\!:\!1$ (C),
$5\!:\!4\!:\!1$ (D), $5\!:\!4\!:\!2$ (E) for $b=0$, shaded points
for $b=0.2$ and light shaded points for $b=-2$.}
\end{figure*}

\begin{figure*}[!tbp]
\subfigure{\includegraphics[width=.48\hsize]{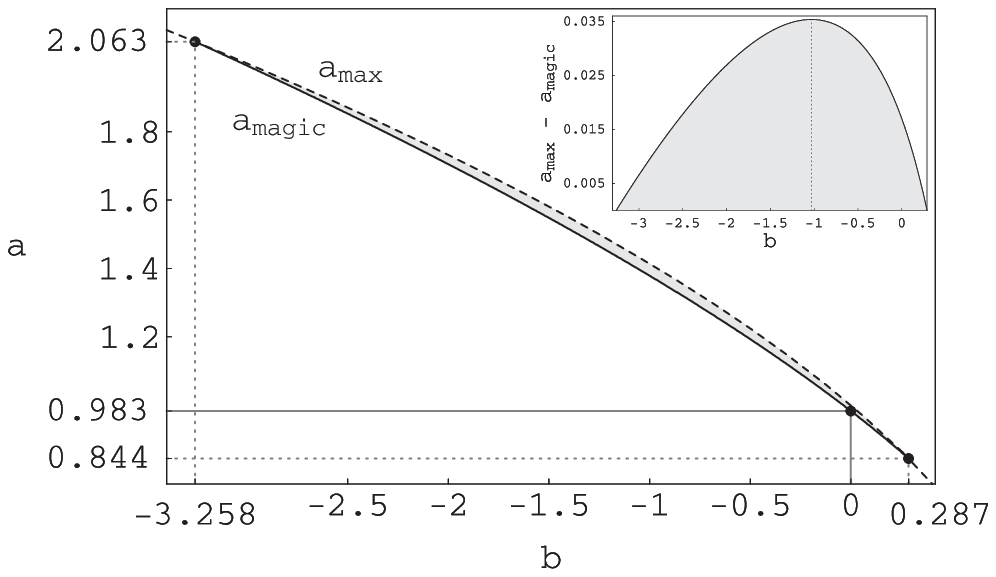}}\quad
\subfigure{\includegraphics[width=.48\hsize]{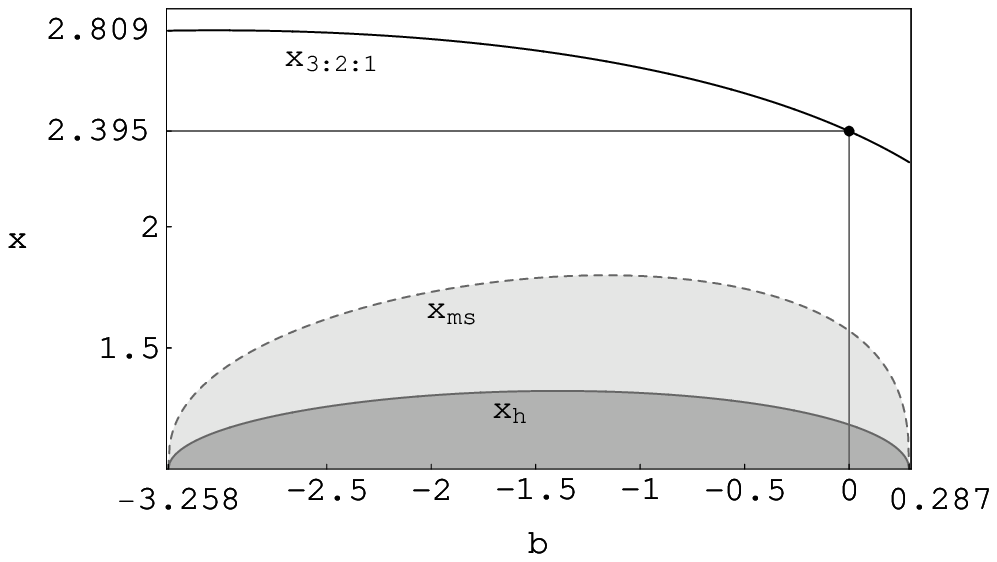}}
\caption{\label{magic-spin}The ``magic'' black hole spin $a$
(\textit{left panel}) and the shared resonance radius $x_{3:2:1}$
(\textit{right panel}) as the function of the ``magic'' brany
parameter $b$ that imply the frequency ratio set
$\nu_{\mathrm{K}}\!:\!\nu_{\theta}\!:\!\nu_{\mathrm{r}} = $
$3\!:\!2\!:\!1$ arising at the shared radius $x_{3:2:1}$. Dashed
line in the \textit{left panel} represents $a_\mathrm{max}$,
corresponding to the extreme black holes. We can see that for all
possible values of the ``magic'' brany parameter
$a_{\mathrm{magic}}\rightarrow a_{\mathrm{max}}$. In the
\textit{right panel}, there is also shown the radius of the outer
event black hole horizon $x_\mathrm{h}$ (gray solid line) and the
marginally stable circular orbit $x_\mathrm{ms}$ (dashed line) of
a rotating black hole carrying a given value of the ``magic''
brany parameter $b$ and ``magic'' black hole spin $a$.}
\end{figure*}

\begin{figure*}[!tbp]
\subfigure{\includegraphics[width=.48\hsize]{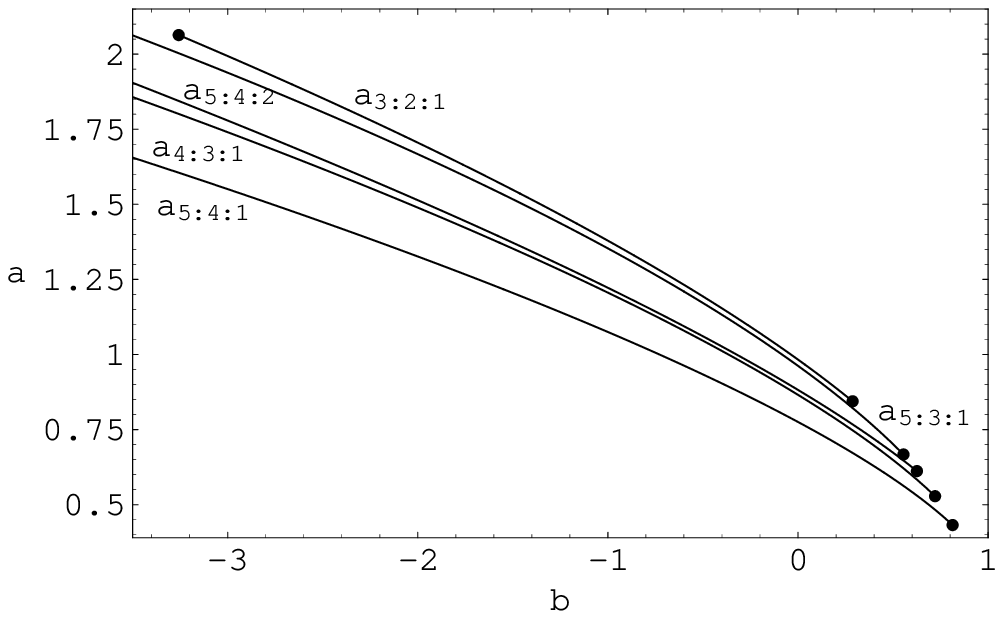}}\quad
\subfigure{\includegraphics[width=.48\hsize]{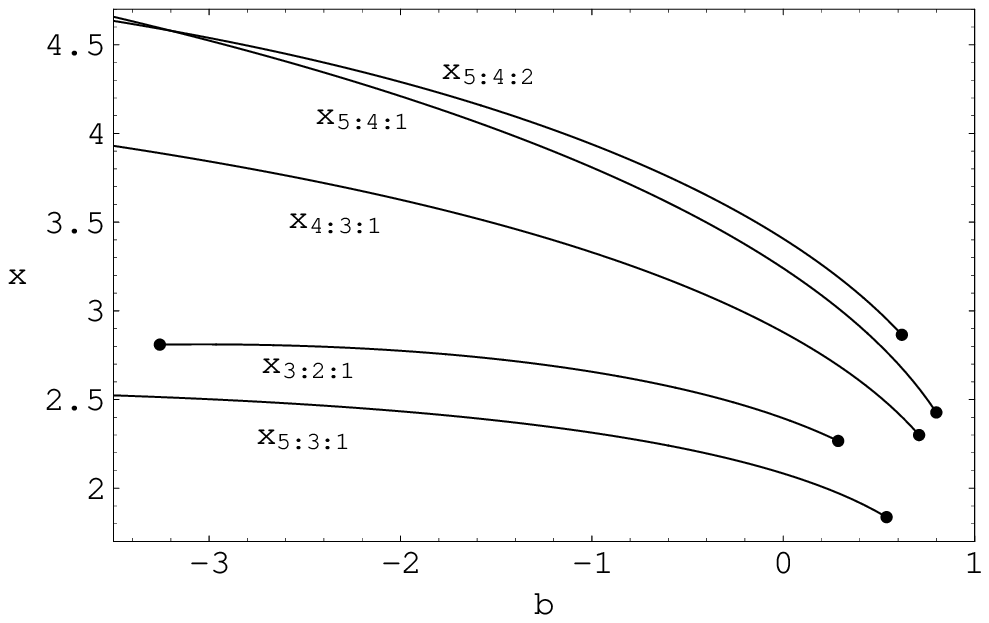}}
\caption{\label{magic-spin-vseV1}The black hole spin (\textit{left
panel}) and the shared resonance radius (\textit{right panel}) for
the ratios $\nu_{\mathrm{K}}\!:\!\nu_{\theta}\!:\!\nu_{\mathrm{r}}
= s\!:\!t\!:\!u$ with integers $\leq 5$ allowed in the field of
brany Kerr black holes.}
\end{figure*}

Clearly, the condition $t^2+2u^2 \leq 3s^2$ is always satisfied.
The corresponding special resonant black hole spin $a$ is then
determined, e.g., by Eq.\,(\ref{spin}) giving
$a^{\theta}(x(s,t,u),t/s)$. Of course, we consider only the black
hole cases when $a \leq a_{\mathrm{max}}$. This condition puts a
restriction on allowed values of $s,t,u$. The function $x(s,t,u)$
for $b=0$ is illustrated in Fig.\,\ref{x-stu} and the resonant
points are given for $5\geq s>t>u$. For $b\neq 0$ the solutions
are determined numerically and the results are given in
Table~\ref{magic-a-b} and
Figs\,\ref{magic-spin}\,--\,\ref{magic-spin-vseV1}.

\tabcolsep=8pt
\begin{table}[b]
\centering \caption{\label{magic-a-b}The intervals of the allowed
values of the brany parameter $b$ and the black hole spin $a$ that
imply the frequency ratio set
$\nu_{\mathrm{K}}\!:\!\nu_{\theta}\!:\!\nu_{\mathrm{r}} =
s\!:\!t\!:\!u$ with $5\geq s>t>u$ at the shared radius. For some
ratios such situation is possible only for naked singularity
spacetimes (NaS).}
\begin{tabular}{ccc}
\hline\noalign{\smallskip}
    $s:t:u$ & $b_{s:t:u}$ & $a_{s:t:u}$
    \\
\noalign{\smallskip}\hline\noalign{\smallskip}
   $3:2:1$ & $-3.258$ -- $+0.287$ & $+0.844$ -- $+2.063$ \\ \noalign{\smallskip}
   $4:2:1$ & \multicolumn{2}{c}{only NaS} \\ \noalign{\smallskip}
   $4:3:1$ & $-21.575$ -- $+0.721$ & $+0.528$ -- $+4.751$ \\ \noalign{\smallskip}
   $4:3:2$ & \multicolumn{2}{c}{only NaS} \\ \noalign{\smallskip}
   $5:2:1$ & \multicolumn{2}{c}{only NaS} \\ \noalign{\smallskip}
   $5:3:1$ & $-9.721$ -- $+0.555$ & $+0.667$ -- $+3.274$ \\ \noalign{\smallskip}
   $5:3:2$ & \multicolumn{2}{c}{only NaS} \\ \noalign{\smallskip}
   $5:4:1$ & $\leq 0.813$ & $\geq 0.432$  \\ \noalign{\smallskip}
   $5:4:2$ & $-9.182$ -- $+0.626$ & $+0.612$ -- $+3.191$ \\ \noalign{\smallskip}
   $5:4:3$ & \multicolumn{2}{c}{only NaS} \\
\noalign{\smallskip}\hline
\end{tabular}
\end{table}

\begin{figure*}[!tbp]
\subfigure{\includegraphics[width=.47\hsize]{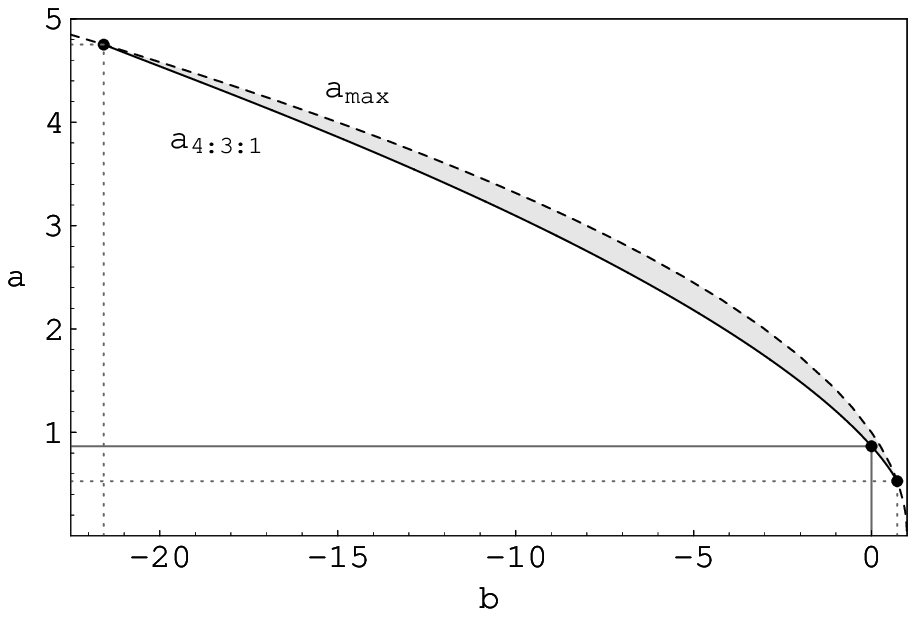}}\quad
\subfigure{\includegraphics[width=.47\hsize]{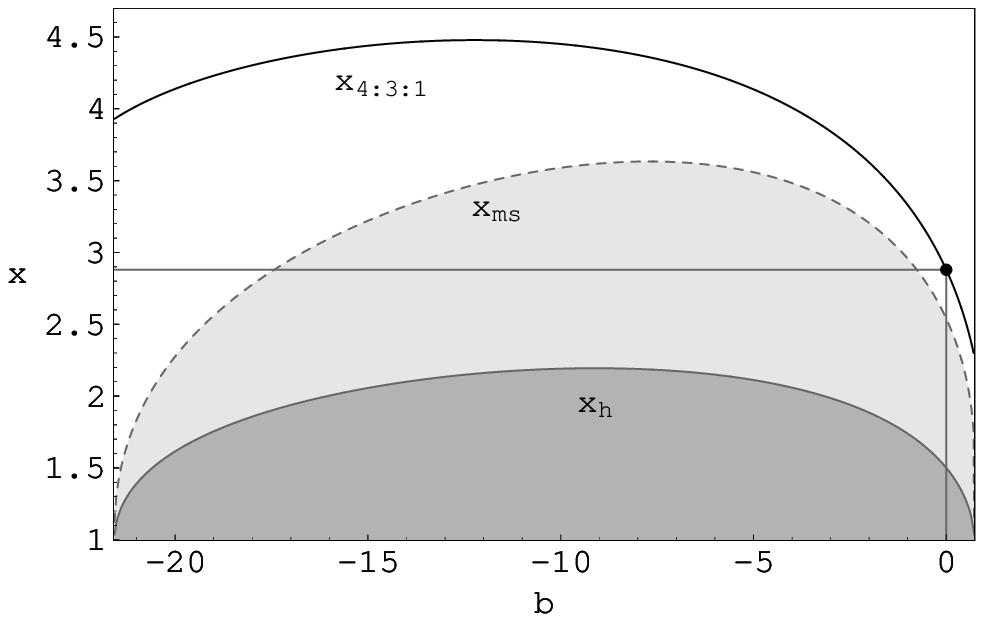}}\\
\subfigure{\includegraphics[width=.47\hsize]{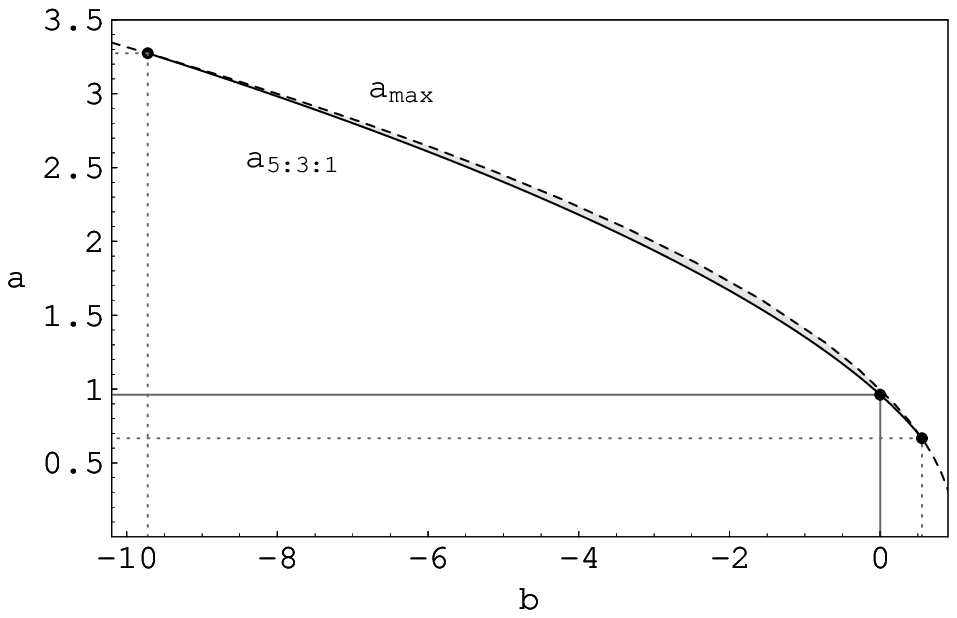}}\quad
\subfigure{\includegraphics[width=.47\hsize]{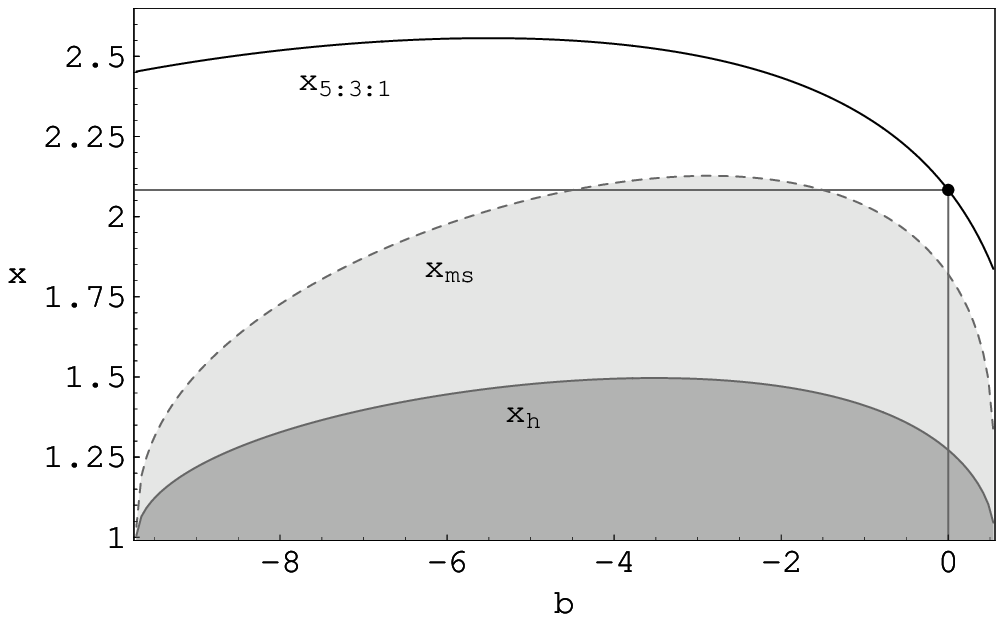}}\\
\subfigure{\includegraphics[width=.47\hsize]{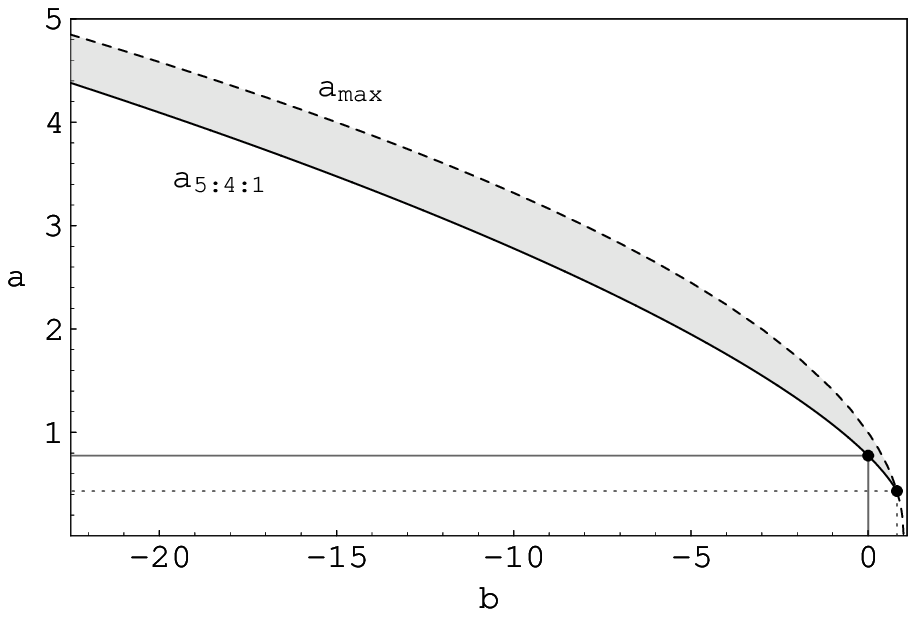}}\quad
\subfigure{\includegraphics[width=.47\hsize]{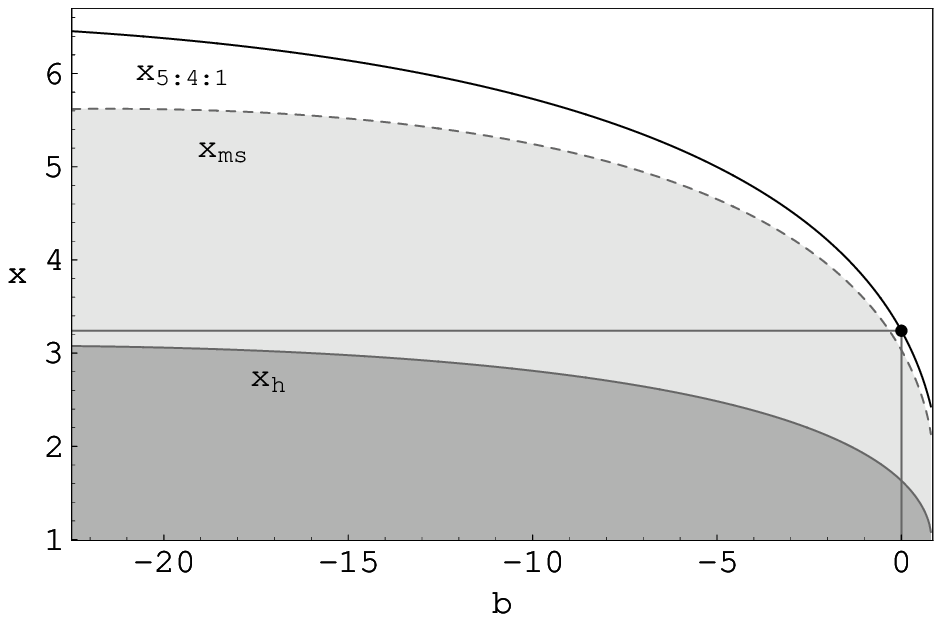}}\\
\subfigure{\includegraphics[width=.47\hsize]{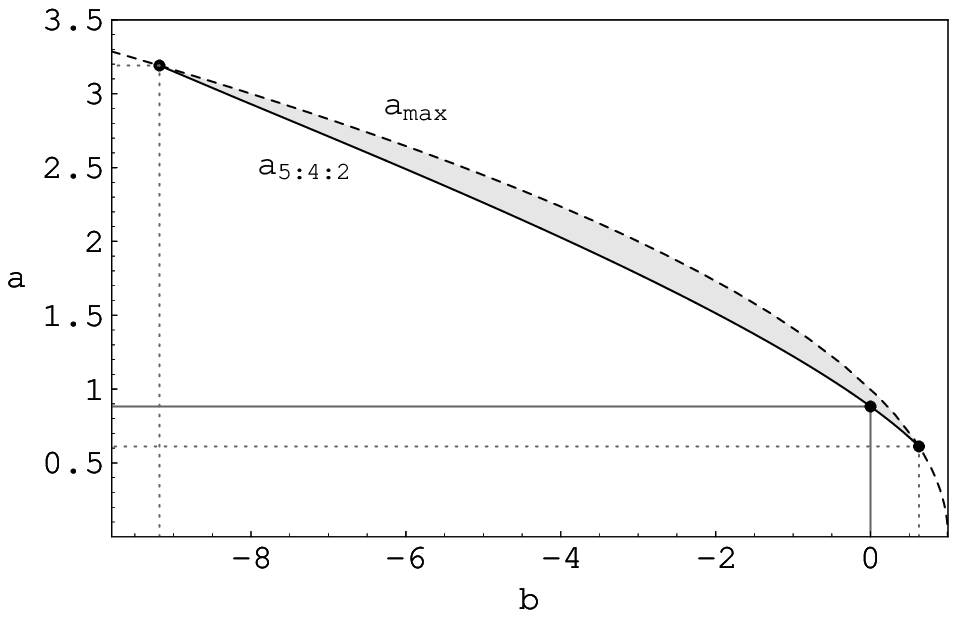}}\quad
\subfigure{\includegraphics[width=.47\hsize]{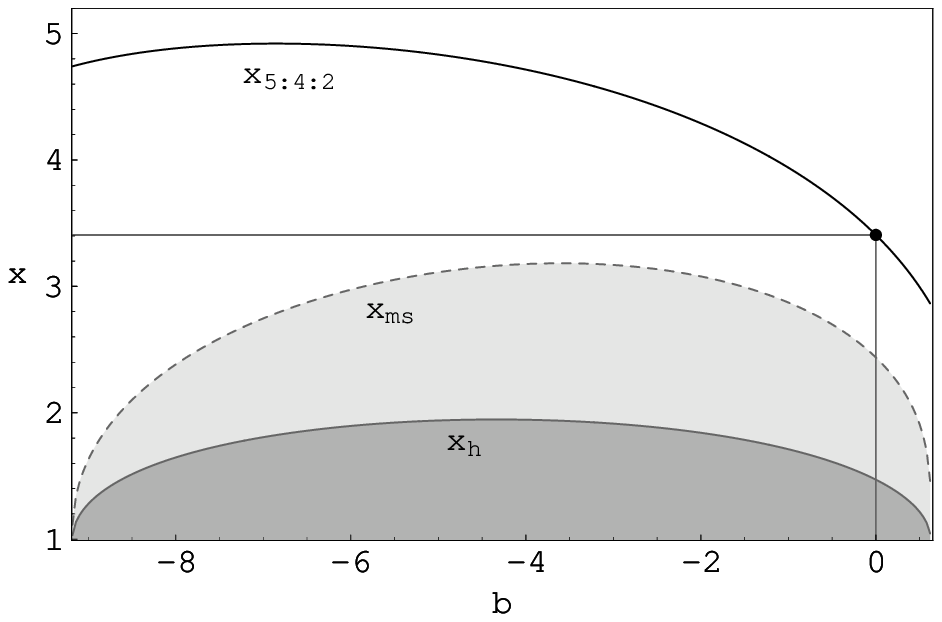}}
\caption{\label{magic-spin-triplety}The black hole spin $a$
(\textit{left panel}) and the shared resonance radius $x_{s:t:u}$
(\textit{right panel}) as the function of the brany parameter $b$
that imply the frequency ratio set
$\nu_{\mathrm{K}}\!:\!\nu_{\theta}\!:\!\nu_{\mathrm{r}} =
s\!:\!t\!:\!u$ arising at the same radius $x_{s:t:u}$ for
$s\!:\!t\!:\!u = 4\!:\!3\!:\!1$, $5\!:\!3\!:\!1$, $5\!:\!4\!:\!1$
and $5\!:\!4\!:\!2$.}
\end{figure*}

A detailed discussion of the black holes admitting strong resonant
phenomena is for small integer ($s\leq 5$) given in
\cite{SKT:2007:kratky:}. Of special interest seems to be the case
of the ``magic'' spin, when the Keplerian and epicyclic
frequencies are in the ratio
$\nu_{\mathrm{K}}\!:\!\nu_{\theta}\!:\!\nu_{\mathrm{r}} =
3\!:\!2\!:\!1$ at the shared resonance radius $x_{3:2:1}$. In
fact, this case involves rather extended structure of resonances
with $\nu_{\mathrm{K}}\!:\!\nu_{\mathrm{r}}=3\!:\!1$,
$\nu_{\mathrm{K}}\!:\!\nu_{\theta}=3\!:\!2$,
$\nu_{\theta}\!:\!\nu_{\mathrm{r}}=2\!:\!1$. Notice that in this
case also the simple combinational frequencies could be in this
small integer ratio as
  \begin{equation}
    \frac{\nu_\mathrm{K}}{\nu_{\theta}-\nu_{\mathrm{r}}}=\frac{3}{1},\qquad
    \frac{\nu_{\mathrm{K}}}{\nu_{\mathrm{K}}-\nu_\mathrm{r}}=\frac{3}{2},\qquad
    \frac{\nu_{\theta}}{\nu_{\theta}-\nu_\mathrm{r}}=\frac{2}{1}.
  \end{equation}
Of course we obtain the strongest possible resonances when the
beat frequencies enter the resonance satisfying the conditions
 \begin{equation}
   \frac{\nu_{\theta}+\nu_{\mathrm{r}}}{\nu_\mathrm{K}}=\frac{3}{3}=1,\qquad
   \frac{\nu_{\theta}}{\nu_\mathrm{K}-\nu_{\mathrm{r}}}=\frac{2}{2}=1,\qquad
   \frac{\nu_{\mathrm{r}}}{\nu_\mathrm{K}-\nu_{\theta}}=1,\quad
   \frac{\nu_{\theta}-\nu_\mathrm{r}}{\nu_{\mathrm{r}}}=1.
 \end{equation}

\begin{figure*}[!tbp]
\subfigure{\includegraphics[width=.48\hsize]{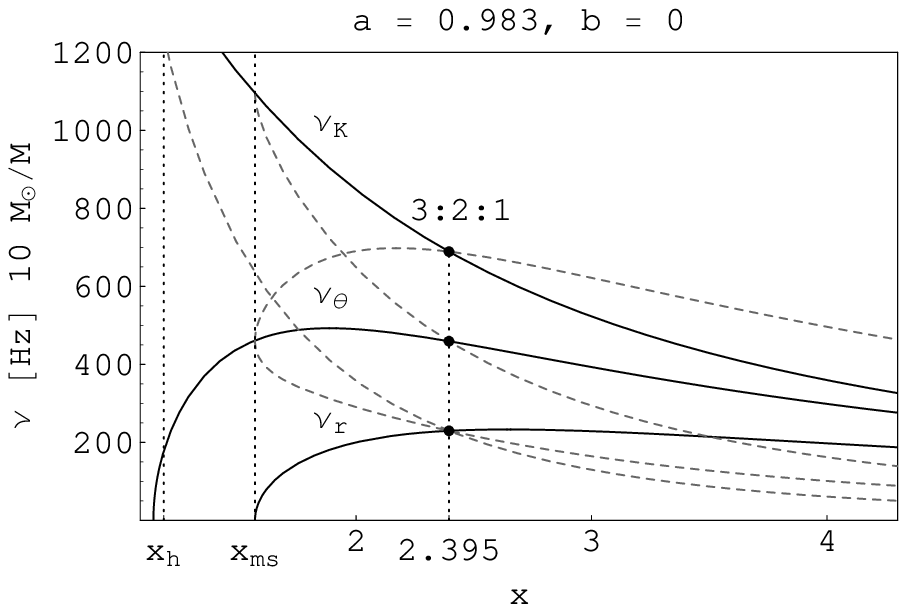}}\quad
\subfigure{\includegraphics[width=.48\hsize]{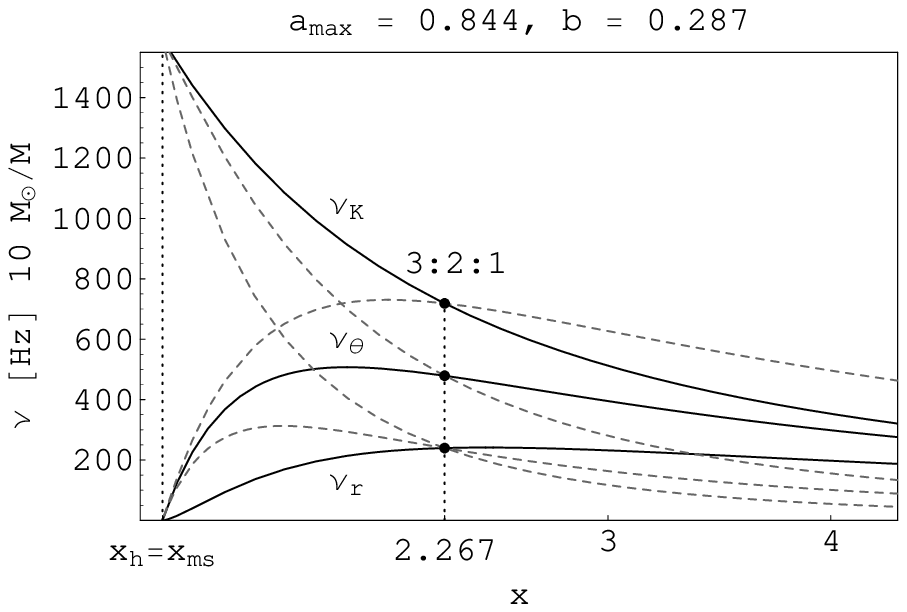}}
\caption{\label{magic-rezy} The special cases of the ``magic''
black hole spin $a$ enabling presence of strong resonant phenomena
at the radius where
$\nu_{\mathrm{K}}\!:\!\nu_{\theta}\!:\!\nu_{\mathrm{r}}=
3\!:\!2\!:\!1$ 
for Kerr spacetime (\textit{left panel}) and for braneworld black
hole with the extreme $b=0.287$ (\textit{right panel}). For
completeness we present the relevant simple combinational
frequencies $\nu_{\theta}-\nu_{\mathrm{r}}$,
$\nu_{\theta}+\nu_{\mathrm{r}}$, $\nu_{\mathrm{K}}-\nu_{\theta}$,
$\nu_{\mathrm{K}}-\nu_{\mathrm{r}}$ (grey dashed lines). Notice
that the ``magic'' spin represents the only case when the
combinational and direct orbital frequencies coincide at the
shared resonance radius.}
\end{figure*}

In the Kerr spacetimes, where $b=0$, we obtain the ``magic'' spin
$a_{\mathrm{magic}}=0.983$ and the shared resonance radius
$x_{3:2:1} = 2.395$ (see Fig.\,\ref{magic-rezy}). Assuming all
possible values of brany parameter $b$ we can conclude that this
special case of the ``magic'' spin could occur only for brany
parameter from the interval
\begin{equation}
    b_{\mathrm{magic}}\in \langle -3.258;0.287 \rangle ,
\end{equation}
that implies the ``magic'' spin from the interval
\begin{equation}
    a_{\mathrm{magic}}\in \langle 0.844;2.063 \rangle .
\end{equation}
Only for these values of $a$ and $b$ we have $a\leq
a_{\mathrm{max}}$, where $a_{\mathrm{max}}$ corresponds to the
extreme black hole (see Fig.\,\ref{magic-spin}).

\subsection{\label{sec:osm:2}Sgr\,A$^*$ black hole parameters}

The Galaxy center source Sgr\,A$^*$ can serve as a proper
candidate system, since three QPOs were reported (but not fully
accepted by the astrophysical community) for the system
\cite{Asch:2004:ASTRA:,Ter:2005:astro-ph0412500:} with frequency
ratio corresponding to the ``magic'' spin
\begin{equation}
(1/692) : (1/1130) : (1/2178) \approx 3 : 2 : 1
\end{equation}
and with the upper frequency being observed with a rather high
error
\begin{equation}
      \nu_{\mathrm{upp}} = (1.445 \pm 0.16)~\mathrm{mHz}.
\end{equation}

Considering a black hole with the spin comparable to the ``magic''
value $a \sim a_{\mathrm{magic}}$, with the frequency ratio
$\nu_{\mathrm{K}}\!:\!\nu_{\theta}\!:\!\nu_{\mathrm{r}} =
3\!:\!2\!:\!1$ at the shared resonance radius $x_{3:2:1}$, and
identifying $\nu_{\mathrm{upp}} = \nu_{\mathrm{K}}$, we obtain for
all possible values of ``magic'' brany parameter
$b_{\mathrm{magic}}\in \langle -3.258;0.287 \rangle$ the black
hole mass of Sgr\,A$^*$ in the interval
\begin{equation}
3.82\times 10^6~\mathrm{M}_{\odot} < M < 5.59\times
10^6~\mathrm{M}_{\odot},
\end{equation}
which meets the allowed range of the Sgr\,A$^*$ mass coming from
the analysis of the orbits of stars moving within 1000 light hours
of Sgr\,A$^*$ \cite{Ghe-etal:2005:ASTRJ2:}
\begin{equation}
2.8 \times 10^6~\mathrm{M}_{\odot} < M < 4.6 \times
10^6~\mathrm{M}_{\odot}
\end{equation}
at its higher mass end. In all these cases, the black hole spin $a
\rightarrow a_{\mathrm{max}}$, in agreement with the assumption
that Galactic center black hole should be fast rotating. The
results are summarized in Table~\ref{Sgr}.

\begin{table}[t]
\centering \caption{\label{Sgr}Determining of the black hole spin
and mass in Sgr\,A$^*$ with assumed observed characteristic
frequency ratio set $\nu_{\mathrm{K}} \!:\! \nu_{\theta} \!:\!
\nu_{\mathrm{r}} = 3\!:\!2\!:\!1$ at the shared resonance radius
$x_{3:2:1}$ for various values of brany parameter $b$;
$\nu_{\mathrm{upp}}=(1.445\pm 0.16)\,\mathrm{mHz}$ is used to
determine the black hole mass.}
\begin{tabular}{ccccc}
\hline\noalign{\smallskip}
    $b_{\mathrm{magic}}$ & $a_{\mathrm{magic}}$ & $x_{\mathrm{ms}}$ &
    $x_{\mathrm{3:2:1}}$ &  $M\,\left[10^6\,\mathrm{M}_{\odot}\right]$\\
\noalign{\smallskip}\hline\noalign{\smallskip} $0.287150$ &
$0.844304$ & $1$ & $2.26663$ & $4.477$ -- $5.592$
\\ \noalign{\smallskip}
$0 $ & $0.983043$ & $1.57081$ & $2.39467$ & $4.293$ -- $5.362$
\\ \noalign{\smallskip}
$-1$ & $1.378867$ & $1.79706$ & $2.65656$ & $3.971$ -- $4.959$
\\ \noalign{\smallskip}
$-2$ & $1.705166$ & $1.73104$ & $2.77498$ & $3.849$ -- $4.808$
\\ \noalign{\smallskip}
$-2.969659$ & $1.985138$ & $1.42803$ & $2.81093$ & $3.819$ --
$4.769$ \\ \noalign{\smallskip}
$-3.257659$ & $2.063410$ & $1$ & $2.80953$ & $3.821$ -- $4.773$ \\
\noalign{\smallskip}\hline
\end{tabular}
\end{table}

\begin{figure*}[!tbp]
\centering
\includegraphics[width=.7\hsize]{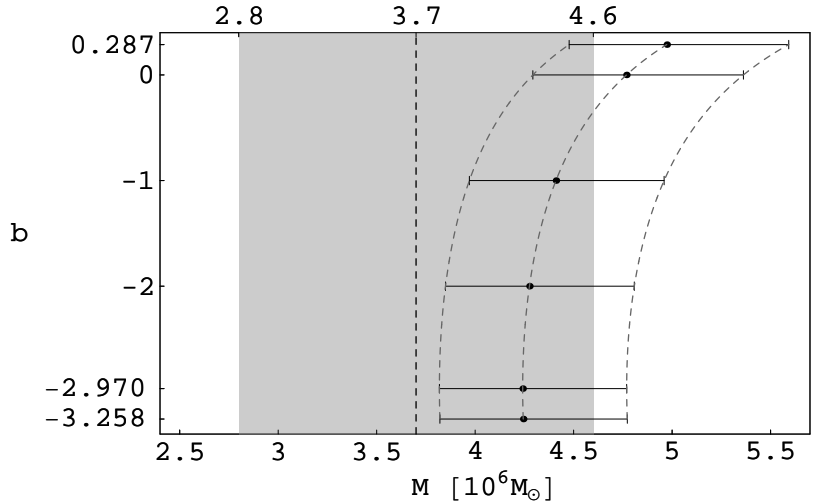}
\caption{\label{offsety} Mass of Sgr\,A$^*$: strong resonant model
with the frequency ratio
$\nu_{\mathrm{K}}\!:\!\nu_{\theta}\!:\!\nu_{\mathrm{r}} =
3\!:\!2\!:\!1$ for various values of brany parameter $b$. The
observational restrictions from the orbital motion of stars in
vicinity of Sgr\,A$^*$ \cite{Ghe-etal:2005:ASTRJ2:} are
illustrated here by the gray rectangle.}
\end{figure*}

From Fig.\,\ref{offsety} we can see that the best fit is obtained
for the brany parameter $b\sim-2.97$ that implies ``magic'' spin
$a_{\mathrm{magic}}\sim 1.99$ and the radius $x_{3:2:1}=2.81$.

The model should be further tested and more precise frequency
measurements are very important for better fits of the models and
observational data.

\section{\label{sec:devet}Conclusions}

The orbital resonance model and its simple generalization to
multiresonance model with strong resonances is formulated for the
brany Kerr black holes, when the bulk-space influence is described
by a single, brany tidal charge parameter.

In the limit of strong gravitational field, the brany parameter
$b$ can be, in principle, high in its magnitude, therefore, we put
no restriction on the values of $b$. We describe the properties of
the radial and vertical epicyclic frequencies related to the
oscillatory motion in the equatorial plane of the Kerr spacetimes.
While their behaviour is qualitatively similar for Kerr and brany
Kerr black holes, there are strong differences in the case of
naked singularities -- in some range of their parameters, the
vertical epicyclic frequency could be even lower than the radial
one. Such a situation is impossible in standard Kerr spacetimes.
Further, in the field of brany Kerr naked singularities, the
structure of the radial profiles is much richer than in the
standard case, namely the number of local extrema could be higher
in comparison with the standard Kerr naked singularities. In a
special family of the brany naked singularity spacetimes, the
radial epicyclic frequency has no zero point since there is no
marginally stable circular geodesic in these spacetimes.

Assuming the parametric resonance acting directly between the
oscillations with radial and vertical epicyclic frequency
\cite{Ter-Abr-Klu:2005:ASTRA:QPOresmodel}, we give the rule for
the resonant radius with a given frequency ratio. The rule is
tested for the uppermost twin frequencies observed in the GRS
1915+105 microquasar; and limits on the spin and brany parameters
are obtained and compared with the estimates for $b=0$, given in
\cite{Ter-Abr-Klu:2005:ASTRA:QPOresmodel}.\footnote{However, it
should be noted that whole the five frequency pattern can be
explained in the framework of the extended resonance model
assuming near-extreme Kerr black holes
\cite{Bla-etal:2006:ASTRJ2:,Sra:2005:ASTRN:,Stu-Sla-Ter:2007:ASTRA:,Stu-Sla-Ter:2006:ASTRA:Humpy,Stu-Sla-Tor:2007:RAGtime8and9CrossRef:XORMHump}.}
 Further, it is shown that the relativistic precession model
\cite{Ste-Vie:1998:ASTRJ2L:} and the trapped oscillations of
warped disc model \cite{Kato:2007:PASJ:,Kat:2004:PUBASJ:QPOsmodel}
are not able to fit the observational data of GRS 1915+105, with
high spin limits, as these two models work well for small or
mediate black hole spin.

In the Galaxy center source Sgr\,A$^*$, three frequencies were
reported
\cite{Asch:2004:ASTRA:,Asc:2007:,Ter:2005:astro-ph0412500:,Tor:2005:ASTRN:}
 that could be treated in the scope of the strong resonance model.
The model predicts an exact value of the black hole spin and puts
limits on its mass. It is shown that the black hole mass estimate
given by the strong resonance model is in the best agreement with
the value of $M \sim 3.7\times 10^6\,\mathrm{M}_{\odot}$
\cite{Ghe-etal:2005:ASTRJ2:} for negative brany parameter $b\sim
-2.97$, with the ``magic'' spin $a\approx 1.99$. Generally, for
negative values of $b$ the fit of the model with observational
data is better than for $b=0$, while for positive values of $b$
the inverse holds.

Our resonance models could be to some extend applied to slowly
rotating neutron stars, when the Kerr metric could be relevant,
and there are some indications that orbital resonance models could
be relevant to explain also the data from binary neutron star
systems
\cite{Tor:2007:ASTRA:RevTwPk6,Tor-Stu-Bak:2007:CEURJP:,Tor-Bak-Stu-Cec:2007::prep,Tor-etal:2008:AA:}.

We can conclude that the orbital resonance model and its
generalization to the multiresonant model is able to put some
astrophysically interesting limits on the values of the brany
parameter and could be useful in estimating influence of
hypothetical external dimension to the properties of the brany
universe.

\begin{acknowledgements}
This work was supported by the Czech grant MSM~4781305903. One of
the authors (Zden\v{e}k Stuchl\'{\i}k) would like to express his
gratitude to the Czech Committee for Collaboration with CERN for
support and Theory Division CERN for perfect hospitality.
\end{acknowledgements}


\begin{thebibliography}{10}

\bibitem{Abr-etal:2005:RAGtime6and7:CrossRef}
Abramowicz, M.A., Barret, D., Bursa, M., Hor{\'{a}}k, J.,
Klu{\'z}niak, W.,
  Rebusco, P., T{\"{o}}r{\"{o}}k, G.: {A note on the slope-shift
  anticorrelation in the neutron star kHz QPOs data}.
\newblock In: Hled{\'{\i}}k and Stuchl{\'{\i}}k
  \cite{Hle-Stu:2005:RAGtime6and7:Proceedings}, pp. 1--9

\bibitem{Abr-etal:2005:ASTRN:}
Abramowicz, M.A., Barret, D., Bursa, M., Hor{\'{a}}k, J.,
Klu{\'z}niak, W.,
  Rebusco, P., T{\"{o}}r{\"{o}}k, G.: {The correlations and anticorrelations in
  QPO data}.
\newblock Astronom. Nachr. \textbf{326}(9), 864--866 (2005), arXiv:astro-ph/0510462v1

\bibitem{Abr-etal:ADAFs:1995:}
Abramowicz, M.A., Chen, X., Kato, S., Lasota, J.P., Regev, O.:
{Thermal
  equilibria of accretion disks}.
\newblock Astrophys. J. \textbf{438}, L37--L39 (1995), arXiv:astro-ph/9409018v1

\bibitem{Abr-Klu:2001:ASTRA:}
Abramowicz, M.A., Klu{\'z}niak, W.: {A precise determination of
black hole spin
  in GRO~J1655$-$40}.
\newblock Astronomy and Astrophysics \textbf{374}, L19 (2001), arXiv:astro-ph/0105077v1

\bibitem{Abr-etal:2004:RAGtime4and5:CrossRef}
Abramowicz, M.A., Klu{\'z}niak, W., Stuchl{\'{\i}}k, Z.,
T{\"{o}}r{\"{o}}k, G.:
  {Twin peak QPOs frequencies in microquasars and Sgr\,A$^*$. The resonance and
  other orbital models}.
\newblock In: Hled{\'{\i}}k and Stuchl{\'{\i}}k
  \cite{Hle-Stu:2004:RAGtime4and5:Proceedings}, pp. 1--23

\bibitem{Ali-Gal:1981:}
Aliev, A.N., Galtsov, D.V.: {Radiation from relativistic particles
in
  nongeodesic motion in a strong gravitational field}.
\newblock General Relativity and Gravitation \textbf{13}, 899--912 (1981)

\bibitem{Ali-Gum:2004:}
Aliev, A.N., {G{\"u}mr{\"u}k{\c c}{\"u}o{\u g}lu}, A.E.:
{Gravitational Field
  Equations on and off a 3-Brane World}.
\newblock Classical Quantum Gravity \textbf{21}, 5081 (2004), arXiv:hep-th/0407095v1

\bibitem{Ali-Gum:2005:}
Aliev, A.N., {G{\"u}mr{\"u}k{\c c}{\"u}o{\u g}lu}, A.E.: {Charged
rotating
  black holes on a 3-brane}.
\newblock Phys. Rev. D \textbf{71}(10), 104,027 (2005), arXiv:hep-th/0502223v2

\bibitem{Ark-Dim-Dva:1998:}
Arkani-Hamed, N., Dimopoulos, S., Dvali, G.: {The Hierarchy
Problem and New
  Dimensions at a Millimeter}.
\newblock Phys. Lett. B \textbf{429}, 263--272 (1998), arXiv:hep-ph/9803315v1

\bibitem{Arn-Des-Mis:1962:}
Arnowitt, R., Deser, S., Misner, C.W.: {The Dynamics of General
Relativity}.
\newblock L. Witten ed. (Wiley 1962) \textbf{7}, 227--265 (1962), arXiv:gr-qc/0405109v1

\bibitem{Asch:2004:ASTRA:}
Aschenbach, B.: {Measuring mass and angular momentum of black
holes with
  high-frequency quasi-periodic oscillations}.
\newblock Astronomy and Astrophysics \textbf{425}, 1075--1082
(2004), arXiv:astro-ph/0406545v1

\bibitem{Asc:2007:}
Aschenbach, B.: {Measurement of Mass and Spin of Black Holes with
QPOs}.
\newblock {Chin. J. Astron. Astrophys.}  (2007),
\newblock accepted, arXiv:0710.3454v1 [astro-ph]

\bibitem{Asc-etal:2004:ASTRA:}
Aschenbach, B., Grosso, N., Porquet, D., Predehl, P.: {X-ray
flares reveal mass
  and angular momentum of the Galactic Center black hole}.
\newblock Astronomy and Astrophysics \textbf{417}, 71--78 (2004), arXiv:astro-ph/0401589v2

\bibitem{Bak-etal:2008:magn:}
Bakala, P., {\v S}r{\'{a}}mkov{\'{a}}, E., Stuchl{\'{\i}}k, Z.,
  T{\"{o}}r{\"{o}}k, G.: {On magnetic-field induced non-geodesic corrections to
  the relativistic precession QPO model}.
\newblock Astrophys. J.  (2008),
\newblock submitted

\bibitem{Bar:1973:BlaHol:}
Bardeen, J.M.: Timelike and null geodesics in the {K}err metric.
\newblock In: C.D. Witt, B.S.D. Witt (eds.) {Black Holes}, p. 215. Gordon and
  Breach, New York--London--Paris (1973)

\bibitem{Bla-etal:2006:ASTRJ2:}
Blaes, O.M., {\v S}r{\'{a}}mkov{\'{a}}, E., Abramowicz, M.A.,
Klu{\'z}niak, W.,
  Torkelsson, U.: {Epicyclic Oscillations of Fluid Bodies: Newtonian Nonslender
  Torus}.
\newblock Astrophys. J. \textbf{665}, 642--653 (2007), arXiv:0706.4483v1 [astro-ph]

\bibitem{Boh-etal:2008:SolarSys:}
B{\"o}hmer, C.G., Harko, T., Lobo, F.S.N.: {Solar system tests of
brane world
  models}.
\newblock Classical Quantum Gravity \textbf{25}(4), 5015 (2008), arXiv:0801.1375v2 [gr-qc]

\bibitem{Car:1968:PHYSREV:}
Carter, B.: {Global Structure of the Kerr Family of Gravitational
Fields}.
\newblock Phys. Rev. \textbf{174}(5), 1559--1571 (1968)

\bibitem{Car:1973:BlaHol:}
Carter, B.: Black hole equilibrium states.
\newblock In: C.D. Witt, B.S.D. Witt (eds.) {Black Holes}, pp. 57--214. Gordon
  and Breach, New York--London--Paris (1973)

\bibitem{Cha-Haw-Rea:2000:}
Chamblin, A., Hawking, S.W., Reall, H.S.: {Brane-World Black
Holes}.
\newblock Phys. Rev. D \textbf{61}(6), 065,007 (2000), arXiv:hep-th/9909205v2

\bibitem{Cha-etal:2001:}
Chamblin, A., Reall, H.S., Shinkai, H.A., Shiromizu, T.: {Charged
Brane-World
  Black Holes}.
\newblock Phys. Rev. D \textbf{63}(6), 064,015 (2001), arXiv:hep-th/0008177v2

\bibitem{Dad-Kal:1977:}
Dadhich, N., Kale, P.P.: {Equatorial circular geodesics in the
Kerr--Newman
  geometry}.
\newblock J. Math. Phys. \textbf{18}, 1727--1728 (1977)

\bibitem{Dad-etal:2000:}
Dadhich, N., Maartens, R., Papadopoulos, P., Rezania, V.: {Black
holes on the
  brane}.
\newblock Phys. Lett. B \textbf{487}, 1 (2000), arXiv:hep-th/0003061v3

\bibitem{Dam-etal:1978:}
Damour, T., Hanni, R.S., Ruffini, R., Wilson, J.R.: {Regions of
magnetic
  support of a plasma around a black hole}.
\newblock Phys. Rev. D \textbf{17}(6), 1518--1523 (1978)

\bibitem{Dim-Lan:2001:}
Dimopoulos, S., Landsberg, G.: {Black Holes at the Large Hadron
Collider}.
\newblock Phys. Rev. Lett. \textbf{87}(16), 161,602 (2001), arXiv:hep-ph/0106295v1

\bibitem{Dov-Kar-Mar:2004:RAGtime4and5:CrossRef}
Dov\v{c}iak, M., Karas, V., Martocchia, A., Matt, G., Yaqoob, T.:
{An XSPEC
  model to explore spectral features from black-hole sources}.
\newblock In: Hled{\'{\i}}k and Stuchl{\'{\i}}k
  \cite{Hle-Stu:2004:RAGtime4and5:Proceedings}, pp. 33--73, arXiv:astro-ph/0407330v1

\bibitem{Emp-Mas-Rat:2002:}
Emparan, R., Masip, M., Rattazzi, R.: {Cosmic Rays as Probes of
Large Extra
  Dimensions and TeV Gravity}.
\newblock Phys. Rev. D \textbf{65}, 064,023 (2002), arXiv:hep-ph/0109287v2

\bibitem{Fab-Min:2005:XSpectraKerr:Book}
Fabian, A.C., Miniutti, G.: {Kerr Spacetime: Rotating Black Holes
in General
  Relativity}.
\newblock Cambridge University Press, Cambridge (2005).
\newblock Eprint arXiv:astro-ph/0507409v1 is a part of this book

\bibitem{Ger-Maa:2001:}
Germani, C., Maartens, R.: {Stars in the braneworld}.
\newblock Phys. Rev. D \textbf{64}, 124,010 (2001), arXiv:hep-th/0107011v3

\bibitem{Ghe-etal:2005:ASTRJ2:}
Ghez, A.M., Salim, S., Hornstein, S.D., Tanner, A., Lu, J.R.,
Morris, M.,
  Becklin, E.E., Duch{\^e}ne, G.: {Stellar Orbits around the Galactic Center
  Black Hole}.
\newblock Astrophys. J. \textbf{620}, 744--757 (2005), arXiv:astro-ph/0306130v2

\bibitem{Gre-Laf:1993:}
Gregory, R., Laflamme, R.: {Black Strings and p-Branes are
Unstable}.
\newblock Phys. Rev. Lett. \textbf{70}(19), 2837--2840 (1993), arXiv:hep-th/9301052v2

\bibitem{Hle-Stu:2004:RAGtime4and5:Proceedings}
Hled{\'{\i}}k, S., Stuchl{\'{\i}}k, Z. (eds.): Proceedings of
RAGtime 4/5:
  Workshops on black holes and neutron stars, Opava, 14--16/13--15 October
  2002/2003. Silesian University in Opava, Opava (2004)

\bibitem{Hle-Stu:2005:RAGtime6and7:Proceedings}
Hled{\'{\i}}k, S., Stuchl{\'{\i}}k, Z. (eds.): Proceedings of
RAGtime 6/7:
  Workshops on black holes and neutron stars, Opava, 16--18/18--20 September
  2004/2005. Silesian University in Opava, Opava (2005)

\bibitem{Hle-Stu:2007:RAGtime8and9:Proceedings}
Hled{\'{\i}}k, S., Stuchl{\'{\i}}k, Z. (eds.): Proceedings of
RAGtime 8/9:
  Workshops on black holes and neutron stars, Opava, Hradec nad Moravic\'{i},
  15--19/19--21 September 2006/2007. Silesian University in Opava, Opava (2007)

\bibitem{Hor-Mae:2001:}
Horowitz, G.T., Maeda, K.: {Fate of the Black String Instability}.
\newblock Phys. Rev. Lett. \textbf{87}(13), 131,301 (2001), arXiv:hep-th/0105111v2

\bibitem{Kar-Vok-Pol:1992:ASTRJ2L:}
Karas, V., Vokrouhlick{\'{y}}, D., Polnarev, A.G.: {In the
vicinity of a
  rotating black hole~-- A fast numerical code for computing observational
  effects}.
\newblock Monthly Notices Roy. Astronom. Soc. \textbf{259}, 569--575 (1992)

\bibitem{Kat:2004:PUBASJ:QPOsmodel}
Kato, S.: {Resonant Excitation of Disk Oscillations by Warps: A
Model of kHz
  QPOs}.
\newblock Publ. Astronom. Soc. Japan \textbf{56}(5), 905--922
(2004), arXiv:astro-ph/0409051v2

\bibitem{Kato:2007:PASJ:}
Kato, S.: {Frequency Correlations of QPOs Based on a Disk
Oscillation Model in
  Warped Disks}.
\newblock Publ. Astronom. Soc. Japan \textbf{59}, 451--455 (2007), arXiv:astro-ph/0701085v1

\bibitem{Kat-Fuk-Min:1998:BHAccDis:}
Kato, S., Fukue, J., Mineshige, S.: {Black-hole accretion disks}.
\newblock In: S.~Kato, J.~Fukue, S.~Mineshige (eds.) {Black-hole accretion
  disks}. Kyoto University Press, Kyoto, Japan (1998)

\bibitem{Kee-Pet:2006:brane-lensing:}
Keeton, C.R., Petters, A.O.: {Formalism for testing theories of
gravity using
  lensing by compact objects. III. Braneworld gravity}.
\newblock Phys. Rev. D \textbf{73}(10), 104,032 (2006), arXiv:gr-qc/0603061v3

\bibitem{Kli:2000:ARASTRA:}
van~der Klis, M.: {Millisecond Oscillations in X-ray Binaries}.
\newblock Annual Review of Astronomy and Astrophysics \textbf{38}, 717--760
  (2000), arXiv:astro-ph/0001167v1

\bibitem{Kli:2004:}
van~der Klis, M.: {Rapid X-ray Variability}.
\newblock In: W.H.G. Lewin, M.~van~der Klis (eds.) {Compact Stellar X-Ray
  Sources}, pp. 39--112. Cambridge University Press, Cambridge (2006)

\bibitem{Klu-etal:Mexico:2007:}
Klu{\'z}niak, W., Abramowicz, M.A., Bursa, M., T{\"{o}}r{\"{o}}k,
G.: {QPOs and
  Resonance in Accretion Disks}.
\newblock In: W.H. Lee, E.~Ram\'{i}rez-Ruiz (eds.) Revista Mexicana de
  Astronomia y Astrofisica (Serie de Conferencias), vol.~27, pp. 18--25 (2007)

\bibitem{Koz-Jar-Abr:1978:ASTRA:}
Koz{\l}owski, M., Jaroszy{\'n}ski, M., Abramowicz, M.A.: {The
analytic theory
  of fluid disks orbiting the Kerr black hole}.
\newblock Astronomy and Astrophysics \textbf{63}(1--2), 209--220 (1978)

\bibitem{Kro-Haw:2002:ASTRJ2:}
Krolik, J.H., Hawley, J.F.: {Where Is the Inner Edge of an
Accretion Disk
  around a Black Hole?}
\newblock Astrophys. J. \textbf{573}(2), 754--763 (2002), arXiv:astro-ph/0203289v1

\bibitem{Lac-Cze-Abr:2006:astro-ph0607594:}
Lachowicz, P., Czerny, B., Abramowicz, M.A.: {Wavelet analysis of
MCG-6-30-15
  and NGC~4051: a possible discovery of QPOs in $2\!:\!1$ and $3\!:\!2$
  resonance}.
\newblock Monthly Notices Roy. Astronom. Soc.  (2006),
\newblock submitted, arXiv:astro-ph/0607594v1

\bibitem{Lan-Lif:1976:Mech:}
Landau, L.D., Lifshitz, E.M.: Mechanics, Course of
Theoretical Physics,
  vol.~I, 3rd edn.
\newblock Elsevier Butterworth-Heinemann, Oxford (1976)

\bibitem{Laor:1991:ASTRJ2:}
Laor, A.: {Line profiles from a disk around a rotating black
hole}.
\newblock Astrophys. J. \textbf{376}, 90--94 (1991)

\bibitem{Maa:2004:}
Maartens, R.: {Brane-world gravity}.
\newblock Living Rev. Rel. \textbf{7}, 7 (2004), arXiv:gr-qc/0312059v2

\bibitem{McCli-Nar-Sha:2007:}
McClintock, J.E., Narayan, R., Shafee, R.: {Estimating the Spins
of
  Stellar-Mass Black Holes}.
\newblock In: M.~Livio, A.~Koekemoer (eds.) {Black Holes}. Cambridge University
  Press, Cambridge (2007), in press, arXiv:0707.4492v1 [astro-ph]

\bibitem{McCli-Rem:2004:CompactX-Sources:}
McClintock, J.E., Remillard, R.A.: {Black Hole Binaries}.
\newblock In: W.H.G. Lewin, M.~van~der Klis (eds.) {Compact Stellar X-Ray
  Sources}. Cambridge University Press, Cambridge (2004), arXiv:astro-ph/0306213v4

\bibitem{McCli-etal:2006:astro-ph/0606076:}
McClintock, J.E., Shafee, R., Narayan, R., Remillard, R.A., Davis,
S.W., Li,
  L.X.: {The Spin of the Near-Extreme Kerr Black Hole GRS 1915+105}.
\newblock Astrophys. J. \textbf{652}, 518--539 (2006), arXiv:astro-ph/0606076v2

\bibitem{Mid-etal:2006:}
Middleton, M., Done, C., Gierli{\'n}ski, M., Davis, S.W.: {Black
hole spin in
  GRS 1915+105}.
\newblock Monthly Notices Roy. Astronom. Soc. \textbf{373}, 1004--1012
(2006), arXiv:astro-ph/0601540v2

\bibitem{Mis-Tho-Whe:1973:Gra:}
Misner, C.W., Thorne, K.S., Wheeler, J.A.: Gravitation.
\newblock W. H. Freeman and Co, New York, San Francisco (1973)

\bibitem{Mod-Pan-Sen:2002:}
Modgil, M.S., Panda, S., Sengupta, G.: {Rotating Brane World Black
Holes}.
\newblock Modern Phys. Lett. A \textbf{17}, 1479--1487 (2002), arXiv:hep-th/0104122v2

\bibitem{Nar-Yi:1994:ASTRJ2:}
Narayan, R., Yi, I.: {Advection-dominated accretion: A
self-similar solution}.
\newblock Astrophys. J. \textbf{428}, L13--L16 (1994), arXiv:astro-ph/9403052v1

\bibitem{Noj-etal:2000:}
Nojiri, S., Obregon, O., Odintsov, S.D., Ogushi, S.: {Dilatonic
Brane-World
  Black Holes, Gravity Localization and Newton's Constant}.
\newblock Phys. Rev. D \textbf{62}(6), 064,017 (2000), arXiv:hep-th/0003148v1

\bibitem{Nov-Tho:1973:BlaHol:}
Novikov, I.D., Thorne, K.S.: Black hole astrophysics.
\newblock In: C.D. Witt, B.S.D. Witt (eds.) {Black Holes}, p. 343. Gordon and
  Breach, New York--London--Paris (1973)

\bibitem{Ran-Sun:1999:}
Randall, L., Sundrum, R.: {An Alternative to Compactification}.
\newblock Phys. Rev. Lett. \textbf{83}(23), 4690--4693 (1999), arXiv:hep-th/9906064v1

\bibitem{Rem:2005:ASTRN:}
Remillard, R.A.: {X-ray spectral states and high-frequency QPOs in
black hole
  binaries}.
\newblock Astronom. Nachr. \textbf{326}(9), 804--807 (2005), arXiv:astro-ph/0510699v1

\bibitem{Rem-McCli:2006:ARASTRA:}
Remillard, R.A., McClintock, J.E.: {X-Ray Properties of Black-Hole
Binaries}.
\newblock Annual Review of Astronomy and Astrophysics \textbf{44}(1), 49--92
  (2006), arXiv:astro-ph/0606352v1

\bibitem{Rez-etal:2003:MONNR:}
Rezzolla, L., Yoshida, S., Maccarone, T.J., Zanotti, O.: {A new
simple model
  for high-frequency quasi-periodic oscillations in black hole candidates}.
\newblock Monthly Notices Roy. Astronom. Soc. \textbf{344}(3), L37--L41
(2003), arXiv:astro-ph/0307487v1

\bibitem{Sas-Shi-Mae:2000:}
Sasaki, M., Shiromizu, T., Maeda, K.-i.: {Gravity, stability, and
energy
  conservation on the Randall--Sundrum brane world}.
\newblock Phys. Rev. D \textbf{62}, 024,008 (2000), arXiv:hep-th/9912233v3

\bibitem{Sch-Stu:2007:RAGtime8and9CrossRef:OEBraK}
Schee, J., Stuchl{\'{\i}}k, Z.: {Optical effects in brany Kerr
spacetimes}.
\newblock In: Hled{\'{\i}}k and Stuchl{\'{\i}}k
  \cite{Hle-Stu:2007:RAGtime8and9:Proceedings}, pp. 221--256

\bibitem{Sch-Stu:2007:RAGtime8and9CrossRef:SpBranKBH}
Schee, J., Stuchl{\'{\i}}k, Z.: {Spectral line profile of
radiating ring
  orbiting a brany Kerr black hole}.
\newblock In: Hled{\'{\i}}k and Stuchl{\'{\i}}k
  \cite{Hle-Stu:2007:RAGtime8and9:Proceedings}, pp. 209--220

\bibitem{Sha:2006::astro-ph/0508302}
Shafee, R., McClintock, J.E., Narayan, R., Davis, S.W., Li, L.X.,
Remillard,
  R.A.: {Estimating the Spin of Stellar-Mass Black Holes by Spectral Fitting of
  the X-Ray Continuum}.
\newblock Astrophys. J. \textbf{636}, L113--L116 (2006), arXiv:astro-ph/0508302v2

\bibitem{Shi-Mae-Sas:2000:}
Shiromizu, T., Maeda, K.-i., Sasaki, M.: {The Einstein Equations
on the 3-Brane
  World}.
\newblock Phys. Rev. D \textbf{62}, 024,012 (2000), arXiv:gr-qc/9910076v3

\bibitem{Sra:2005:ASTRN:}
{\v S}r{\'{a}}mkov{\'{a}}, E.: {Epicyclic oscillation modes of a
Newtonian,
  non-slender torus}.
\newblock Astronom. Nachr. \textbf{326}(9), 835--837 (2005)

\bibitem{Ste-Vie:1998:ASTRJ2L:}
Stella, L., Vietri, M.: {Lense--Thirring Precession and
Quasi-periodic
  Oscillations in Low-Mass X-Ray Binaries}.
\newblock Astrophys. J. Lett. \textbf{492}, L59--L62 (1998), arXiv:astro-ph/9709085v1

\bibitem{Str:2007:astro-ph/0701390:}
Strohmayer, T.E., Mushotzky, R., Winter, L., Soria, R., Uttley,
P., Cropper,
  M.: {Quasi-periodic Variability in NGC 5408 X-1}.
\newblock Astrophys. J. \textbf{660}, 580--586 (2007), arXiv:astro-ph/0701390v1

\bibitem{Stu-Kot-Tor:2007:RAGtime8and9CrossRef:MrmQPO}
Stuchl{\'{\i}}k, Z., Kotrlov{\'{a}}, A., T{\"{o}}r{\"{o}}k, G.:
  {Multi-resonance models of QPOs}.
\newblock In: Hled{\'{\i}}k and Stuchl{\'{\i}}k
  \cite{Hle-Stu:2007:RAGtime8and9:Proceedings}, pp. 363--416

\bibitem{SKT:2007:kratky:}
Stuchl{\'{\i}}k, Z., Kotrlov{\'{a}}, A., T{\"{o}}r{\"{o}}k, G.:
{Black holes
  admitting strong resonant phenomena}.
\newblock Acta Astronom. (2008), accepted, arXiv:0812.4418v1 [astro-ph]

\bibitem{Stu-Sla-Tor:2007:RAGtime8and9CrossRef:XORMHump}
Stuchl{\'{\i}}k, Z., Slan{\'{y}}, P., T{\"{o}}r{\"{o}}k, G.:
{Extended orbital
  resonance model with hump-induced oscillations}.
\newblock In: Hled{\'{\i}}k and Stuchl{\'{\i}}k
  \cite{Hle-Stu:2007:RAGtime8and9:Proceedings}, pp. 449--467

\bibitem{Stu-Sla-Ter:2006:ASTRA:Humpy}
Stuchl{\'{\i}}k, Z., Slan{\'{y}}, P., T{\"{o}}r{\"{o}}k, G.:
{Humpy
  LNRF-velocity profiles in accretion discs orbiting almost extreme Kerr black
  holes~-- A possible relation to quasi-periodic oscillations}.
\newblock Astronomy and Astrophysics \textbf{463}(3), 807--816
(2007), arXiv:astro-ph/0612439v1

\bibitem{Stu-Sla-Ter:2007:ASTRA:}
Stuchl{\'{\i}}k, Z., Slan{\'{y}}, P., T{\"{o}}r{\"{o}}k, G.:
{LNRF-velocity
  hump-induced oscillations of a Keplerian disc orbiting near-extreme Kerr
  black hole: a possible explanation of high-frequency QPOs in GRS 1915+105}.
\newblock Astronomy and Astrophysics \textbf{470}, 401--404
(2007), arXiv:0704.1252v2 [astro-ph]

\bibitem{Ter:2005:astro-ph0412500:}
T{\"{o}}r{\"{o}}k, G.: {A possible $3\!:\!2$ orbital epicyclic
resonance in
  QPOs frequencies of Sgr\,A$^*$}.
\newblock Astronomy and Astrophysics \textbf{440}(1), 1--4 (2005), arXiv:astro-ph/0412500v1

\bibitem{Tor:2005:ASTRN:}
T{\"{o}}r{\"{o}}k, G.: {QPOs in microquasars and Sgr\,A$^*$:
measuring the
  black hole spin}.
\newblock Astronom. Nachr. \textbf{326}(9), 856--860 (2005), arXiv:astro-ph/0510669v1

\bibitem{Tor:2007:ASTRA:RevTwPk6}
T{\"{o}}r{\"{o}}k, G.: {Reverse of twin peak QPO amplitude
relationship in six
  atoll sources}.
\newblock Astronomy and Astrophysics  (2008),
\newblock submitted

\bibitem{Tor-etal:2008:AA:}
T{\"{o}}r{\"{o}}k, G., Abramowicz, M.A., Bakala, P., Bursa, M.,
Hor{\'{a}}k,
  J., Rebusco, P., Stuchl{\'{\i}}k, Z.: {On the origin of clustering of
  frequency ratios in the atoll source 4U~1636$-$53}.
\newblock Acta Astronom. \textbf{58}, 113--119 (2008), arXiv:0802.4026v2 [astro-ph]

\bibitem{Ter-Abr-Klu:2005:ASTRA:QPOresmodel}
T{\"{o}}r{\"{o}}k, G., Abramowicz, M.A., Klu{\'z}niak, W.,
Stuchl{\'{\i}}k, Z.:
  {The orbital resonance model for twin peak kHz quasi periodic oscillations in
  microquasars}.
\newblock Astronomy and Astrophysics \textbf{436}(1), 1--8 (2005), arXiv:astro-ph/0401464v1

\bibitem{Ter-etal:2007:spin-problem}
T{\"{o}}r{\"{o}}k, G., Abramowicz, M.A., Stuchl{\'{\i}}k, Z., {\v
  S}r{\'{a}}mkov{\'{a}}, E.: {QPOs in microquasars: the spin problem}.
\newblock In: W.I. {Hartkopf}, E.F. {Guinan}, P.~{Harmanec} (eds.) IAU
  Symposium, vol. 240, pp. 724--726 (2007), arXiv:astro-ph/0610497v1

\bibitem{Tor-Bak-Stu-Cec:2007::prep}
T{\"{o}}r{\"{o}}k, G., Bakala, P., Stuchl{\'{\i}}k, Z.,
{\v{C}}ech, P.:
  {Modelling the twin peak QPO distribution in the atoll source 4U~1636$-$53}.
\newblock Acta Astronom. \textbf{58}, 1--14 (2008)

\bibitem{Tor-Stu:2005:RAGtime6and7:CrossRef}
T{\"{o}}r{\"{o}}k, G., Stuchl{\'{\i}}k, Z.: {Epicyclic frequencies
of Keplerian
  motion in Kerr spacetimes}.
\newblock In: Hled{\'{\i}}k and Stuchl{\'{\i}}k
  \cite{Hle-Stu:2005:RAGtime6and7:Proceedings}, pp. 315--338

\bibitem{Ter-Stu:2005:ASTRA:}
T{\"{o}}r{\"{o}}k, G., Stuchl{\'{\i}}k, Z.: {Radial and vertical
epicyclic
  frequencies of Keplerian motion in the field of Kerr naked singularities.
  Comparison with the black hole case and possible instability of naked
  singularity accretion discs}.
\newblock Astronomy and Astrophysics \textbf{437}(3), 775--788 (2005), arXiv:astro-ph/0502127v1

\bibitem{Tor-Stu-Bak:2007:CEURJP:}
T{\"{o}}r{\"{o}}k, G., Stuchl{\'{\i}}k, Z., Bakala, P.: {A remark
about
  possible unity of the neutron star and black hole high frequency QPOs}.
\newblock Central European J. Phys. \textbf{5}(4), 457--462 (2007)

\bibitem{Zak:2003:}
Zakharov, A.F.: {The iron $K_{\alpha}$ line as a tool for analysis
of black
  hole characteristics}.
\newblock {Publications of the Astronomical Observatory of Belgrade}
  \textbf{76}, 147--162 (2003), arXiv:astro-ph/0411611v1

\bibitem{Zak-Rep:2006:}
Zakharov, A.F., Repin, S.V.: {Different types of Fe $K_{\alpha}$
lines from
  radiating annuli near black holes}.
\newblock New Astronomy \textbf{11}, 405--410 (2006), arXiv:astro-ph/0510548v1

\bibitem{Zel-Nov:1971:RelAstr:}
Ze{l'}dovich, Y.B., Novikov, I.D.: Relativistic Astrophysics,
vol.~1.
\newblock University of Chicago Press, Chicago (1971)

\end{thebibliography}

\end{document}